\documentclass[11pt]{article}

\usepackage[margin=1in]{geometry}

\usepackage{amsmath,amssymb}
\usepackage{graphicx}
\usepackage{booktabs}
\usepackage{xcolor}
\usepackage{algorithm}
\usepackage{algpseudocode}
\usepackage{caption}
\usepackage{subcaption}
\usepackage{comment}
\usepackage{tabularx}
\usepackage{array}

\usepackage[numbers,sort&compress]{natbib}
\usepackage[
    colorlinks=true,
    linkcolor=blue!50!black,
    citecolor=blue!50!black,
    urlcolor=blue!50!black,
    backref=page
]{hyperref}

\title{\LARGE\textbf{Learned Regularization for Sparse-View CT:}\\[4pt]
\large\textbf{Accuracy, Lesion Detectability and Distribution Shift of a\\
Plug-and-Play Prior against Tuned Total Variation}}

\author{Mohammed Abo Oraig \\ 
\small Independent research \\ 
\small \texttt{mohammed-abo-oraig@etud.univ-tln.fr}}

\date{}

\begin{document}
\maketitle

\begin{abstract}
Learned image priors are often proposed to replace hand-crafted regularizers in sparse-view computed tomography (CT), yet they are usually judged by PSNR on data resembling their training set. We evaluate one noise-conditioned DnCNN denoiser used as a Plug-and-Play prior against filtered back-projection (FBP), Tikhonov and total-variation (TV) regularization, with every method tuned on validation data and all differences paired. On synthetic $28\times 28$ digit images, the prior beats box-constrained TV by about $3.4$\,dB at moderate noise ($\sigma=0.05$, 20 angles), but the advantage shrinks at low noise and reverses under distribution shift: an MNIST-trained prior is $3.1$\,dB (Fashion-MNIST) and $4.3$--$4.6$\,dB (Shepp--Logan phantom) below TV, whereas a Fashion-MNIST-trained prior beats TV by $2.2$\,dB on Fashion-MNIST, consistent with prior mismatch rather than a failure of PnP itself. In a lesion-insertion test, the prior detects lesions worse than TV in distribution despite its higher PSNR, and under mismatch it produces lesion-like false blobs far more often ($0.43$ against $0.03$ per image). On real LoDoPaB-CT slices (patient-disjoint split; 30, 60 and 120 angles; three dose levels; 40 test patients), the prior does not consistently outperform a well-tuned TV in PSNR ($-0.93$ to $+0.33$\,dB) and has lower SSIM in eight of nine conditions. Its lesion-detection AUC is higher than this TV's in all nine conditions ($+0.03$ to $+0.16$; significant in six), but never significantly higher than that of Hann-filtered FBP, which the PSNR-tuned TV falls below in detection in all nine conditions (seven significant). On non-anatomical phantoms its PSNR falls below TV in all six conditions. PSNR and lesion detectability thus rank the methods differently. Our conclusions concern one small denoiser, and the CT analysis was developed after seeing results, without pre-specified primary comparisons.
\end{abstract}

\tableofcontents

\section{Introduction}
\label{sec:intro}

Reconstructing a computed tomography (CT) image from few projection angles or at low photon counts is an
ill-posed inverse problem. From measurements
\begin{equation}
y = Ax + n,
\end{equation}
where $A$ is the forward (Radon) operator and $n$ the noise, the image $x$ cannot be recovered stably when
$A$ is rank-deficient or ill-conditioned, as it is for sparse-view data, without prior knowledge about $x$.
The classical remedy is explicit regularization,
\begin{equation}
\hat{x} = \arg\min_x \; \tfrac{1}{2}\lVert Ax - y \rVert_2^2 + \lambda R(x),
\end{equation}
with a hand-crafted penalty $R$ such as Tikhonov or total variation (TV) \citep{rudin1992tv}.

Learned priors replace $R$ by statistics estimated from data. End-to-end networks, including unrolled
iterations, are trained for a given operator and noise level \citep{jin2017,adler2018}. Denoiser-based methods,
Plug-and-Play (PnP) \citep{venkatakrishnan2013plugandplay} and regularization by denoising \citep{romano2017}
(reviewed in \citep{kamilov2023}), instead insert a trained denoiser such as DnCNN \citep{zhang2017dncnn}
in place of the proximal operator of $R$ in an iterative scheme. Learned methods can be much more accurate than
classical reconstruction on large benchmarks: in the LoDoPaB-CT dataset paper, FBP followed by a U-Net improves the PSNR by several dB over FBP \citep{leuschner2021}. We study PnP with a DnCNN prior because
the prior is trained once, independently of the operator, so the same prior can be placed in the same
data-fidelity iteration as the classical baselines and tested when the geometry, the dose or the image
distribution changes. We make no claim about end-to-end networks trained for a specific task, and our
conclusions concern one small denoiser architecture.

Reconstruction methods are usually compared by PSNR and SSIM on test images drawn from the distribution of the
training set. Several concerns are known to be left open by such comparisons. Learned reconstructions can be
unstable under small perturbations \citep{antun2020instabilities} and sensitive to distribution shift
in accelerated MRI \citep{darestani2021shift}; they can contain plausible structure that is not in the object
\citep{bhadra2021}; and denoiser-based iterations with an unconstrained CNN denoiser come without convergence
guarantees \citep{ryu2019pnp,chan2016pnp}. Moreover, image quality for a diagnostic purpose is better measured
by performance at a task than by a pixel-wise error \citep{barrett1993,barrett2004}. Finally, the verdict depends
on the baseline: a regularizer whose weight and iteration count are not tuned for each condition is a weak
comparison.

This paper is an evaluation study, not a new method. It asks three questions about a PnP-DnCNN prior compared with
tuned FBP, Tikhonov and TV reconstruction:
\begin{enumerate}
\item[\textbf{Q1}] Does the learned prior keep its advantage over TV when the image distribution changes?
\item[\textbf{Q2}] Can we find cases where it creates lesion-like structure that the ground truth does not
support, as measured by a location-known local-contrast statistic?
\item[\textbf{Q3}] Do the conclusions hold over a range of noise levels and reconstruction conditions?
\end{enumerate}
We answer them at two scales. In a controlled setting of $28\times28$ digit images (MNIST-derived), noise level,
number of angles, forward-model error, denoiser seeds and the training distribution of the prior can all be varied
and repeated; the same questions are then examined with a matrix-free operator and Poisson noise. We then ask which
conclusions survive on real $362\times362$ slices of LoDoPaB-CT with a patient-disjoint split, three angle counts
and three dose levels. In every experiment the data are split into disjoint training, validation and test sets,
each method is tuned on the validation set for each condition, and differences are paired (Section~\ref{sec:method}).
Besides PSNR and SSIM, we measure the detectability of synthetic lesions inserted into the images (paired AUC,
lesion retention and false lesion-like blobs). The first digit experiments (E1--E5) use inverse-crime data; later
experiments simulate the measurements on a finer grid \citep{wirgin2004}, and
Section~\ref{sec:crime-results} quantifies the difference.

In brief: on digits a matched prior beats TV by about $3.4$\,dB at moderate noise, and a mismatched prior loses to
it by $3.1$--$4.6$\,dB. On real CT, one PnP-DnCNN prior is roughly on par with a well-tuned TV in PSNR and slightly
below it in SSIM, yet it detects small high-contrast lesions better than that TV (significantly in six of nine
conditions), which itself detects them worse than plain FBP.

Our contributions are:
\begin{itemize}
\item An evaluation protocol for comparing a PnP prior with classical regularization: per-condition tuning of
every method, a TV control run for twice the iterations (TV-long), paired statistics, and data generation on a
finer grid with a sensitivity analysis (Section~\ref{sec:crime-results}).
\item Robustness evidence on digit images: forward-model mismatch, distribution shift with matched-prior and seed
controls, the effect of the iteration count, and an error decomposition along the singular vectors of the forward
operator (Sections~\ref{sec:e3}--\ref{sec:seeds-results}).
\item A lesion-insertion evaluation with an operating-point study of whether re-tuning can restore detectability
(Sections~\ref{sec:lesion-results} and \ref{sec:oppoint-results}).
\item A matrix-free operator with exact adjoint, validated against the dense one, and a Poisson-noise study
(Sections~\ref{sec:mf-results}--\ref{sec:lesion-poisson}).
\item A patient-disjoint study on LoDoPaB-CT over three angle counts and three doses, with a simulator check and a
shift to non-anatomical phantoms (Sections~\ref{sec:e14}--\ref{sec:e19}). The study reports negative and mixed
results, an early comparison that we discarded, and the analysis steps taken after seeing results
(Section~\ref{sec:history3}); nothing was pre-specified.
\end{itemize}

Section~\ref{sec:related} reviews related work, Section~\ref{sec:method} the methods, Section~\ref{sec:experiments}
reports experiments E1--E19, and Section~\ref{sec:discussion} answers the three questions.

\section{Related Work}
\label{sec:related}

\paragraph{Variational regularization.} Tikhonov regularization and total variation (TV) \citep{rudin1992tv}
remain standard baselines for ill-posed inverse problems because they are convex, well understood and
simple to tune. TV is a strong prior for piecewise-constant images, which makes it a competitive baseline
for digit-like data and a natural reference for CT. TV has long been used for sparse-view CT~\citep{sidky2008tv}.

\paragraph{Learned CT reconstruction.} Deep networks have been used to post-process filtered
back-projection images \citep{jin2017}, to unroll iterative reconstruction \citep{adler2018}, and, more
recently, as generative priors for sparse-view CT and accelerated MRI \citep{song2022}. These methods, and
the LoDoPaB-CT benchmark results cited in the Introduction \citep{leuschner2021}, are trained for a given
operator and noise level. We instead study the Plug-and-Play family, in which the prior is trained once and
the data-fidelity step stays explicit, so that the prior can be compared with classical regularizers inside
the same iteration; our conclusions concern this family and a small denoiser, not end-to-end networks.

\paragraph{Plug-and-Play and denoiser-based priors.}
\citet{venkatakrishnan2013plugandplay} proposed replacing the proximal operator of a variational regularizer by
an off-the-shelf denoiser inside an ADMM or proximal-gradient iteration; regularization by denoising
\citep{romano2017} is a related formulation, and \citet{kamilov2023} review the field. CNN denoisers were used
as plug-in priors by \citet{zhang2017ircnn}. DnCNN \citep{zhang2017dncnn} uses residual learning for Gaussian
denoising and, in its blind form, is trained over a range of noise levels; FFDNet \citep{zhang2018ffdnet}
instead receives the noise level as an input. We use a DnCNN-style residual network with the noise level as an
additional input channel, so that one network serves every denoising strength $\sigma_d$ tried in the PnP
iteration, where $\sigma_d$ is a tuned hyperparameter. Convergence results require assumptions on the denoiser,
such as boundedness or a Lipschitz constraint \citep{chan2016pnp,ryu2019pnp}; an unconstrained CNN denoiser
comes with no such guarantee, which is relevant to the peak-then-decay behaviour we observe in
Section~\ref{sec:e5}. Convergent variants constrain the denoiser, for example as a gradient step~\citep{hurault2022gradient}, and strong deep denoisers with noise-level schedules~\citep{zhang2022dpir} are the usual practical choice; we use neither.

\paragraph{Instabilities, robustness and distribution shift.} \citet{antun2020instabilities} reported, for
MRI and CT networks, severe artefacts from tiny perturbations, missed small structural changes, and no such
instability for the sparsity-based methods (including TV) in their examples.
\citet{darestani2021shift} measured the robustness of trained and untrained reconstruction methods in
accelerated MRI and found that both suffer similarly from distribution shifts and that the better-performing
deep methods do not lose robustness. Our shifts are natural (another image class, a different kind of
phantom) rather than adversarial, our prior is a small PnP denoiser, and the shifts are larger than
within-domain variation; we report a reversal of the advantage over TV for a mismatched prior
(Sections~\ref{sec:seeds-results} and \ref{sec:discussion}), and we do not claim that it holds for other
learned methods.

\paragraph{Hallucinations, the inverse crime and task-based assessment.}
\citet{bhadra2021} define hallucinations in tomographic reconstruction as errors in image components that the
measurements cannot stably determine, which can come only from the prior; in a stylised MRI setting they
found more such errors for a U-Net on out-of-distribution data. We adopt this view and extend the decomposition
to the weakly determined subspace of a finite-precision operator (Section~\ref{sec:subspace}). Our setting
differs in the modality (CT), the prior (a PnP denoiser), the controls (a prior trained on the shifted
distribution, denoiser seeds, a longer-running TV) and the evidence (a lesion-insertion test, and real data over
a grid of doses and angle counts). Generating data with the discrete model that is later inverted is known as
the inverse crime \citep{wirgin2004,kaipio2005} and tends to overstate the accuracy of methods that exploit the
exact model; we reduce it in Section~\ref{sec:crimefree}. Finally, we follow the task-based view of image
quality \citep{barrett1993,barrett2004}: fidelity metrics such as PSNR need not predict whether a feature of
interest survives reconstruction, so we also measure the detection of inserted lesions with a location-known
statistic and count lesion-like false blobs in lesion-free reconstructions.

\section{Methodology}
\label{sec:method}

\subsection{Problem setting and experimental settings}\label{sec:settings}
We reconstruct an image $x\in\mathbb{R}^n$ from measurements $y=Ax+n$, where $A$ is a parallel-beam Radon transform with $K$ projection angles evenly spaced in $[0^\circ,180^\circ)$.
The study uses three settings, summarised in Table~\ref{tab:settings}; Stages~1 and~2 use synthetic $28\times28$ digit images so that every factor can be varied and repeated, and Stage~3 uses real CT slices.
All methods in a comparison receive the same data, the same operator and the same validation images for tuning; the number of tuned parameters differs between methods (Section~\ref{sec:protocol12}).

\begin{table}[htbp]
\centering\small\setlength{\tabcolsep}{4pt}
\caption{The three settings of the study. Datasets: MNIST~\citep{lecun1998mnist},
Fashion-MNIST~\citep{xiao2017fashion}, EMNIST letters~\citep{cohen2017emnist},
the Shepp--Logan phantom~\citep{shepp1974} and LoDoPaB-CT~\citep{leuschner2021}.}\label{tab:settings}
\begin{tabular}{p{2.3cm}p{2.6cm}p{3.6cm}p{3.2cm}p{3.0cm}}
\toprule
Setting & Images & Operator and noise & Data used & Experiments\\
\midrule
Digits, Gaussian noise (Stage 1) & $28\times28$ MNIST; Fashion-MNIST, EMNIST letters, Shepp--Logan for shift & dense Radon matrix, 40 bins, $K\in\{5,10,20,40,90\}$; additive Gaussian noise $\sigma\in\{0,0.01,0.05,0.1,0.2\}$ on the normalised sinogram & prior: MNIST training split (59\,500 images); tuning: 100 validation images; test: 200 images & E1--E10\\
Digits, Poisson noise (Stage 2) & as above & matrix-free operator (same geometry); Poisson photon noise, $I_0\in\{1000,300,100,30,10\}$ & as above & E11--E13\\
LoDoPaB-CT (Stage 3) & $362\times362$ slices & matrix-free parallel-beam, 513 bins, $K\in\{30,60,120\}$ of 1000 angles; Poisson noise, $N_0=4096$ photons per detector bin and fractions $1,\tfrac14,\tfrac1{10}$ of it & patient-disjoint: 8\,960 training slices, 20 validation slices, 40 test slices (one per patient) & E14--E19\\
\bottomrule
\end{tabular}
\end{table}

\subsection{Forward operator and classical baselines (Stages 1--2)}
For the digit images, $A\in\mathbb{R}^{m\times n}$ is stored as an explicit dense matrix: the discrete Radon transform of scikit-image~\citep{vanderwalt2014skimage} (\texttt{radon}, \texttt{circle=False}) is applied to each of the $n=784$ canonical basis images and the sinograms are stacked as columns.
The detector has 40 bins, so $m=40K$ ($800\times784$ for $K=20$). $A$ is divided by the image size $28$ so that sinogram values are of order one; the Gaussian noise standard deviation $\sigma$ is defined on this normalised sinogram, $y=Ax+\sigma\varepsilon$ with $\varepsilon\sim\mathcal{N}(0,I)$.
The dense matrix gives the exact adjoint $A^\top$, which matters for the PnP gradient step, and allows the reconstruction operator to be perturbed independently of the data-generating operator in the mismatch experiments.

\textbf{Filtered back-projection (FBP)} uses the ramp filter of scikit-image (\texttt{iradon}); it has no tunable parameter and is not regularised. \textbf{Tikhonov regularization} solves
\begin{equation}
\hat{x} = \arg\min_x \tfrac{1}{2}\lVert Ax-y\rVert_2^2 + \tfrac{\lambda}{2}\lVert x\rVert_2^2
= (A^\top A + \lambda I)^{-1} A^\top y
\end{equation}
in closed form. \textbf{Total variation} solves
\begin{equation}
\hat{x} = \arg\min_x \tfrac{1}{2}\lVert Ax - y\rVert_2^2 + \lambda \, \mathrm{TV}(x)
\end{equation}
by ADMM~\citep{boyd2011admm} (\emph{TV-ADMM}) with an isotropic TV penalty, forward differences $D$ with Neumann boundary conditions, the splitting $z=Dx$, a fixed penalty parameter $\rho=0.05$, 100 iterations and the $x$-update computed with a precomputed inverse.
The Tikhonov and TV-ADMM solutions are not constrained to $[0,1]$; all reconstructions in Stages~1--2 are clipped to $[0,1]$ before the metrics are computed.

\subsection{Learned prior: PnP proximal gradient}\label{sec:pnp}
\textbf{Denoiser.} $D_\sigma$ is a DnCNN-style~\citep{zhang2017dncnn} residual network that also receives the noise level: the input has two channels, the image and a constant map equal to $\sigma$ (as in FFDNet~\citep{zhang2018ffdnet}), followed by 8 convolutional layers of $3\times3$ kernels with 64 channels, ReLU activations and batch normalisation in the hidden layers; the output is $x-\mathrm{net}(x,\sigma)$ (residual learning).
For Stages~1--2 it is trained on clean MNIST images (the training split without 500 held-out images, 59\,500 images) corrupted by Gaussian noise with $\sigma\sim\mathcal{U}(0,\sigma_{\max})$, $\sigma_{\max}=0.3$, with the $\ell_2$ loss, Adam~\citep{kingma2015adam} (learning rate $10^{-3}$), batch size 128 and 8 epochs with a cosine schedule.
One trained network serves all PnP experiments; $\sigma_d$, the noise level passed to the network at inference, is a tuned hyperparameter that is distinct from the training noise range. At inference the network is in evaluation mode and its output is clipped to $[0,1]$.

\textbf{Iteration.} Reconstruction replaces the proximal operator of an explicit regularizer by the denoiser,
\begin{equation}
x_{t+1} = D_{\sigma_d}\big(x_t - \eta \, A^\top(Ax_t - y)\big), \qquad
\eta = 1/L(A^\top A),
\end{equation}
where $L(A^\top A)$ is the largest eigenvalue of $A^\top A$, estimated by power iteration, and the output is $x_k$ after $k$ iterations (a tuned hyperparameter). Algorithm~\ref{alg:pnp} summarises the procedure; in Stages~1--2 the initial image is the FBP reconstruction.
Both the initial image and every iterate are clipped to $[0,1]$ (the denoiser output is clipped, which acts as a box constraint). The TV-ADMM and Tikhonov baselines of Stages~1--2 have no such constraint inside the iteration, and TV has a fixed iteration count (100 ADMM iterations) while the PnP count $k$ is tuned. As a control we re-ran the key digit-image comparisons with a box-constrained TV (Chambolle--Pock~\citep{chambolle2011}, $0\le x\le 1$, 1000 iterations, $\lambda$ tuned on the same validation set; \emph{TV-box}). The control covers the noise and angle sweeps (E2b), the distribution-shift and matched-prior matrix (E4/E8) and the lesion study (E9); the other Stage~1--2 experiments (E3, E5, E7, E10 and the Stage~2 dose sweep) use TV-ADMM only. Every claim below states which TV it refers to. In Stage~3 the TV baseline is box-constrained.

\begin{algorithm}[htbp]
\caption{PnP proximal gradient (PnP-PGD)}
\label{alg:pnp}
\begin{algorithmic}[1]
\Require measurements $y$, operator $A$, denoiser $D$, strength $\sigma_d$, step $\eta$, number of iterations $k$, initial image $x_0$
\State $x_0 \gets \mathrm{clip}_{[0,1]}(x_0)$
\For{$t = 0, \dots, k-1$}
  \State $g_t \gets A^\top(Ax_t - y)$
  \State $x_{t+1} \gets \mathrm{clip}_{[0,1]}\bigl(D_{\sigma_d}(x_t - \eta \, g_t)\bigr)$
\EndFor
\State \Return $x_k$
\end{algorithmic}
\end{algorithm}

\subsection{Evaluation protocol (Stages 1--2)}\label{sec:protocol12}
Data are split into disjoint training, validation and test subsets (Table~\ref{tab:settings}). The hyperparameters, $\lambda$ for Tikhonov and TV and $(\sigma_d,k)$ for PnP, are selected by grid search maximising the mean validation PSNR, independently for each experimental condition (noise level, number of angles or dose) in E1, E2 and E12, and then evaluated once on the test set.
The grids are $\lambda_{\rm Tik}\in10^{[-5,0]}$ (21 log-spaced values), $\lambda_{\rm TV}\in10^{[-5,-1]}$ (17 values), $\sigma_d\in\{0.005,0.01,0.02,0.05,0.1,0.15,0.2,0.3\}$ (extended to 0.5 for the denoisers of Section~\ref{sec:seeds}) and $k\in\{0,1,2,3,5,10,25,50,100,200,400\}$; PnP therefore has 88 grid points against 17 for TV. The value $k=0$ is a special case that applies the denoiser once to the FBP image, without a data-fidelity step.
For the robustness experiments (E3, E4) the hyperparameters are tuned once at a nominal condition and \emph{frozen} across all mismatch and shift conditions, as in a deployed pipeline that is not retuned for every operating condition.
PSNR (peak 1) and SSIM~\citep{wang2004} (scikit-image implementation, $7\times7$ window, data range 1) are computed on the clipped reconstruction. Every condition uses one noise realisation per image (validation and test noise are drawn with different seeds).
We report the mean with a 95\% interval, $\pm1.96$ standard errors over test images, and use a two-sided paired Wilcoxon signed-rank test between PnP and TV on the same images and noise realisations. Where a table reports a paired gap with a bootstrap interval, or seed-to-seed variation as a standard deviation over denoisers, its caption says so.

\subsection{Data simulated on a finer grid}\label{sec:crimefree}
In E1--E5 the data are generated with the matrix $A$ used for reconstruction (an ``inverse crime''~\citep{wirgin2004,kaipio2005}), except where a finer-grid repeat is stated (Table~\ref{tab:e1e2gap}). For the later digit experiments (Sections~\ref{sec:crime-results}--\ref{sec:oppoint-results}) we instead simulate measurements on a finer grid, which reduces the inverse crime but does not remove it. The $28\times28$ image $x$ is upsampled by an odd integer
factor $f=5$ (to $140\times140$) with bicubic interpolation and clipped to $[0,1]$, projected with the same
discrete Radon transform at the same angles, and every $f$ adjacent detector bins are summed onto the 40-bin
detector of the reconstruction operator, giving $y=\mathcal{S}(x)+n$ with $\mathcal{S}\neq A$. Reconstructions use
$A$, and all metrics use the original $28\times28$ image as ground truth. The model error
$\|\mathcal{S}(x)-Ax\|_2$ averages $0.09$ at 20 angles ($0.05$ at 5 and $0.17$ at 90 angles), i.e.\ 6\% of the
noise norm $\sigma\sqrt{m}$, $m=40\,n_{\rm angles}$, at $\sigma=0.05$ and 31\% at $\sigma=0.01$. Because bicubic
upsampling smooths the object, we also report a nearest-neighbour variant (object unchanged, model error $0.03$) and
a bilinear variant (model error $\approx0.10$, a stronger smoothing) as a sensitivity band. All hyperparameters are
re-tuned on validation data simulated in the same way.

\subsection{Denoiser seeds, noise range and matched-prior control}\label{sec:seeds}
Besides the single MNIST denoiser of Section~\ref{sec:pnp} we train, with the same architecture and
optimiser: (i) three MNIST denoisers with $\sigma_{\max}=0.3$ (seeds 0--2);
(ii) three MNIST denoisers with $\sigma_{\max}=0.5$, which allows $\sigma_d$ up to $0.5$ and removes the grid-edge
tuning at $\sigma=0.2$ observed in E1; (iii) three denoisers trained on the Fashion-MNIST~\citep{xiao2017fashion} training split ($\sigma_{\max}=0.3$). In the matched-prior control, all methods (Tikhonov, TV, PnP) are tuned on held-out
Fashion-MNIST training images and tested on the Fashion-MNIST test split. The $\sigma_d$ grid is extended to
$\{0.005,\dots,0.3,0.4,0.5\}$ for all denoisers.

\subsection{Weakly determined subspaces and prior-driven content}\label{sec:subspace}
Let $A=U\Sigma V^{\top}$ be the SVD of the dense operator, with singular values $s_1\ge\dots\ge s_n$. For a
threshold $\tau$ let $W_\tau=\mathrm{span}\{v_i : s_i<\tau s_1\}$ and let $P_\tau$ be the orthogonal projector onto
$W_\tau$. For 20 angles the operator has an exact null space of dimension 79 (singular values at machine precision),
followed by a continuum of very small singular values with no gap; we therefore use $\tau=0.05$ as the headline
choice, giving $\dim W_{0.05}=319$ of the 784 dimensions (and report $\tau=0.10$ and the exact null space as
sensitivity checks). A gradient step $x\mapsto x-\eta A^\top(Ax-y)$ never changes the component of $x$ in the null
space of $A$, so null-space content of a PnP reconstruction is supplied by the denoiser alone (up to the clipping to $[0,1]$, which is nonlinear). For a reconstruction
$\hat x$ and ground truth $x$ we report
\begin{align}
\|\hat x-x\|_2^2 &= \|P_\tau(\hat x-x)\|_2^2+\|(I-P_\tau)(\hat x-x)\|_2^2 \quad\text{(orthogonal split)},\\
\iota &= \frac{\sum\|P_\tau\hat x\|_2^2}{\sum\|P_\tau x\|_2^2}\quad\text{(injection)},\\
c &= \frac{\sum\langle P_\tau\hat x,P_\tau x\rangle}{\sqrt{\sum\|P_\tau\hat x\|_2^2\,\sum\|P_\tau x\|_2^2}}\quad\text{(alignment)},
\end{align}
with sums over test images (pooled ratios). $\iota>1$ means that the reconstruction places more energy in
$W_\tau$ than the true images contain there; $c$ measures how much of that content points in the right direction.
These statistics cannot by themselves distinguish prior-driven structure from amplified noise, so we interpret them
only in comparisons with identical noise, data and solver settings.

\subsection{Lesion-insertion protocol}\label{sec:lesion}
A lesion is an additive Gaussian blob, $x_{\rm les}=\mathrm{clip}_{[0,1]}\bigl(x+c\,e^{-\|p-p_0\|^2/(2\sigma_l^2)}\bigr)$,
inserted at a random integer location $p_0$ of the central $16\times16$ region of a test image, with
$\sigma_l\in\{0.5,0.75,1.25,2.0\}$\,px and contrast $c\in\{0.25,0.5\}$ (eight lesion types). We reconstruct the
lesion-present image and the lesion-absent original with independent noise realizations ($\sigma=0.05$, 20 angles,
finer-grid data), using the same locations for every method. A lesion is \emph{valid} if its inserted contrast in the
ground truth is at least 50\% of nominal (a lesion placed on a saturated stroke is clipped away); invalid cases are
excluded for all methods, leaving 157--197 of 200 images per type. The observer is a location-known
statistic, the \emph{local contrast}: the mean of the image in a disc of radius $\rho=\lceil1.5\sigma_l\rceil$ minus its
mean in the surrounding annulus of width 3 px. We report
\begin{itemize}
\item the \emph{paired AUC}: the fraction of images for which the statistic is larger in the lesion-present than in the
      lesion-absent reconstruction of the same base image (the background cancels; ties count one half);
\item the \emph{retention} $R=\mathrm{E}[t(\hat x_{\rm present})-t(\hat x_{\rm absent})]/\mathrm{E}[t(x_{\rm les})-t(x)]$,
      which is $1$ if the lesion contrast is preserved and below 1 if it is attenuated;
\item the number of \emph{false lesions} per lesion-absent image: local maxima of the contrast map of the
      reconstruction whose response is at least $\theta$, half the response of the nominal lesion, and for which the
      ground truth has no contrast above $\theta/2$ within 2\,px (Section~\ref{sec:observer3} generalises $\theta$ to a
      multiple of the nominal response for the CT study); counts are compared across methods within a lesion type, not across types;
\item the lesion's \emph{observability}, the fraction of its energy outside $W_{0.05}$.
\end{itemize}
For the operating-point study (Section~\ref{sec:oppoint-results}) we sweep 45 PnP settings ($\sigma_d\in\{0.02,0.05,0.1,0.15,0.2\}$,
$k\in\{1,2,3,5,10,25,50,100,200\}$) and 9 values of the TV weight, choose operating points on 200 held-out
\emph{in-distribution} validation images (disjoint from the images used for PSNR tuning) and report them on the
test sets, so that the choice is neither made on the evaluation images nor informed by the shifted distribution.

\subsection{Matrix-free operator}\label{sec:mf}
Up to this point the forward operator $A$ was stored as a dense $800\times784$ matrix, which allows exact adjoints,
closed-form Tikhonov inverses and a full singular value decomposition, but does not scale to clinical image sizes. We
therefore implemented a matrix-free operator that is never assembled. The forward projection embeds the image in a
$40\times40$ canvas (the diagonal of the image, as in the dense operator), rotates it by each angle with bilinear
interpolation about the pixel used by the reference implementation, and sums along the detector rows. The adjoint is the
exact transpose of this discretisation, obtained by reverse-mode differentiation, so $\langle Ax,y\rangle=\langle x,A^{\top}y\rangle$
holds to floating-point precision by construction. The classical baselines use only forward
and adjoint applications: Tikhonov by conjugate gradients on $A^{\top}A+\lambda I$, and TV by the same ADMM as before with the $x$-update solved by
conjugate gradients. The singular subspaces used in Section~\ref{sec:subspace} are obtained with a randomized SVD
\citep{halko2011} (power iterations and oversampling, forward and adjoint products only); the error split in
Section~\ref{sec:nullspace-results} then uses the well-determined subspace $\{i: s_i/s_1\ge0.05\}$ and its complement.
The matrix-free operator was validated against the dense one on the digit images only (E11); the CT study of Stage~3 uses a different matrix-free parallel-beam operator (Section~\ref{sec:lodopab}).

\subsection{Poisson noise}\label{sec:poisson}
Low-dose CT measures photon counts. With $p=Ax$ the (normalised) line integrals, the expected counts are
$I=I_0\exp(-\mu p)$ with $I_0$ incident photons per ray and attenuation scale $\mu=2$; the counts are
$N\sim\mathrm{Poisson}(I)$ and the data used for reconstruction are $\hat p=-\mu^{-1}\log(\max(N,1)/I_0)$.
The noise in $\hat p$ is signal dependent, with variance approximately $e^{\mu p}/(I_0\mu^2)$, and is larger along
rays that pass through more material. We use $I_0\in\{1000,300,100,30,10\}$, which gives an effective noise
standard deviation of $0.018,0.032,0.056,0.106,0.198$ on the sinogram and therefore spans the Gaussian range
$\sigma\in[0.02,0.2]$ of the earlier experiments. The data are generated from the finer-grid object
(Section~\ref{sec:crimefree}, bicubic, factor 5), so the inverse crime is reduced as before. The reconstruction methods are
unchanged and treat $\hat p$ with an unweighted least-squares data term (a Gaussian approximation); for PnP we also
tested a variant with weights $w=N/I_0$ (an inverse-variance weighting up to a constant). All hyperparameters are tuned at each dose on 100 held-out MNIST validation images
and frozen for every test set.

\subsection{LoDoPaB-CT setting}\label{sec:lodopab}
LoDoPaB-CT~\citep{leuschner2021} provides $362\times362$ slices with values $\mu/\mu_{\max}\in[0,1]$ and simulated low-dose observations of 1000 parallel-beam angles on 513 detector bins. The ground-truth slices derive from the LIDC-IDRI database~\citep{armato2011lidc,armato2015lidcdata},
distributed through The Cancer Imaging Archive~\citep{clark2013tcia}.
The official split has 35\,820 training slices from 632 patients, 3\,522 validation slices from 60 patients and 3\,553 test slices from 60 patients; we verified that the three sets of patient identifiers
are disjoint. We use 8\,960 training slices (from the 632 training patients) to train the prior, 20 validation slices, one from each of 20 patients, for all tuning, and 40 test slices, one from each of 40
patients (the same 40 in every experiment), for all evaluation; the patient is therefore the unit of statistical analysis.
The projector is a matrix-free parallel-beam operator with a unit-tested adjoint;
its conventions (rotation sign, detector flip, image transposition, angle offset) were fixed by projecting the ground truth and comparing with the supplied observations over 16 variants: the selected one has a relative residual of $0.038$ (correlation $0.9979$), the best alternative $0.166$,
and the fitted pixel size is $1.0044$ times the nominal $0.26\,\mathrm{m}/362$. Sparse-view data use $K\in\{30,60,120\}$ evenly spaced angles out of the 1000.

\subsection{Dose model and simulation}\label{sec:sim3}
The observations follow $y=-\ln(N/N_0)/\mu_{\max}$ with $N\sim\mathrm{Poisson}(N_0e^{-\mu_{\max}Ax})$, $N_0=4096$ photons per detector bin, $\mu_{\max}=81.35858\,\mathrm{m^{-1}}$, and zero counts replaced by $0.1$.
We checked this model on the data: the counts recovered from the observations are integers in $100\%$ of the bins for $N_0=4096$ (and 8192), against $9.7$--$10.4\%$ for perturbed values of $\mu_{\max}$ (chance level $10\%$) and $24.7\%$ and $49.9\%$ for $N_0=1024$ and 2048.
The dispersion index $\overline{(N-\lambda)^2/\lambda}$ around the counts expected from the ground truth is $1.06$, $1.25$ and $1.81$ for expected counts in $[20,200)$, $[200,1000)$ and $[1000,5000)$; values above one include discretisation and ground-truth error, so the real data may contain somewhat more variance than the simulation.
Lower doses $\phi\in\{1,0.25,0.1\}$ (4096, 1024 and about 410 photons per detector bin) are obtained from the real observations by exact binomial thinning of the recovered counts (a thinned Poisson variable is Poisson with $\phi$ times the mean); these real thinned observations are used for \emph{tuning}.
For \emph{evaluation} the data are simulated without the inverse crime: the slice is up-sampled by a factor two (bicubic), projected on the $724\times724$ grid with half-size pixels and 1026 detector bins, pairs of bins are averaged, and Poisson noise is added exactly as above.
Hyperparameters are therefore tuned on real thinned data and applied to simulated data. Section~\ref{sec:e14} checks the simulator against real data for the TV baseline only.

\subsection{Methods and tuning in the CT study}\label{sec:methods3}
\emph{FBP-Hann}: filtered back-projection with the ramp filter multiplied by a Hann window whose cut-off (a fraction of Nyquist) is tuned; it is a linear regularised reconstruction. \emph{TV}: isotropic total variation with the box constraint $[0,1]$ (forward differences, Neumann boundary), solved by the primal--dual algorithm of Chambolle and Pock~\citep{chambolle2011} for 150 iterations
from the FBP-Hann image. The problem is normalised by $\lVert A\rVert^2$, i.e.\ $\tfrac{1}{2\lVert A\rVert^2}\lVert Ax-y\rVert_2^2+\lambda\,\mathrm{TV}(x)$, so TV weights are in these normalised units and are not comparable with those of Stage~1.
\emph{TV-long} is the same with 300 iterations and a tuned weight; it was added to test whether TV's lower iteration count explains the comparison with PnP (Section~\ref{sec:history3}).
\emph{PnP-const}: Algorithm~\ref{alg:pnp} with $\eta=1/L$ ($L$ by power iteration), started from the FBP-Hann image, with
$(\sigma_d,k)$ tuned; \emph{PnP-TVinit}: the same started from the TV solution. In the dose $\times$ angle study the tuned $k$ is at most 12 (Table~\ref{tab:params3}), so PnP-TVinit performs at most 162 iterations in total and PnP-const at most 12.
An iteration of each method has a different cost, and computation time was not matched.
$D_\sigma$ is the noise-level-conditioned DnCNN of Section~\ref{sec:pnp} (8 layers, 64 channels, $\sigma$ as a second input channel, residual output), here trained on the 8\,960 LoDoPaB slices for 15\,000 steps (batch 16, random $128\times128$ crops and flips, $\sigma\sim U(0,0.1)$, Adam, cosine schedule).
A 17-layer variant (20\,000 steps) was also trained and evaluated at $K=60$.
Every method and condition is tuned separately on the 20 validation slices (real observations, thinned and angle-subsampled) by maximising the mean PSNR (peak 1, reconstructions clipped to $[0,1]$).
The initial grids of the dose $\times$ angle study are cut-off $\{0.08,0.12,0.16,0.2,0.3,0.4\}$, TV weight $\{10^{-3},3\cdot10^{-3},10^{-2},3\cdot10^{-2}\}$, $\sigma_d\in\{0.01,0.02,0.03,0.05,0.08\}$ and $k\in\{1,2,4,8,12,20,30,45\}$ (the $K=60$ study used similar grids, listed in the code repository); whenever the optimum falls on the boundary of a grid, the grid is extended (up to twice; cut-off $\times0.75$ or $+0.1$, TV weight $\times3$ or $\div3$,
$\sigma_d$ $\times1.6$ or $\div1.6$). The final choices are listed in Table~\ref{tab:params3}. Two PnP-TVinit optima remain on a grid boundary after the extensions, and one tuned $\sigma_d$ ($0.128$) exceeds the training range of the denoiser $(0,0.1)$; results from these cells are flagged where they are used.

\subsection{Lesion observer}\label{sec:observer3}
The protocol of Section~\ref{sec:lesion} is applied to $362\times362$ slices. Lesions are additive Gaussian blobs, clipped to $[0,1]$, with $\sigma_l\in\{2,4\}$\,px ($1.4$ and $2.9$\,mm at $0.72$\,mm per pixel) and contrast $c=0.06$ on the $\mu/\mu_{\max}$ scale
(in the step at $K=60$ also $c=0.03$). If water is taken to be about $0.25$ on this scale, which is an assumption, $c=0.06$ and $0.03$ correspond to roughly $240$ and $120$\,HU; the lesions are small and noisy at low dose but are not low-contrast in the clinical sense.
Four lesions of one type are inserted into each slice at random positions at least $10\sigma_{l,\max}+4=44$\,px apart, inside the body eroded by 22\,px (CT slices) or in the central square (phantoms); the lesion-free and lesion-present images are reconstructed from independent noise realisations.
A lesion is valid if its inserted response is at least half the nominal one. The observer is the location-known, zero-sum local-contrast statistic of Section~\ref{sec:lesion} (disc of radius $\lceil1.5\sigma_l\rceil$ minus a 3-px annulus), and the detection score is the paired AUC pooled over valid lesions (up to $40\times4=160$ lesions per lesion type on CT slices, $20\times4=80$ on phantoms).
A \emph{false blob} is a local maximum (in a $5\times5$ window) of the response map of a lesion-\emph{free} reconstruction whose response is at least $\gamma\,r_{\rm nom}$ ($r_{\rm nom}$: response to the nominal lesion on an empty background; $\gamma\in\{0.5,0.75,1,1.5\}$) and for which the truth has no response above $0.25\,r_{\rm nom}$ within 2\,px.
$\gamma=1$ and $1.5$ count structures that a reader of this statistic would take for at least a full lesion. The statistic measures lesion-like response at the scale of the kernel; we did not run a reader study, so ``false blob'' refers only to this response. Because the LoDoPaB ground truth is itself a reconstruction, ``false'' means unsupported by this reference, not absent from the patient.

\subsection{Statistics (Stage 3)}\label{sec:stats3}
All comparisons are paired: two methods are evaluated on the same patients and the same simulated data, and the difference is summarised by its mean over patients with a $95\%$ percentile bootstrap interval over patients (2\,000 resamples in the $K=60$ lesion experiments, 5\,000 elsewhere), with the same resamples for both methods.
Lesion types are pooled with equal weight; per-type rows are exploratory. For the shift comparison, the CT and phantom differences are bootstrapped independently (different images) and their difference is the \emph{shift effect}.
No analysis was designated primary in advance (Section~\ref{sec:history3}), and no correction for multiplicity is applied, so all intervals are descriptive: we report the numbers of cells with significant positive and negative differences
and read the sign pattern across conditions rather than any single interval. Each condition uses one noise realisation, which limits the precision of a single cell (for AUC about $\pm0.05$); the same 40 patients, and the same 20 phantoms, are used in all conditions, so conditions are not independent.

\subsection{Analysis history}\label{sec:history3}
The analysis was developed in steps after seeing results, and some steps changed the conclusions; none of the choices below was pre-specified.
(i)~A first comparison at $K=60$ (tuning on 6 validation slices, Hann cut-off $0.2$) found PnP $0.85$ and $0.33$\,dB below TV at full and quarter dose.
Re-tuning on 20 slices (cut-off $0.3$ at full dose) raised PnP-const by $1.06$\,dB ($30.68\to31.74$) and $0.32$\,dB ($29.40\to29.72$) but TV by only $0.38$ and $0.09$\,dB, and the gap became $-0.17$ and $-0.10$\,dB (Table~\ref{tab:e15}); the initial gap was an artefact of the starting image and the small tuning set, and we do not use it.
(ii)~The lesion experiments were first tuned on 10 slices with narrow grids; they were repeated with 20 slices, extended grids and a separately tuned TV-long. At full dose the parameters of FBP, TV, TV-long and PnP-const did not change (the PSNR of PnP-TVinit changed by $0.07$\,dB) and no pooled verdict changed; at quarter dose one of 20 pooled verdicts changed (PnP-TVinit $-$ TV-long AUC, from significant to not significant).
(iii)~The PnP variants of Section~\ref{sec:e15} (PnP-TVinit, annealing, TV+denoise, the 17-layer denoiser) were introduced after PnP-const first fell below TV.
(iv)~Paired statistics, the threshold sweep and the false-blob definition at several strengths were added after seeing the first lesion results, and the dose $\times$ angle grid and the shift extension to $K=30,120$ were run afterwards.

\section{Experiments}
\label{sec:experiments}
\subsection{E1--E2: In-distribution accuracy under noise and undersampling}\label{sec:e1e2}
\paragraph{Setting.}
Test images are 200 MNIST digits; all hyper-parameters are tuned on 100 separate validation images (Section~\ref{sec:protocol12}). The
noise sweep fixes 20 projection angles and varies $\sigma\in\{0,0.01,0.05,0.1,0.2\}$; the angle sweep fixes
$\sigma=0.05$ and uses $5,10,20,40,90$ angles. Each cell is one noise realisation. Tables~\ref{tab:e1e2} and~\ref{tab:e1e2gap} first use data generated with the reconstruction
operator itself (``inverse crime''), the setting of the original study; Table~\ref{tab:e1e2gap} then repeats the PnP--TV
comparison with data simulated on a finer grid (bicubic, factor~5; the default finer-grid setting, Section~\ref{sec:crime-results}).
Two total-variation baselines are reported: \emph{TV-ADMM} is the unconstrained ADMM solver used throughout Stages~1--2, and
\emph{TV-box} is the same functional solved with the constraint $0\le x\le1$ by Chambolle--Pock (1000 iterations, $\lambda$ re-tuned on the validation images).
TV-box is the fairer comparison, because the PnP iterates are clipped to $[0,1]$ at every iteration; it is a control added after the main experiments (E2b), and the other
Stage~1--2 experiments (E3, E5, E7, E10 and the Stage~2 dose sweep) were run against TV-ADMM only.

\begin{table}[htbp]
\centering\small\setlength{\tabcolsep}{5pt}
\caption{Test PSNR (dB), mean over 200 images, data generated with the reconstruction operator (inverse crime). Bold: highest mean in the row (not a significance statement; see Table~\ref{tab:e1e2gap} for paired intervals).
$\dagger$: the TV-box optimum lies below the initial $\lambda$ grid ($\lambda_{\mathrm{norm}}=3.7\times10^{-7}$, i.e.\ almost no regularisation), so $42.5$\,dB
is a data-model artefact and should not be read as a property of TV.}\label{tab:e1e2}
\begin{tabular}{lcccccc}
\toprule
Condition & FBP & Tikhonov & TV-ADMM & TV-box & PnP-DnCNN \\
\midrule
$\sigma{=}0$, 20 angles     & 24.2 & 28.9 & 36.0 & \textbf{42.5}$^\dagger$ & 29.6 \\
$\sigma{=}0.01$, 20 angles  & 23.1 & 24.1 & 29.0 & \textbf{29.8} & 29.1 \\
$\sigma{=}0.05$, 20 angles  & 15.5 & 17.1 & 20.1 & 20.7 & \textbf{24.0} \\
$\sigma{=}0.1$, 20 angles   & 10.4 & 14.7 & 16.9 & 17.2 & \textbf{19.9} \\
$\sigma{=}0.2$, 20 angles   & 6.6  & 12.9 & 14.3 & 14.6 & \textbf{15.3} \\
\midrule
5 angles, $\sigma{=}0.05$   & 9.1  & 14.2 & 15.5 & 16.5 & \textbf{17.7} \\
10 angles, $\sigma{=}0.05$  & 12.1 & 15.9 & 18.4 & 18.9 & \textbf{21.4} \\
40 angles, $\sigma{=}0.05$  & 18.1 & 18.2 & 21.8 & 22.4 & \textbf{25.6} \\
90 angles, $\sigma{=}0.05$  & 20.6 & 20.0 & 24.1 & 24.6 & \textbf{27.1} \\
\bottomrule
\end{tabular}
\end{table}

\begin{table}[htbp]
\centering\small\setlength{\tabcolsep}{4pt}
\caption{PnP$-$TV PSNR gap (dB) with 95\% bootstrap intervals over the 200 test images; positive means PnP is better. Left: data generated with the reconstruction operator (the PSNR values of Table~\ref{tab:e1e2}).
Right: data simulated on a finer grid, which reduces the inverse crime. Hyper-parameters are re-tuned for each data model. $\dagger$: see Table~\ref{tab:e1e2}.}\label{tab:e1e2gap}
\begin{tabular}{lcccc}
\toprule
 & \multicolumn{2}{c}{Inverse crime} & \multicolumn{2}{c}{Finer grid (bicubic)}\\
\cmidrule(lr){2-3}\cmidrule(lr){4-5}
Condition & vs TV-ADMM & vs TV-box & vs TV-ADMM & vs TV-box \\
\midrule
$\sigma{=}0$, 20 angles    & $-6.43$ [$-6.70,-6.18$] & $-12.93$ [$-13.30,-12.57$]$^\dagger$ & $-3.29$ [$-3.47,-3.10$] & $-5.13$ [$-5.32,-4.94$] \\
$\sigma{=}0.01$, 20 angles & $+0.07$ [$-0.07,+0.21$] & $-0.75$ [$-0.89,-0.60$] & $+1.02$ [$+0.87,+1.16$] & $+0.35$ [$+0.21,+0.49$] \\
$\sigma{=}0.05$, 20 angles & $+3.86$ [$+3.70,+4.02$] & $+3.33$ [$+3.19,+3.47$] & $+4.02$ [$+3.85,+4.18$] & $+3.48$ [$+3.34,+3.63$] \\
$\sigma{=}0.1$, 20 angles  & $+3.01$ [$+2.79,+3.24$] & $+2.65$ [$+2.43,+2.88$] & $+3.10$ [$+2.88,+3.33$] & $+2.73$ [$+2.51,+2.96$] \\
$\sigma{=}0.2$, 20 angles  & $+1.03$ [$+0.78,+1.29$] & $+0.65$ [$+0.40,+0.89$] & $+1.08$ [$+0.83,+1.33$] & $+0.69$ [$+0.45,+0.93$] \\
\midrule
5 angles, $\sigma{=}0.05$  & $+2.13$ [$+1.86,+2.41$] & $+1.14$ [$+0.90,+1.39$] & $+2.36$ [$+2.07,+2.64$] & $+1.35$ [$+1.10,+1.61$] \\
10 angles, $\sigma{=}0.05$ & $+3.03$ [$+2.81,+3.26$] & $+2.51$ [$+2.28,+2.75$] & $+3.29$ [$+3.09,+3.50$] & $+2.77$ [$+2.56,+2.98$] \\
40 angles, $\sigma{=}0.05$ & $+3.83$ [$+3.69,+3.97$] & $+3.26$ [$+3.13,+3.39$] & $+3.97$ [$+3.82,+4.11$] & $+3.39$ [$+3.26,+3.52$] \\
90 angles, $\sigma{=}0.05$ & $+3.04$ [$+2.90,+3.18$] & $+2.57$ [$+2.44,+2.69$] & $+3.24$ [$+3.10,+3.39$] & $+2.79$ [$+2.67,+2.91$] \\
\bottomrule
\end{tabular}
\end{table}

\begin{figure}[htbp]
\centering
\includegraphics[width=0.95\textwidth]{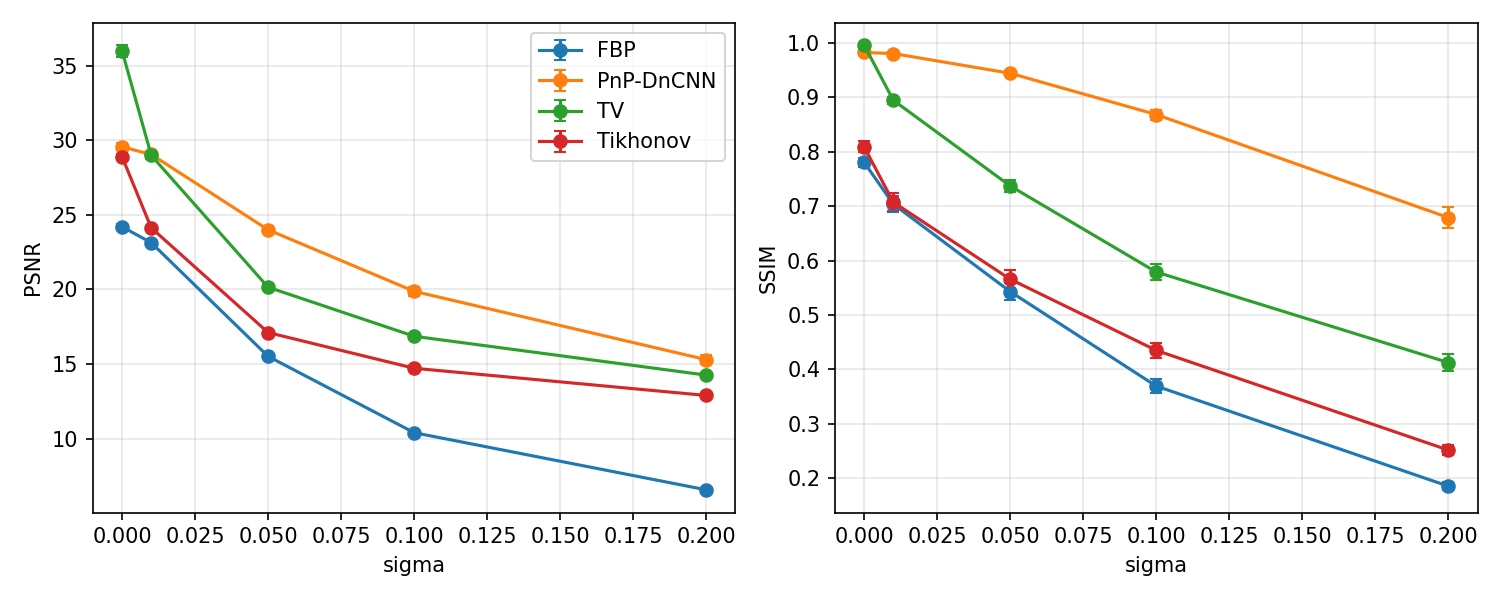}
\caption{Reconstruction accuracy (PSNR, left; SSIM, right) against the measurement noise $\sigma$ at 20 projection angles, inverse-crime data, TV-ADMM as the TV baseline. At $\sigma=0$ TV is clearly best in PSNR; at $\sigma=0.01$ TV and PnP-DnCNN are within noise of each other in PSNR (Table~\ref{tab:e1e2gap}), while PnP has the higher SSIM; for $\sigma\ge0.05$ PnP-DnCNN is best in both. Error bars are the mean $\pm1.96$ standard errors over 200 test images and reflect image-to-image variability only, not tuning or training variability.}
\label{fig:comparison_noise}
\end{figure}

\begin{figure}[htbp]
\centering
\includegraphics[width=0.95\textwidth]{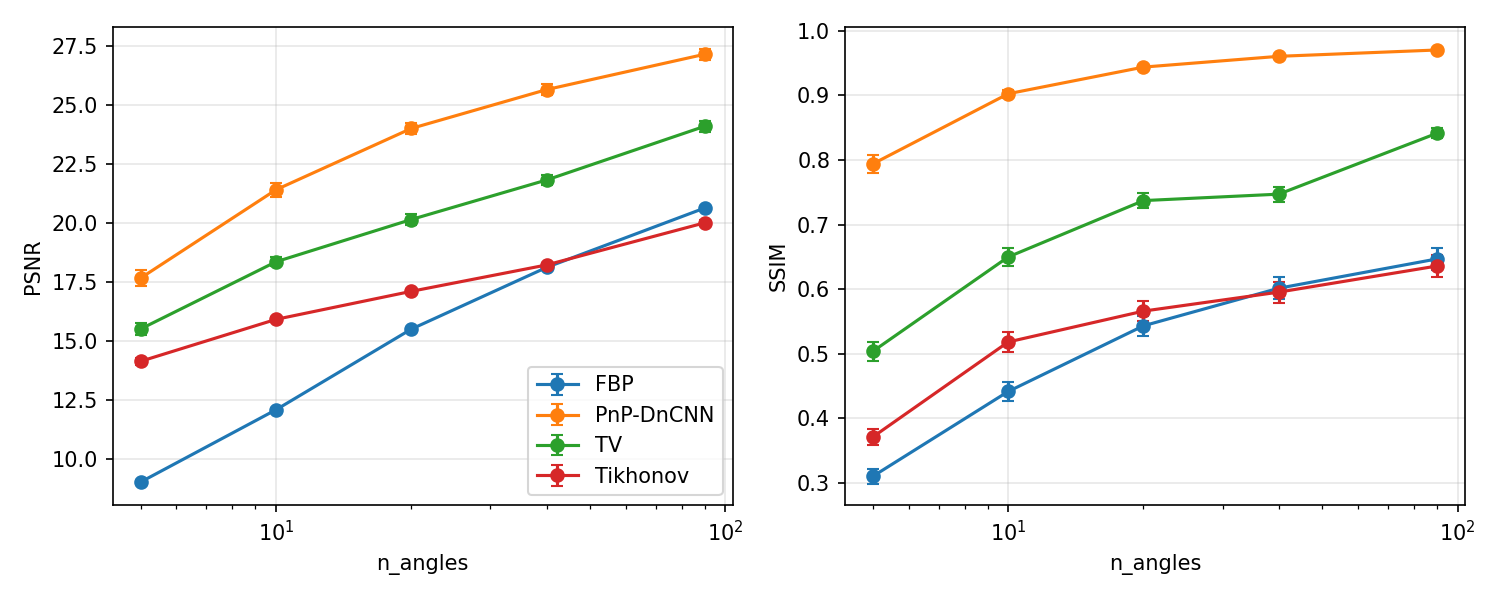}
\caption{As Figure~\ref{fig:comparison_noise}, against the number of projection angles at $\sigma=0.05$ (logarithmic axis). PnP-DnCNN is best at every angle count tested.}
\label{fig:comparison_angles}
\end{figure}

\begin{figure}[htbp]
\centering
\includegraphics[width=0.95\textwidth]{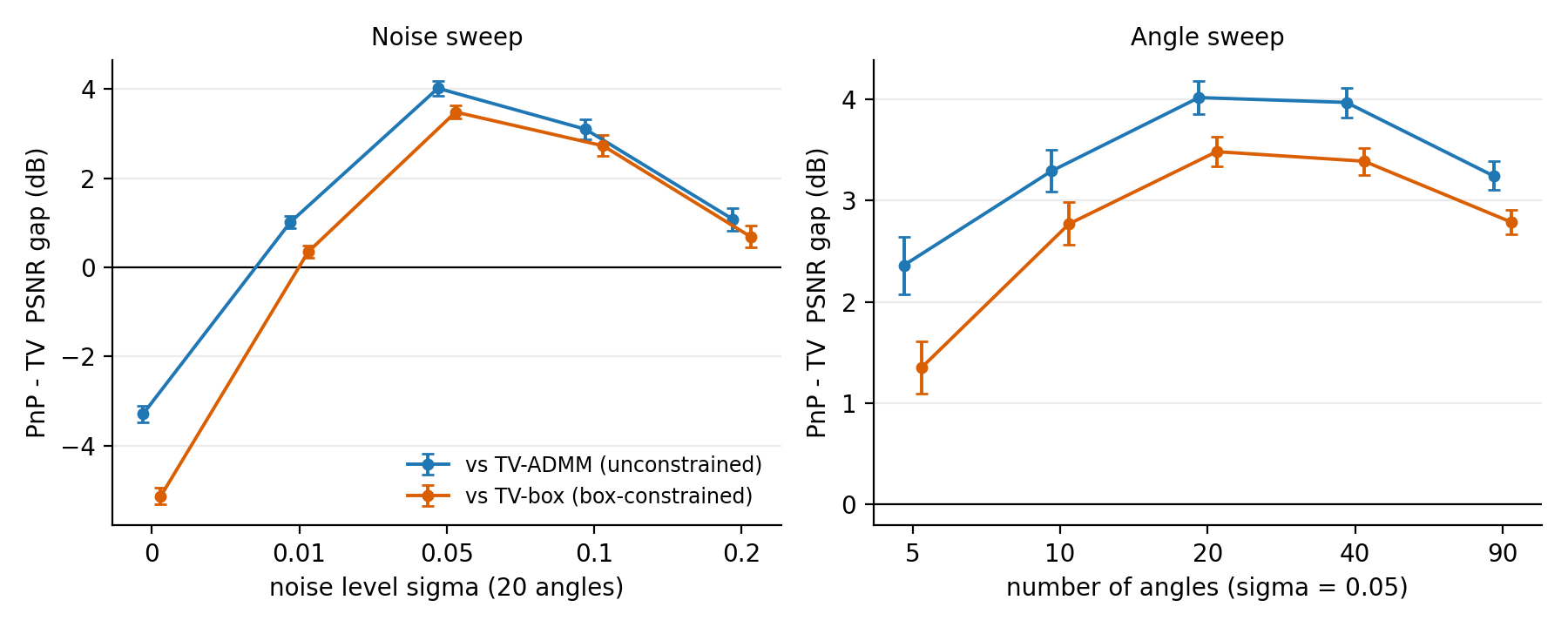}
\caption{PnP$-$TV PSNR gap on finer-grid (bicubic) data against the unconstrained TV-ADMM and the box-constrained TV-box. Bars are 95\% bootstrap intervals over 200 images. The box constraint lowers the PnP advantage by $0.4$--$1.0$\,dB at $\sigma\ge0.05$ and, in the angle sweep, most at 5 angles; on these data it changes no sign (on inverse-crime data it does at $\sigma=0.01$, Table~\ref{tab:e1e2gap}).}
\label{fig:e1e2gap}
\end{figure}

\paragraph{Results.}
\emph{(i) Where the learned prior wins.} For $\sigma\ge0.05$ PnP-DnCNN has the highest PSNR at every angle count and noise level tested, and the gap
over TV is significantly positive against both TV variants and on both data models (Table~\ref{tab:e1e2gap}). The gap is largest at $\sigma=0.05$ with 20--40 angles:
$+3.8$ to $+3.9$\,dB against TV-ADMM, but $+3.3$\,dB against TV-box on inverse-crime data and $+3.4$ to $+3.5$\,dB on finer-grid data. We take the box-constrained, finer-grid value as the headline
($\approx3.5$\,dB for this denoiser; $3.4$\,dB averaged over three independently trained denoisers in E8). Against TV-ADMM, PnP has the higher PSNR in 100\% of the 200 images at $\sigma=0.05$ with 20, 40 and 90 angles, in 96\% at $\sigma=0.1$, but only
in 68\% at $\sigma=0.2$ and 87\% at 5 angles, where the average gap is small ($+1.0$ and $+2.1$\,dB). Against TV-box the gap shrinks fastest in the hardest regimes: from $+2.1$ to $+1.1$\,dB at 5 angles and
from $+1.03$ to $+0.65$\,dB at $\sigma=0.2$ (inverse-crime data).

\emph{(ii) SSIM.} The SSIM gain is larger than the PSNR gain suggests (TV-ADMM $0.74\to$ PnP $0.94$ at $\sigma=0.05$ with 20 angles; $0.50\to0.79$ at 5 angles; $0.41\to0.68$ at $\sigma=0.2$) and PnP has the higher SSIM, significantly so (paired Wilcoxon), in every condition except $\sigma=0$, where TV is significantly higher ($0.996$ vs.\ $0.982$). We did not compute SSIM for TV-box. SSIM is later shown to be lower for PnP on real CT (Stage~3); we attribute the high SSIM here to the match between
the denoiser's training distribution and the test digits, but did not test this.

\emph{(iii) The noise-free and low-noise end depends on the data model and the TV solver.} At $\sigma=0$ the learned prior is below TV in every variant ($-6.4$\,dB on inverse-crime data, $-3.3$\,dB finer-grid against TV-ADMM; larger still against TV-box).
At $\sigma=0.01$ the sign of the PnP--TV gap changes with these choices: not distinguishable from zero against TV-ADMM on inverse-crime data ($+0.07$\,dB), $-0.75$\,dB against TV-box on the same data, and $+0.35$\,dB against TV-box on finer-grid data.
No claim about PnP versus TV at $\sigma\le0.01$ should therefore be made from E1--E2 alone; Section~\ref{sec:crime-results} analyses the data-model dependence.

\emph{(iv) Classical baselines.} TV (either variant) is the best classical method in every condition, and Tikhonov is above FBP except at 90 angles (Table~\ref{tab:e1e2}). A possible reason for TV's advantage is that digit images are close to piecewise constant; we did not test this.

\paragraph{Caveats.}
(a)~Each cell uses one noise realisation, a single denoiser and one tuning run; the intervals in Table~\ref{tab:e1e2gap} describe the variability across test images only. The only measure of
training variability is E8 (three denoisers), where the in-distribution gap varies by $0.03$\,dB between seeds, small compared with the gaps above.
(b)~PnP has two tuned parameters (denoiser level $\sigma_d$ and iteration count $k$, $5\le k\le100$ across conditions; 88 grid points) against one for TV (17 grid points), and early stopping is itself a regulariser. The TV iteration count is fixed and not tuned (100 for TV-ADMM, 1000 for TV-box), so the comparison is between a tuned few-iteration PnP and a TV with a fixed, larger iteration count; TV-box addresses the box constraint but not the difference in tuning degrees of freedom.
(c)~All numbers are for $28\times28$ digits whose training distribution matches the test distribution by construction; what happens when it does not is the subject of E4 and E8.

\begin{figure}[htbp]
\centering
\includegraphics[width=0.95\textwidth]{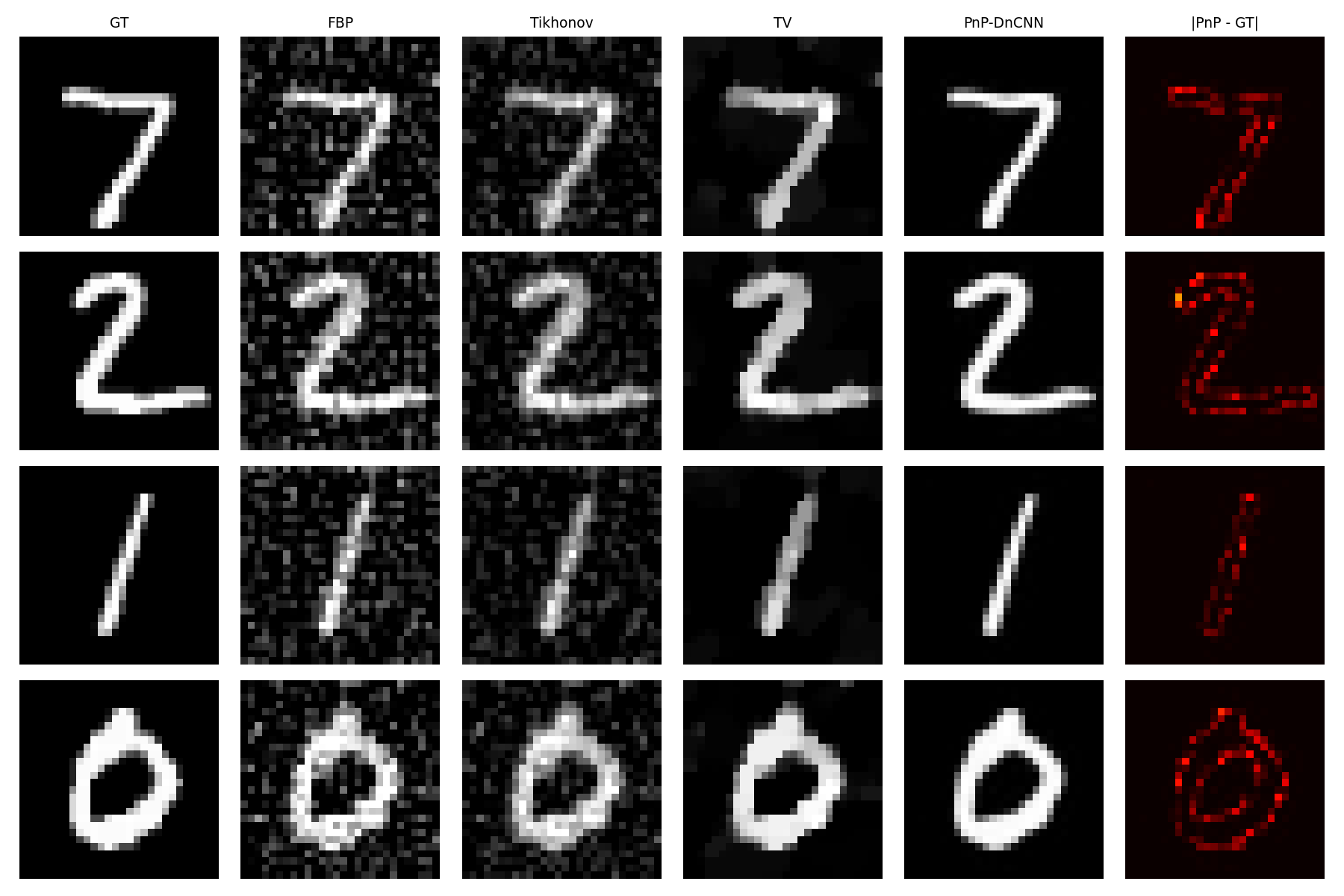}
\caption{In-distribution reconstructions at $\sigma=0.05$, 20 angles (inverse-crime data, one noise draw). Rows are the first four test images (no selection). Columns: ground truth, FBP, Tikhonov, TV-ADMM, PnP-DnCNN and the absolute error $|\hat{x}_{\mathrm{PnP}}-x|$ (brighter is larger). FBP and Tikhonov retain visible noise; TV-ADMM gives flat regions with reduced stroke intensity; PnP-DnCNN is visually closest to the ground truth. The error maps show the residual concentrated along stroke boundaries. A TV-box column and per-image PSNR values are not shown.}
\label{fig:recon_grid}
\end{figure}

\subsection{E3: Forward-model mismatch}\label{sec:e3}
\paragraph{Setting.}
Measurements are generated with the true 20-angle operator at $\sigma=0.05$ (inverse-crime data, 200 test digits), and the reconstruction uses an operator whose projection angles are
perturbed by $\delta\in\{0,0.5,1,2,4\}^\circ$ (the angular spacing is $9^\circ$). Two perturbations are used: a \emph{constant offset}, $\theta_j\to\theta_j+\delta$, and \emph{per-angle jitter},
$\theta_j\to\theta_j+\delta\,\varepsilon_j$ with $\varepsilon_j\sim\mathcal{N}(0,1)$, so that $\delta$ is the standard deviation, not a bound. One jitter vector is drawn per level (fixed seed) and shared by all images and methods.
All hyper-parameters are frozen at their values for the nominal, matched operator (TV-ADMM $\lambda=5.6\times10^{-3}$; PnP $\sigma_d=0.1$, $k=5$; Tikhonov $\lambda=1.8\times10^{-2}$) and are not re-tuned under mismatch.
The TV baseline is the unconstrained TV-ADMM; E3 was not repeated with TV-box.

\begin{table}[htbp]
\centering\small\setlength{\tabcolsep}{4pt}
\caption{Test PSNR (dB) under forward-model mismatch with hyper-parameters frozen at their nominal values, and the PnP$-$TV gap (mean $\pm1.96$ standard errors of the paired per-image difference, 200 images).
PnP has the higher PSNR in 100\% of the images in every row except $\delta=4^\circ$ (96\% for the offset, 99\% for the jitter).}
\label{tab:e3}
\begin{tabular}{lcccccc}
\toprule
 & \multicolumn{3}{c}{Constant offset} & \multicolumn{3}{c}{Per-angle jitter ($\delta$ = s.d.)}\\
\cmidrule(lr){2-4}\cmidrule(lr){5-7}
$\delta$ & TV & PnP & PnP$-$TV & TV & PnP & PnP$-$TV \\
\midrule
$0^\circ$   & 20.15 & 24.00 & $+3.86\pm0.16$ & 20.15 & 24.00 & $+3.86\pm0.16$ \\
$0.5^\circ$ & 20.13 & 23.91 & $+3.78\pm0.17$ & 20.14 & 23.99 & $+3.85\pm0.17$ \\
$1^\circ$   & 20.04 & 23.62 & $+3.58\pm0.16$ & 20.10 & 23.93 & $+3.83\pm0.16$ \\
$2^\circ$   & 19.71 & 22.61 & $+2.90\pm0.16$ & 19.99 & 23.65 & $+3.66\pm0.16$ \\
$4^\circ$   & 18.56 & 20.05 & $+1.49\pm0.14$ & 19.19 & 22.18 & $+2.98\pm0.19$ \\
\midrule
Loss at $4^\circ$ & $1.6$ & $4.0$ & & $1.0$ & $1.8$ & \\
\bottomrule
\end{tabular}
\end{table}

\begin{figure}[htbp]
\centering
\includegraphics[width=0.95\textwidth]{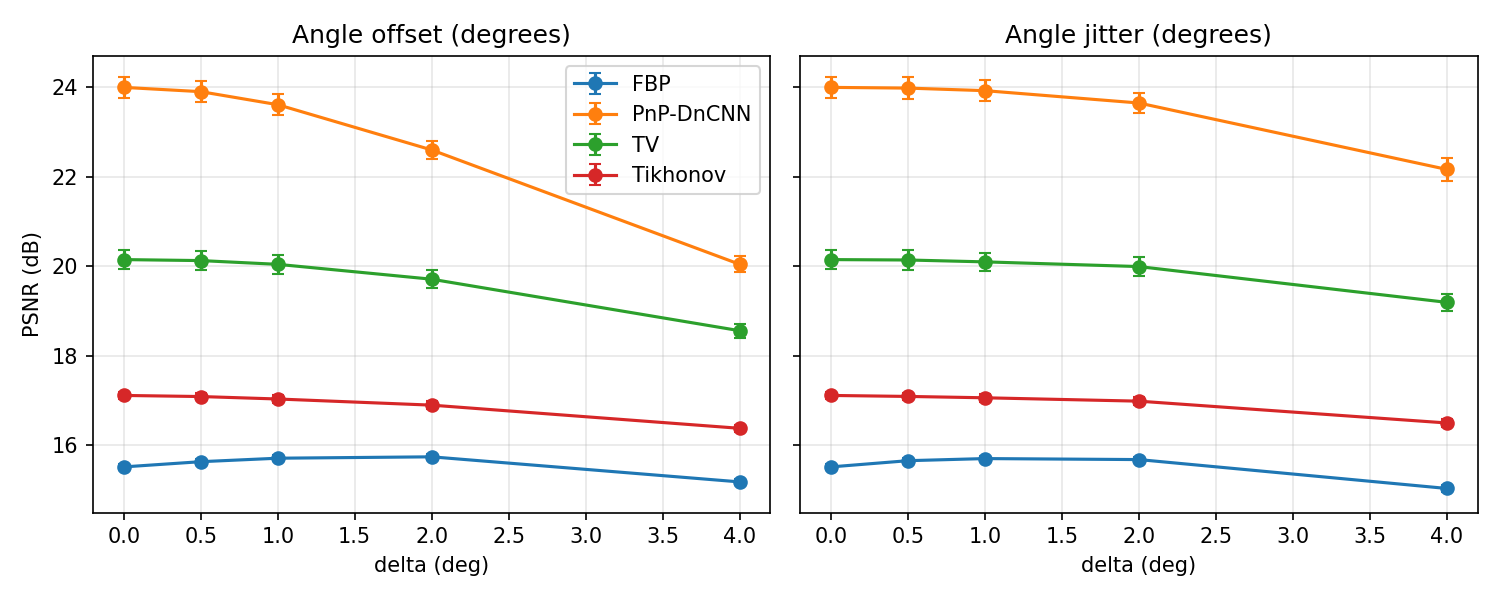}
\caption{Test PSNR under forward-model mismatch as a function of the angular perturbation $\delta$ (degrees): constant offset (left) and per-angle jitter with standard deviation $\delta$ (right), TV-ADMM as the TV baseline.
Hyper-parameters are frozen at their nominal values and are not re-tuned. PnP-DnCNN has the highest mean PSNR at every level, and loses more than TV in absolute terms under both perturbations; the offset curve
mainly reflects a rigid rotation of the reconstruction (Table~\ref{tab:e3rot}). Error bars: mean $\pm1.96$ standard errors over 200 test images, which reflect image variability only (a single jitter draw is used).}
\label{fig:robust_mismatch}
\end{figure}

\paragraph{A constant offset is a rotation.}
Data generated at the angles $\theta_j$ are consistent with an operator at the angles $\theta_j+\delta$ if the image is rotated by $\delta$ about the centre. A reconstruction that is perfect up to this rotation is therefore
the best a method can do, yet a pixel-wise metric against the true image penalises it. To separate the two effects we scored each reconstruction against the true digit and against the true digit rotated by $\delta$ (Table~\ref{tab:e3rot};
the opposite direction gives lower agreement). Against the rotated digit PnP keeps its nominal accuracy (24.1--24.2\,dB at all offsets, against 24.0\,dB at $\delta=0$), so the 4.0\,dB loss in Table~\ref{tab:e3} is
registration error, not a loss of reconstruction quality; TV also loses nothing against the rotated digit (it scores 20.4--21.0\,dB against 20.15\,dB nominal; part of this is because the interpolated reference is smoother than the true digit).
The sharper reconstruction has more to lose under a pixel-wise metric: even a perfect reconstruction rotated by $4^\circ$ would score only 22.1\,dB, and PnP's 20.05\,dB is within about 2\,dB of this ceiling.

\begin{table}[htbp]
\centering\small\setlength{\tabcolsep}{6pt}
\caption{Constant offset: PSNR (dB) of the reconstruction against the true digit and against the true digit rotated by $\delta$ (cubic interpolation). The last column is the PSNR of the true digit rotated by $\delta$ against itself, the best score a perfect but rotated reconstruction can reach.
Computed with \texttt{run\_e3\_rotation\_check.py}; the nominal values reproduce the stored E3 results.}\label{tab:e3rot}
\begin{tabular}{lccccc}
\toprule
 & \multicolumn{2}{c}{TV} & \multicolumn{2}{c}{PnP} & \\
\cmidrule(lr){2-3}\cmidrule(lr){4-5}
$\delta$ & vs true & vs rotated & vs true & vs rotated & Ceiling \\
\midrule
$0^\circ$ & 20.15 & 20.15 & 24.00 & 24.00 & $\infty$ \\
$1^\circ$ & 20.04 & 20.36 & 23.62 & 24.14 & 35.9 \\
$2^\circ$ & 19.71 & 20.66 & 22.61 & 24.23 & 29.0 \\
$4^\circ$ & 18.56 & 20.99 & 20.05 & 24.14 & 22.1 \\
\bottomrule
\end{tabular}
\end{table}

\paragraph{Results.}
\emph{(i) Per-angle jitter is the genuine inconsistency.} With a jitter of $4^\circ$ (s.d.) PnP loses $1.8$\,dB and TV $1.0$\,dB; the advantage falls from $+3.86$ to $+2.98$\,dB and PnP stays ahead on 99\% of the images.
\emph{(ii) Constant offset.} In raw PSNR PnP loses $4.0$\,dB against $1.6$\,dB for TV and the advantage falls to $+1.49$\,dB (still positive on 96\% of the images). Table~\ref{tab:e3rot} shows that this is a registration effect, so it
should not be read as evidence that PnP is the less robust method to a rigid geometric error; it is evidence that PSNR penalises a sharp but shifted reconstruction more.
\emph{(iii) What is not shown.} Neither method collapses within the range tested, but both TV-ADMM and the PnP parameters are frozen, so the experiment measures the sensitivity of nominally tuned methods, not what a re-tuned method would achieve.

\paragraph{Caveats.}
(a)~One jitter draw per level and one noise realisation; the intervals describe image variability only. (b)~Only the projection angles are perturbed (no detector offset, scale or beam-hardening errors), at $\sigma=0.05$,
on inverse-crime data. (c)~Frozen hyper-parameters: a mismatch acts like extra noise, so re-tuning might favour a larger $\sigma_d$ for PnP and a larger $\lambda$ for TV; we did not test this. (d)~TV-ADMM is unconstrained; the TV-box control
was not run for E3.

\subsection{E4: Distribution shift}\label{sec:e4}
\paragraph{Setting.}
The denoiser is trained on MNIST digits only. The complete pipeline, with hyper-parameters tuned once on the MNIST validation images at $\sigma=0.05$ and 20 angles (TV-ADMM $\lambda=5.6\times10^{-3}$; PnP $\sigma_d=0.1$, $k=5$) and then frozen, is applied to 200 Fashion-MNIST~\citep{xiao2017fashion} test images, 200 EMNIST~\citep{cohen2017emnist} letters, and a $28\times28$ Shepp--Logan phantom~\citep{shepp1974}. The phantom is a single image, so its 20 test cases are 20 independent noise draws and its
interval describes noise variability only. Data are generated with the reconstruction operator (inverse crime) with a different noise realisation from E1, so the MNIST row differs slightly from Table~\ref{tab:e1e2}. The labels ``near'' and ``far''
shift are our a priori judgement; no distance between the distributions was computed.

\begin{table}[htbp]
\centering\small\setlength{\tabcolsep}{4pt}
\caption{Test PSNR and SSIM under distribution shift with the pipeline frozen at its MNIST values (one MNIST-trained denoiser, TV-ADMM). Gap $=$ PnP$-$TV PSNR, mean $\pm1.96$ standard errors of the paired difference
(over images; over noise draws for the phantom); ``wins'' is the share of test cases in which PnP has the higher PSNR.}
\label{tab:e4}
\begin{tabular}{lccccccc}
\toprule
 & \multicolumn{3}{c}{PSNR (dB)} & & \multicolumn{2}{c}{SSIM} \\
\cmidrule(lr){2-4}\cmidrule(lr){6-7}
Test set & TV & PnP & gap (wins) & & TV & PnP \\
\midrule
MNIST (in-distribution)       & 20.27 & \textbf{23.97} & $+3.70\pm0.16$ (100\%) & & 0.742 & \textbf{0.944} \\
EMNIST letters (near shift)   & 20.09 & \textbf{22.03} & $+1.94\pm0.18$ (92\%)  & & 0.842 & \textbf{0.938} \\
Fashion-MNIST (far shift)     & \textbf{19.61} & 16.57 & $-3.05\pm0.31$ (11\%) & & 0.680 & 0.675 \\
Shepp--Logan (far shift)      & \textbf{19.19} & 15.03 & $-4.16\pm0.19$ (0\%)  & & \textbf{0.722} & 0.517 \\
\bottomrule
\end{tabular}
\end{table}

\begin{figure}[htbp]
\centering
\includegraphics[width=0.8\textwidth]{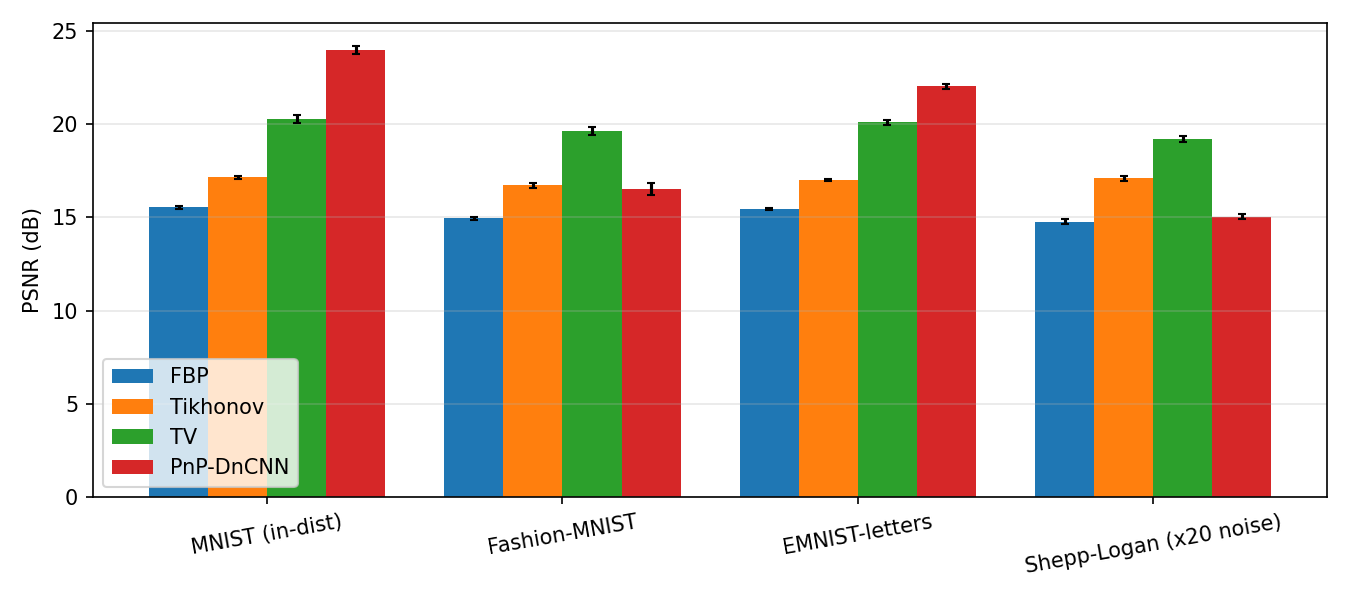}
\caption{Test PSNR under distribution shift for the four methods with the pipeline frozen at its MNIST values. PnP-DnCNN leads on MNIST and on EMNIST letters, which share the stroke-like statistics of digits,
but is below TV on Fashion-MNIST ($-3.0$\,dB) and on the phantom ($-4.2$\,dB). Against Tikhonov it is $0.2$\,dB below on Fashion-MNIST and $2.1$\,dB below on the phantom; against FBP it is $1.6$\,dB above on Fashion-MNIST and only $0.3$\,dB
above on the phantom. Error bars: mean $\pm1.96$ standard errors; for the phantom they describe 20 noise draws of one image, not variation between images.}
\label{fig:robust_shift}
\end{figure}

\begin{table}[htbp]
\centering\small\setlength{\tabcolsep}{4pt}
\caption{Control with a box-constrained TV (TV-box, $0\le x\le1$, 1000 iterations, $\lambda$ tuned on the MNIST validation images). Three independently trained MNIST denoisers (E8), bicubic finer-grid data,
$\sigma=0.05$, 20 angles. Mean test PSNR over the three denoisers and gap $=$ PnP$-$TV, mean $\pm$ s.d.\ over denoisers.}\label{tab:e4box}
\begin{tabular}{lccccc}
\toprule
Test set & TV-ADMM & TV-box & PnP & PnP$-$TV-ADMM & PnP$-$TV-box \\
\midrule
MNIST          & 20.21 & 20.69 & 24.05 & $+3.84\pm0.03$ & $+3.36\pm0.03$ \\
EMNIST letters & 20.28 & 20.35 & 22.07 & $+1.79\pm0.09$ & $+1.72\pm0.09$ \\
Fashion-MNIST  & 19.42 & 19.45 & 16.32 & $-3.10\pm0.30$ & $-3.13\pm0.30$ \\
Shepp--Logan   & 19.28 & 19.50 & 14.96 & $-4.33\pm0.12$ & $-4.55\pm0.12$ \\
\bottomrule
\end{tabular}
\end{table}

\begin{figure}[htbp]
\centering
\includegraphics[width=0.6\textwidth]{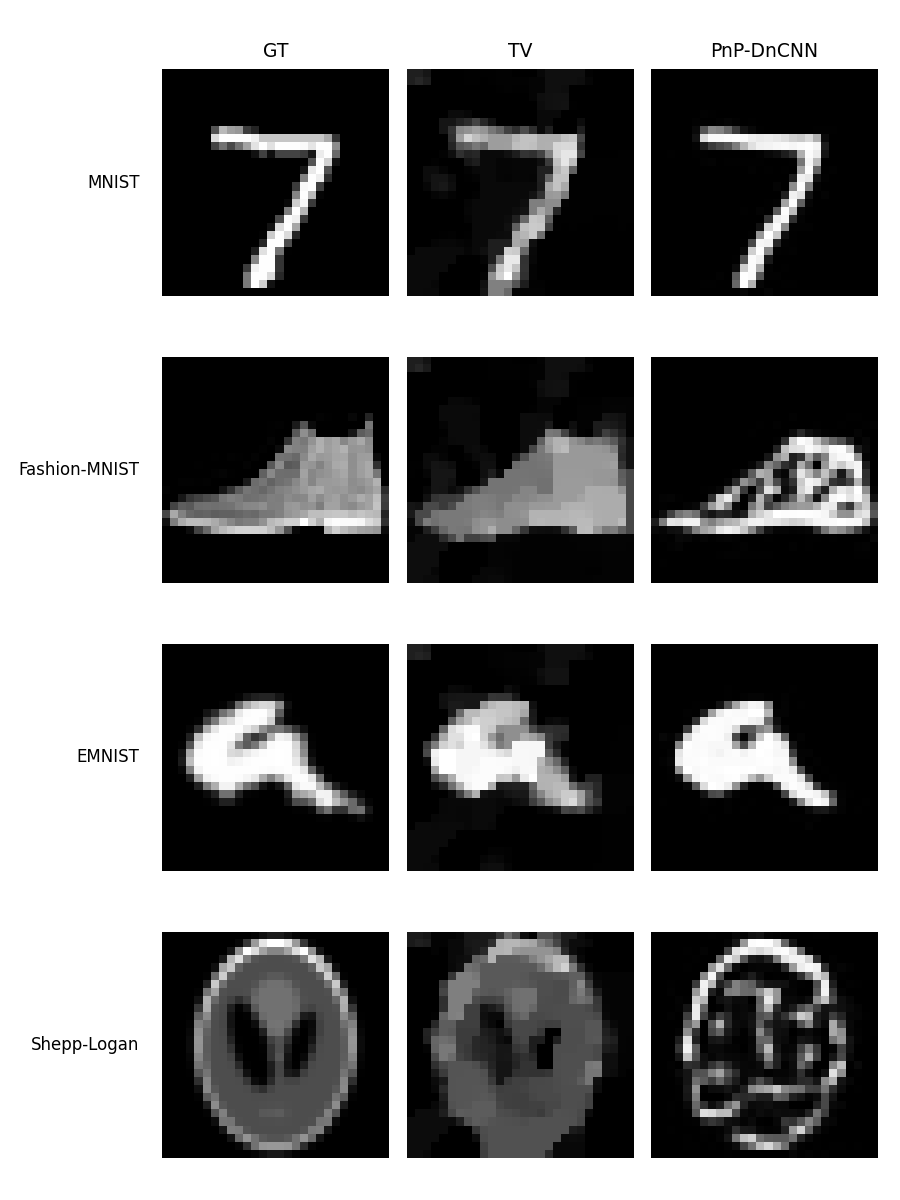}
\caption{Reconstructions at $\sigma=0.05$, 20 angles, with the frozen pipeline: the first test image of each set (no selection), one noise draw, TV-ADMM and the single MNIST-trained PnP denoiser of Table~\ref{tab:e4}.
On Fashion-MNIST and the phantom the PnP output contains spotted or stroke-like fragments that are absent from the ground truth; in these examples the TV output is blurred and blocky but does not introduce such structure. A TV-box column and PSNR values are not shown.}
\label{fig:shift}
\end{figure}

\paragraph{Results.}
\emph{(i) The advantage does not transfer to far-shifted images.} On EMNIST letters PnP keeps about half of its in-distribution advantage ($+1.9$ against $+3.7$\,dB). On Fashion-MNIST and the phantom it is below TV, in 89\% and 100\% of the test cases respectively (Table~\ref{tab:e4}).
The reversal is not an artefact of the unconstrained TV: on the data of Table~\ref{tab:e4box}, the gaps are $-3.10$ and $-4.33$\,dB against TV-ADMM and $-3.13$ and $-4.55$\,dB against the box-constrained TV, slightly larger because the box constraint helps TV on the phantom by $0.2$\,dB and on Fashion-MNIST by $0.03$\,dB. The in-distribution advantage, in contrast,
loses $0.5$\,dB to the box constraint ($+3.84\to+3.36$\,dB).
\emph{(ii) Variability.} Across three independently trained MNIST denoisers the Fashion-MNIST gap has a standard deviation of $0.30$\,dB and the phantom gap $0.12$\,dB (Table~\ref{tab:e4box}); the single denoiser of Table~\ref{tab:e4} gives similar values ($-3.05$ and $-4.16$\,dB, on a different data model) and the sign never changes.
\emph{(iii) PSNR and SSIM disagree on Fashion-MNIST.} PnP is $3.1$\,dB below TV in PSNR but equal in SSIM ($0.675$ against $0.680$), while on the phantom both are clearly lower ($0.517$ against $0.722$). We did not examine why; a plausible reading is that the PnP error on Fashion-MNIST is
mainly in smooth intensity levels that SSIM tolerates, which is untested.
\emph{(iv) Structure rather than blur.} In Figure~\ref{fig:shift}, TV fails by blurring, whereas PnP adds structure not present in the image. This is a qualitative observation on one image per set. It is quantified with inserted lesions in
Section~\ref{sec:lesion-results}: for the MNIST-trained prior on Fashion-MNIST, PnP produces $0.38$ false lesions per lesion-absent image, against $0.14$ for TV-box and $0.05$ for TV-ADMM (means over three denoisers and eight lesion types).
The converse experiment, in which a prior trained on Fashion-MNIST is tested on Fashion-MNIST, is in Section~\ref{sec:seeds-results}.

\paragraph{Caveats.}
(a)~The pipeline is frozen at MNIST-tuned values. The deficit therefore combines a mismatched prior with mismatched hyper-parameters, and we did not test whether re-tuning $\sigma_d$ and $k$ on the target set would reduce it; the matched-prior control of E8 changes the prior and the tuning set together.
(b)~The phantom is one image, so the ``far shift'' evidence on it is one object; Fashion-MNIST supplies the only far-shift evidence over many images. (c)~Table~\ref{tab:e4} uses inverse-crime data and Table~\ref{tab:e4box} finer-grid data; the conclusions agree, but the numbers are not interchangeable.
(d)~Only three datasets are used, so no claim about how the deficit scales with the distance between distributions can be made. MNIST digits are close to binary, whereas Fashion-MNIST images and the phantom contain intermediate grey levels, so the shifts confound structural and intensity statistics; we did not separate the two.

\subsection{E5: Behaviour of PnP beyond the selected iteration count}\label{sec:e5}
\paragraph{Setting.}
Validation tuning selects few iterations: $k=5$ at $\sigma\le0.05$ with 20 or 90 angles, $10$ at $\sigma=0.1$ or 40 angles, $25$ at $\sigma=0.2$ or 10 angles, and $100$ at 5 angles (E1--E2). To see what happens beyond the selected $k$ we
ran the tuned pipeline ($\sigma=0.05$, 20 angles, inverse-crime data, 200 test digits) with $\sigma_d\in\{0.02,0.05,0.1,0.2\}$ and recorded the iterate at $k\in\{1,2,5,10,25,50,100,200,400\}$ (so the location of a peak is resolved only to this grid).
For each iterate we computed the PSNR, the data residual $\lVert Ax_k-y\rVert$ relative to the noise norm $\sigma\sqrt{m}$, and the squared error in bands of the singular values $s_i$ of $A$ relative to the largest, $s_i/s_1$ (the decomposition of Section~\ref{sec:subspace}).
The tuned setting is $(\sigma_d,k)=(0.1,5)$.

\begin{figure}[htbp]
\centering
\includegraphics[width=\textwidth]{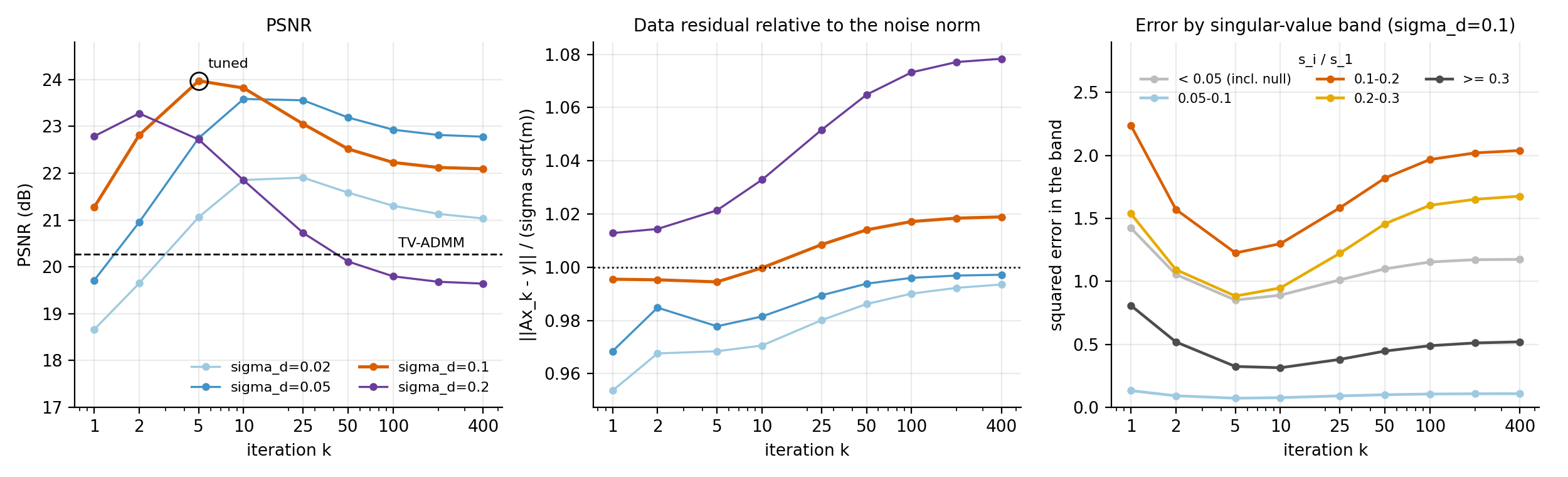}
\caption{PnP iterates on 200 MNIST test digits ($\sigma=0.05$, 20 angles). Left: PSNR against the iteration count $k$ for four denoiser strengths; the circle marks the tuned setting and the dashed line is TV-ADMM.
Middle: data residual relative to the expected noise norm $\sigma\sqrt{m}$. Right: squared error in bands of $s_i/s_1$ for $\sigma_d=0.1$ (the bands contain 319, 46, 253, 129 and 37 of the 784 directions, in the order of the legend). Every point is one run with one noise draw.}
\label{fig:convergence}
\end{figure}

\paragraph{Results.}
\emph{(i) Early stopping is necessary at every $\sigma_d$ tested.} For $\sigma_d=0.1$ the PSNR peaks at $k=5$ ($23.97$\,dB) and falls to $22.09$\,dB at $k=400$ (Figure~\ref{fig:convergence}, left). The decay is stronger for larger $\sigma_d$ ($\sigma_d=0.2$: $23.28$\,dB at $k=2$ to $19.64$\,dB at $k=400$, below TV's $20.27$\,dB) and weaker for smaller ones. Among the four $\sigma_d$ the best converged
setting (chosen on the test images) is $\sigma_d=0.05$ at $22.78$\,dB: $1.2$\,dB below the early-stopped optimum, but still $2.5$\,dB above TV. Early stopping is therefore worth about 1\,dB here and is not what makes PnP better than TV in distribution.

\emph{(ii) The data residual stays at the noise level.} For $\sigma_d=0.1$ the ratio $\lVert Ax_k-y\rVert/(\sigma\sqrt m)$ moves from $0.994$ at the selected iterate to $1.019$ at $k=400$ (at most $1.08$ for $\sigma_d=0.2$) (Figure~\ref{fig:convergence}, middle). The iterates are
data-consistent to within a few percent of the noise level at every $k$, and the residual changes by 2.5\% between the best and the converged iterate, so a discrepancy-principle stopping rule could not pick $k=5$ from it. The prior therefore does not increasingly override the data term: the iteration converges to a point that fits the data as well as the early iterates do but has a larger error.

\emph{(iii) Where the error grows.} Between $k=5$ and $k=400$ ($\sigma_d=0.1$) the squared error rises from $3.36$ to $5.52$. Of this increase of $2.16$, $74\%$ lies in the directions with $0.1\le s_i/s_1<0.3$ (382 of 784 directions), $17\%$ in the directions below $0.1$ (including the exact null space), $10\%$ in $0.3\le s_i/s_1<0.5$, and none in the directions with $s_i/s_1\ge0.5$, where the error does not
grow (percentages rounded; Figure~\ref{fig:convergence}, right, shows the bands $<0.05$, $0.05$--$0.1$, $0.1$--$0.2$, $0.2$--$0.3$ and $\ge0.3$, so the split of the last band is from the stored data). Without the denoiser, the gradient step with $\eta=1/L$ would contract the distance between the iterate and the data-fitting (least-squares) solution in direction $i$ by $(1-(s_i/s_1)^2)^k$ (Section~\ref{sec:nullspace-results}),
i.e.\ by $0.82$ for $s_i/s_1=0.2$ after 5 iterations, $0.36$ after 25 and $0.02$ after 100. The data-fitting solution contains the noise, so these directions are those in which further iterations would start to fit it, at about the iterations where the PSNR declines (10 to 100); this is what classical semi-convergence of iterative regularisation predicts. It is consistent with the data, not a test of the mechanism. In particular, the tuned $k$ is also $5$ at $\sigma=0$ (E1), so noise fitting cannot be the only cause; we did not record the iteration curve on noise-free data.

\begin{table}[htbp]
\centering\small\setlength{\tabcolsep}{4pt}
\caption{PSNR (dB) of PnP at the tuned setting ($\sigma_d=0.1$, $k=5$), run to $k=400$ with the same $\sigma_d$, and at the best $(\sigma_d,k)$ over the stored grid ($\sigma_d\in\{0.02,\dots,0.2\}$, $k\in\{1,\dots,400\}$), selected on the test images themselves and therefore optimistic. Same frozen pipeline, noise draws and TV-ADMM as E4; the phantom is one image with 20 noise draws.}\label{tab:e5}
\begin{tabular}{lccccc}
\toprule
Test set & TV-ADMM & PnP $k=5$ & PnP $k=400$ & best $(\sigma_d,k)$ & FBP \\
\midrule
MNIST (in-distribution) & 20.27 & 23.97 & 22.09 & 23.97 \,(0.1, 5) & 15.53 \\
EMNIST letters          & 20.09 & 22.03 & 20.11 & 22.03 \,(0.1, 5) & 15.47 \\
Fashion-MNIST           & 19.61 & 16.57 & 13.80 & 17.86 \,(0.2, 1) & 14.94 \\
Shepp--Logan            & 19.19 & 15.03 & 12.40 & 16.36 \,(0.1, 1) & 14.77 \\
\bottomrule
\end{tabular}
\end{table}

\emph{(iv) Under shift, stopping early limits the damage but cannot remove it.} On the two shifted sets the PSNR is highest at $k=1$, and run to convergence PnP is $5.8$\,dB (Fashion-MNIST) and $6.8$\,dB (phantom) below TV and below FBP (Table~\ref{tab:e5}). Even the best setting of the grid stays $1.8$ and $2.8$\,dB below TV.
The in-distribution benefit of early stopping and the shift deficit may share a cause, namely that more iterations let the prior shape the result further, but this is an interpretation and was not tested.

\paragraph{Caveats.}
(a)~Single noise draw, one denoiser, four $\sigma_d$ values, and checkpoints rather than every iteration, so the peak is located only to the checkpoint grid (a per-iteration run on 50 images, not shown, peaked at about $k=5$ for $\sigma_d=0.1$). (b)~The band decomposition is reported for $\sigma_d=0.1$ only. (c)~The ``best'' setting in Table~\ref{tab:e5} is chosen on test data.
(d)~Inverse-crime data and TV-ADMM only. Section~\ref{sec:nullspace-results} gives the error decomposition on finer-grid data and the iteration sweep on inverse-crime and bilinear data (Table~\ref{tab:nullspace-k}); no bicubic iteration sweep was run.

\subsection{E6: Sensitivity to the inverse crime}\label{sec:crime-results}
\paragraph{Setting.}
E1--E5 generate the data with the reconstruction matrix $A$. Here the measurements are instead simulated on a five-times finer grid with three object resamplings (nearest, bicubic, bilinear; Section~\ref{sec:crimefree}), and compared with the inverse-crime data. The mean model error
$\lVert\mathcal{S}(x)-Ax\rVert_2$ is $0$, $0.03$, $0.09$ and $0.10$ for the inverse crime, nearest, bicubic and bilinear data (20 angles); for comparison, the noise norm $\sigma\sqrt m$ is $0.28$, $1.41$, $2.83$ and $5.66$ for $\sigma=0.01,0.05,0.1,0.2$.
With finer-grid data the case $\sigma=0$ is therefore not noise-free: the only error in the measurements is the model error. Hyper-parameters of all methods are re-tuned for each data model. The PnP$-$TV gaps below are against TV-ADMM unless stated; there is one noise draw and one denoiser per cell.

\begin{table}[htbp]
\centering\small\setlength{\tabcolsep}{5pt}
\caption{PnP$-$TV PSNR gap (dB), 20 angles, 200 test digits, for four data models. TV is TV-ADMM except in the last two rows, which use the box-constrained TV-box (E2b control, only run for the inverse-crime and bicubic data).
$\dagger$: the TV-box optimum lies below the initial $\lambda$ grid (Table~\ref{tab:e1e2}).}\label{tab:crime}
\begin{tabular}{lccccccc}
\toprule
Data model & Model error & $\sigma=0$ & $0.01$ & $0.05$ & $0.1$ & $0.2$\\
\midrule
Same operator (inverse crime)            & 0     & $-6.43$ & $+0.07$ & $+3.86$ & $+3.01$ & $+1.03$\\
Fine grid, nearest                       & 0.03  & $-5.86$ & $+0.16$ & $+3.86$ & $+3.01$ & $+1.02$\\
Fine grid, bicubic (default)             & 0.09  & $-3.29$ & $+1.02$ & $+4.02$ & $+3.10$ & $+1.08$\\
Fine grid, bilinear                      & 0.10  & $-0.97$ & $+1.76$ & $+3.94$ & $+2.87$ & $+0.85$\\
\midrule
Inverse crime, vs TV-box                 & 0     & $-12.93^\dagger$ & $-0.75$ & $+3.33$ & $+2.65$ & $+0.65$\\
Bicubic, vs TV-box                       & 0.09  & $-5.13$ & $+0.35$ & $+3.48$ & $+2.73$ & $+0.69$\\
\bottomrule
\end{tabular}
\end{table}

\begin{table}[htbp]
\centering\small\setlength{\tabcolsep}{5pt}
\caption{PnP$-$TV-ADMM PSNR gap (dB) in the angle sweep ($\sigma=0.05$) for the four data models. The mean model error of the bicubic data is $0.05$, $0.07$, $0.09$, $0.12$, $0.17$ at $5,10,20,40,90$ angles.}\label{tab:crimeang}
\begin{tabular}{lccccc}
\toprule
Data model & 5 & 10 & 20 & 40 & 90 \\
\midrule
Same operator (inverse crime) & $+2.13$ & $+3.03$ & $+3.86$ & $+3.83$ & $+3.04$ \\
Fine grid, nearest            & $+2.19$ & $+3.04$ & $+3.86$ & $+3.84$ & $+3.06$ \\
Fine grid, bicubic (default)  & $+2.36$ & $+3.29$ & $+4.02$ & $+3.97$ & $+3.24$ \\
Fine grid, bilinear           & $+2.17$ & $+3.23$ & $+3.94$ & $+4.04$ & $+3.57$ \\
\bottomrule
\end{tabular}
\end{table}

\begin{table}[htbp]
\centering\small\setlength{\tabcolsep}{5pt}
\caption{As Table~\ref{tab:e1e2}, with measurements simulated on the finer grid (bicubic, $f=5$); mean test PSNR (dB) over 200 digits. The PnP$-$TV-ADMM gaps are the bicubic rows of Tables~\ref{tab:crime} and~\ref{tab:crimeang}; intervals and TV-box results are in Table~\ref{tab:e1e2gap}.}\label{tab:table1fine}
\begin{tabular}{lcccc}
\toprule
Condition & FBP & Tikhonov & TV-ADMM & PnP-DnCNN \\
\midrule
$\sigma=0$, 20 angles    & 24.3 & 27.4 & 33.2 & 29.9 \\
$\sigma=0.01$, 20 angles & 23.2 & 23.9 & 28.4 & 29.4 \\
$\sigma=0.05$, 20 angles & 15.5 & 17.1 & 20.1 & 24.1 \\
$\sigma=0.1$, 20 angles  & 10.4 & 14.7 & 16.8 & 19.9 \\
$\sigma=0.2$, 20 angles  &  6.6 & 12.9 & 14.3 & 15.3 \\
5 angles, $\sigma=0.05$  &  9.1 & 14.1 & 15.5 & 17.8 \\
10 angles, $\sigma=0.05$ & 12.1 & 15.9 & 18.2 & 21.5 \\
40 angles, $\sigma=0.05$ & 18.1 & 18.2 & 21.8 & 25.8 \\
90 angles, $\sigma=0.05$ & 20.7 & 20.0 & 24.1 & 27.3 \\
\bottomrule
\end{tabular}
\end{table}

\begin{figure}[htbp]
\centering
\includegraphics[width=\linewidth]{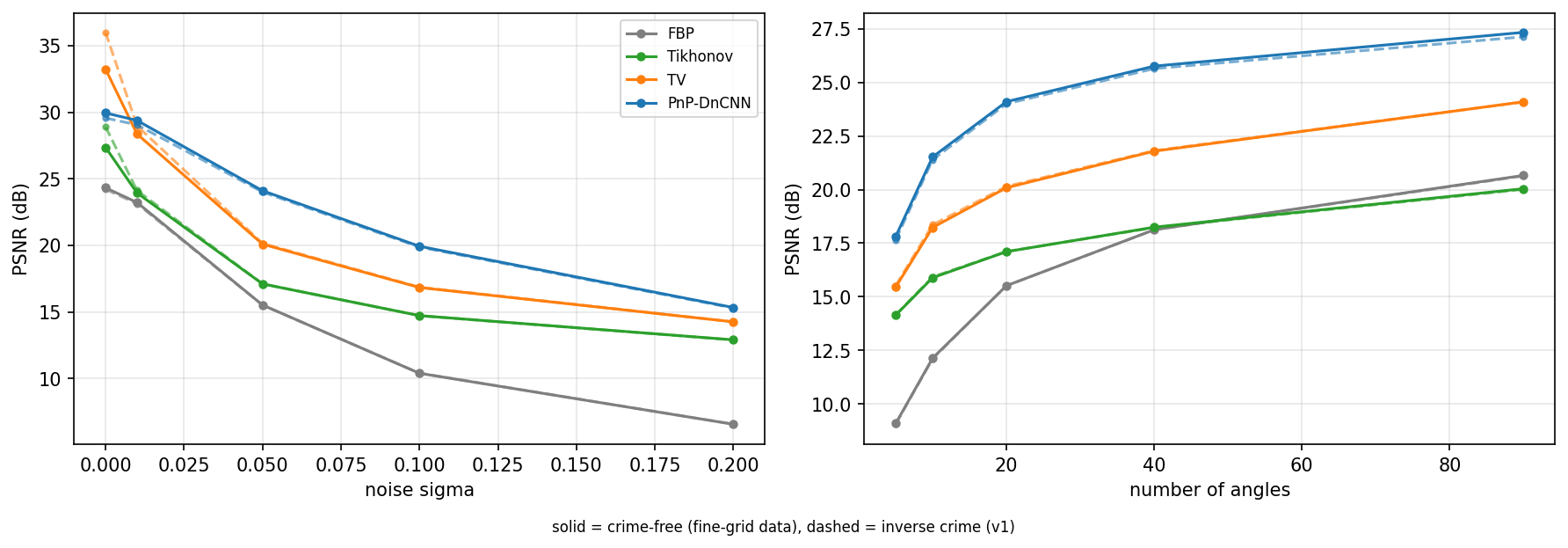}
\caption{PSNR versus noise level (left) and number of angles (right) for data simulated on the bicubic finer grid (solid) and with the inverse crime (dashed), TV-ADMM as the TV baseline. The curves nearly coincide for $\sigma\ge0.05$ and separate at $\sigma\le0.01$, where TV loses $2.8$\,dB at $\sigma=0$ ($36.0\to33.2$\,dB) while PnP gains $0.3$\,dB.}\label{fig:crime}
\end{figure}

\paragraph{Results.}
\emph{(i) The conclusions at $\sigma\ge0.05$ do not depend on the data model.} The model error is at most about 7\% of the noise norm there, and the PnP$-$TV-ADMM gap differs between the inverse-crime and the bicubic data by at most $0.16$\,dB in the noise sweep (Table~\ref{tab:crime}) and $0.26$\,dB in the angle sweep (Table~\ref{tab:crimeang}); between the inverse-crime and the bilinear data it differs by up to $0.53$\,dB (90 angles).
At $\sigma\ge0.05$ the gap against TV-box is about $0.4$ to $0.5$\,dB smaller than against TV-ADMM on both data models but keeps its sign (Table~\ref{tab:crime}, last rows).

\emph{(ii) The low-noise end depends on the data model, and also on the TV solver.} At $\sigma=0$ TV is ahead of PnP in every variant, by $6.4$\,dB (inverse crime), $5.9$ (nearest), $3.3$ (bicubic) and $1.0$\,dB (bilinear); against TV-box the lead is larger ($12.9$\,dB, with the TV-box optimum below the $\lambda$ grid, and $5.1$\,dB). At $\sigma=0.01$ the PnP lead against TV-ADMM ranges from $+0.07$ to $+1.76$\,dB across the four data models, and against TV-box it is $-0.75$\,dB (inverse crime) or $+0.35$\,dB (bicubic).
The ranking at $\sigma\le0.01$ is therefore not a single number: against TV-ADMM it changes with the data model by up to $5.5$\,dB at $\sigma=0$, and at $\sigma=0.01$ it changes sign with the choice of TV solver and data model. We make no claim about PnP versus TV at $\sigma\le0.01$ from the digit experiments. A possible reading is that TV, tuned for $\sigma\approx0$, fits the data almost exactly and is penalised when the data contain model error, whereas PnP is not; we did not test this.

\emph{(iii) Out-of-distribution deficits persist.} With the pipeline of E4 the PnP$-$TV-ADMM gap is $-3.05$ (inverse crime), $-3.04$, $-2.82$ and $-3.88$\,dB on Fashion-MNIST, and $-4.16$, $-4.16$, $-4.21$ and $-5.37$\,dB on the phantom, for the four data models in the order of Table~\ref{tab:crime}; the in-distribution EMNIST advantage stays between $+1.8$ and $+1.9$\,dB. The deficits are largest for the bilinear data, whose object is the smoothest.

\emph{(iv) The default data model.} We use the bicubic data as the default finer-grid setting for the later experiments. The choice was made after seeing all four variants, because its low-noise results lie between those of the nearest and the bilinear data; it is a convention, not a derived optimum, and the results at $\sigma\le0.01$ should be read as the band spanned by the variants.

\paragraph{Caveats.}
(a)~One noise draw and one denoiser per cell; hyper-parameters are re-tuned for each data model, so part of the differences between rows is tuning. (b)~The model error is a deterministic resampling difference, not the unknown mismatch of a real scanner; the real-data setting is addressed in Stage~3 (Section~\ref{sec:e14}).
(c)~TV-box was run for two of the four data models; the other experiments of this section use TV-ADMM.

\subsection{E7: Error decomposition along the singular vectors}\label{sec:nullspace-results}
\paragraph{Setting.}
We apply the decomposition of Section~\ref{sec:subspace} to the shift setting of E4 on finer-grid (bicubic) data: the MNIST-trained denoiser with the pipeline tuned on MNIST and frozen, $\sigma=0.05$, 20 angles, 200 test images per set (the phantom: 20 noise draws of one image).
The weakly determined subspace is $W_{0.05}$ ($s_i<0.05\,s_1$, $\dim W_{0.05}=319$ of 784, containing the 79-dimensional exact null space); $W^\perp$ is its complement. We report the squared error in each part as the ratio PnP/TV-ADMM (below 1: PnP has less error),
with the injection $\iota$ and alignment $c$ statistics. Paired Wilcoxon tests on the per-image errors give $p<10^{-10}$ for every ratio of the three 200-image sets in Table~\ref{tab:nullspace}; for the phantom, with 20 noise draws, $p=1.9\times10^{-6}$ is the smallest value attainable.

\begin{table}[htbp]
\centering\small\setlength{\tabcolsep}{4pt}
\caption{Error decomposition at $\sigma=0.05$, 20 angles, bicubic fine-grid data, $W_{0.05}$ ($\dim=319$).
Ratios are PnP/TV-ADMM (below 1: PnP has less error); $\iota$ is the injection and $c$ the alignment statistic of
Section~\ref{sec:subspace}.}\label{tab:nullspace}
\begin{tabular}{lccccccc}
\toprule
 & \multicolumn{3}{c}{error ratio PnP/TV} & \multicolumn{2}{c}{$\iota$} & \multicolumn{2}{c}{$c$}\\
\cmidrule(lr){2-4}\cmidrule(lr){5-6}\cmidrule(lr){7-8}
Test set & in $W$ & in $W^\perp$ & total & TV & PnP & TV & PnP\\
\midrule
MNIST (in-dist.)  & 0.48 & 0.40 & 0.42 & 0.58 & 0.76 & 0.45 & 0.76\\
EMNIST letters    & 0.70 & 0.64 & 0.65 & 1.22 & 1.55 & 0.43 & 0.67\\
Fashion-MNIST     & 1.27 & 2.32 & 2.05 & 0.29 & 0.80 & 0.36 & 0.37\\
Shepp--Logan      & 2.59 & 2.64 & 2.63 & 0.44 & 3.27 & 0.09 & 0.23\\
\bottomrule
\end{tabular}
\end{table}

\begin{table}[htbp]
\centering\small\setlength{\tabcolsep}{4pt}
\caption{Sensitivity of Table~\ref{tab:nullspace} to the subspace: the exact null space ($\dim=79$), $W_{0.05}$ (319) and $W_{0.10}$ (365). Error ratio PnP/TV in the subspace and PnP injection $\iota$.}\label{tab:nullspace-tau}
\begin{tabular}{lcccccc}
\toprule
 & \multicolumn{2}{c}{exact null} & \multicolumn{2}{c}{$W_{0.05}$} & \multicolumn{2}{c}{$W_{0.10}$}\\
\cmidrule(lr){2-3}\cmidrule(lr){4-5}\cmidrule(lr){6-7}
Test set & ratio & $\iota$ & ratio & $\iota$ & ratio & $\iota$\\
\midrule
MNIST (in-dist.) & 0.49 & 0.72 & 0.48 & 0.76 & 0.47 & 0.76\\
EMNIST letters   & 0.68 & 1.56 & 0.70 & 1.55 & 0.70 & 1.56\\
Fashion-MNIST    & 1.21 & 0.70 & 1.27 & 0.80 & 1.27 & 0.83\\
Shepp--Logan     & 2.64 & 3.68 & 2.59 & 3.27 & 2.21 & 2.74\\
\bottomrule
\end{tabular}
\end{table}

\begin{figure}[htbp]
\centering
\includegraphics[width=0.75\linewidth]{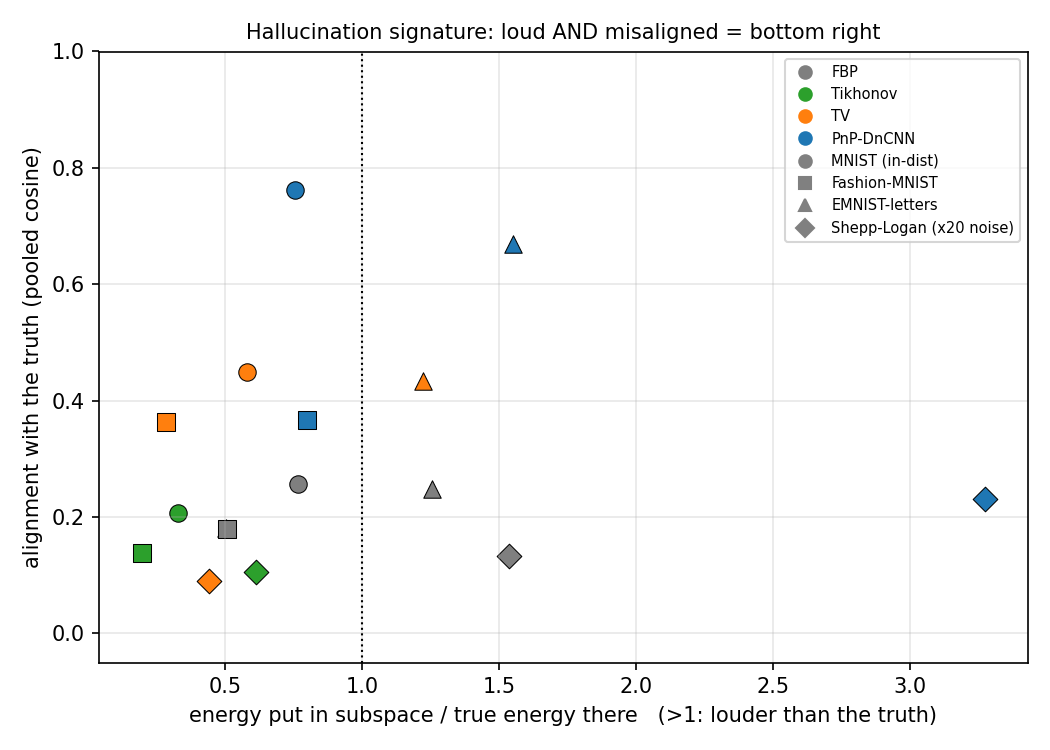}
\caption{Injection (horizontal axis: energy placed in $W_{0.05}$ divided by the true energy there) against alignment (vertical axis: pooled cosine with the truth in $W_{0.05}$) for each method and test set (bicubic fine-grid data).
Points to the right of the dotted line place more energy in $W_{0.05}$ than the true images contain; low alignment means that this energy does not point in the true direction. Only PnP-DnCNN on the phantom is both loud and poorly aligned.}\label{fig:signature}
\end{figure}

\begin{table}[htbp]
\centering\small\setlength{\tabcolsep}{4pt}
\caption{Total error energy ratio PnP/TV-ADMM as a function of the iteration count, with $\sigma_d=0.1$ and the same frozen pipeline: at the tuned $k=5$, at the best $k$ of the grid $\{1,2,5,10,25,50,100,200,400\}$ (selected on the test images, hence optimistic, with the selected $k$ in brackets) and at $k=400$.
Left: data with the inverse crime ($\sigma_d$ grid of E5); right: bilinear fine-grid data.}\label{tab:nullspace-k}
\begin{tabular}{lcccccc}
\toprule
 & \multicolumn{3}{c}{Inverse crime} & \multicolumn{3}{c}{Bilinear fine grid}\\
\cmidrule(lr){2-4}\cmidrule(lr){5-7}
Test set & $k=5$ & best $k$ & $k=400$ & $k=5$ & best $k$ & $k=400$\\
\midrule
MNIST (in-dist.) & 0.43 & 0.43 (5) & 0.71 & 0.43 & 0.42 (5) & 0.65\\
EMNIST letters   & 0.64 & 0.64 (5) & 1.01 & 0.66 & 0.62 (5) & 0.93\\
Fashion-MNIST    & 2.16 & 1.56 (2) & 4.13 & 2.65 & 1.47 (2) & 3.71\\
Shepp--Logan     & 2.60 & 1.92 (1) & 4.77 & 3.44 & 1.89 (1) & 4.58\\
\bottomrule
\end{tabular}
\end{table}

\paragraph{Results.}
\emph{(i) In distribution the advantage is not confined to the unobserved directions.} On MNIST PnP has $0.48$ times TV's error in $W_{0.05}$ and $0.40$ times in its complement, and its content in $W_{0.05}$ is better aligned with the truth ($c=0.76$ against $0.45$). The same holds for EMNIST letters ($0.70$ and $0.64$).
A gradient step never changes the null-space component, so the null-space error reduction (ratio $0.49$ in the exact null space, Table~\ref{tab:nullspace-tau}) is attributable to the denoiser, up to the clipping and the FBP initial image.

\emph{(ii) Under shift the excess error is mostly in the observed directions.} On Fashion-MNIST the ratios are $1.27$ in $W_{0.05}$ and $2.32$ in $W^\perp$, on the phantom $2.59$ and $2.64$. Because $W^\perp$ contains most of the error energy, $93\%$ (Fashion-MNIST) and $89\%$ (phantom) of PnP's excess error over TV lies outside $W_{0.05}$.
The failure under shift is therefore not confined to the weakly determined subspace. These conclusions hold for the exact null space and for $W_{0.10}$ (Table~\ref{tab:nullspace-tau}); the phantom injection falls from $3.7$ to $2.7$ with the threshold but stays above 1 throughout.

\emph{(iii) The injection--alignment signature separates the two shifted sets.} On the phantom PnP places $3.3$ times the true energy in $W_{0.05}$ (TV: $0.44$) with an alignment of only $0.23$ (TV: $0.09$; Figure~\ref{fig:signature}), whereas on MNIST its injection of $0.76$ comes with an alignment of $0.76$.
On Fashion-MNIST the injection is $0.80$ (TV: $0.29$) with an alignment of $0.37$, no different from TV's ($0.36$): PnP's excess error there lies in $W^\perp$ and is not visible in this statistic. The statistic does not by itself distinguish prior-driven structure from amplified noise (Section~\ref{sec:subspace});
the evidence that the phantom injection is driven by the prior is the comparison of E8, where denoisers trained on Fashion-MNIST give an injection of $0.30$--$0.35$ against $2.8$--$3.4$ for MNIST-trained ones, with identical noise and solver.

\emph{(iv) Iterations.} Without the denoiser, the gradient step contracts the distance to the data-fitting solution in direction $i$ after $k$ iterations by $(1-(s_i/s_1)^2)^k$: at $k=5$ a direction with $s_i=0.1\,s_1$ retains $95\%$ of its initial distance, $s_i=0.3\,s_1$ $62\%$ and $s_i=0.5\,s_1$ $24\%$, so the tuned iteration leaves a large share of directions to the denoiser.
On the shifted sets no setting recovers TV's accuracy: even the best $k$ at $\sigma_d=0.1$ has $1.5$--$1.9$ times TV's error, and the error grows with $k$ to $3.7$--$4.8$ times at $k=400$ (Table~\ref{tab:nullspace-k}). In distribution the converged iterate ($k=400$) remains better than TV on MNIST ($0.65$--$0.71$) and is on par with it on EMNIST letters ($0.93$--$1.01$). Section~\ref{sec:e5} shows where along the singular-value axis the error grows with $k$.

\paragraph{Caveats.}
(a)~All statistics are for one denoiser, one noise draw per image and one data model per table; the phantom is a single object. (b)~The threshold $\tau=0.05$ is a convention (there is no gap in the singular values); Table~\ref{tab:nullspace-tau} shows that the conclusions do not depend on it, but the numbers do. (c)~The error ratio is a ratio of means of squared errors; the Wilcoxon tests are paired over images.
(d)~Table~\ref{tab:nullspace-k} uses two data models other than the one of Table~\ref{tab:nullspace} (no bicubic iteration sweep was run), and the best $k$ is selected on the test images. (e)~The comparison is against TV-ADMM only.

\subsection{E8: Denoiser seeds and matched-prior control}\label{sec:seeds-results}
This experiment asks whether the distribution-shift results of E4 depend on the particular denoiser that was trained, and
whether the reversal on Fashion-MNIST is caused by the prior itself or by something else in the pipeline. Table~\ref{tab:seeds}
repeats the shift experiment with three independently trained denoisers per configuration (bicubic data, $\sigma=0.05$, 20 angles).
Three families are used: \emph{m03} (trained on MNIST, $\sigma_{\max}=0.3$, seeds 0--2), \emph{m05} (MNIST, $\sigma_{\max}=0.5$)
and \emph{f03} (Fashion-MNIST, $\sigma_{\max}=0.3$). Within each run all methods are tuned on the validation split of the
family's own training distribution, then frozen; the noise-level grid for $\sigma_d$ was extended to $0.5$.

\paragraph{Seed variation.} Seed-to-seed variation of the gap is at most $0.3$\,dB (standard deviation over three seeds), an order of
magnitude below the in-distribution and shift effects. On Fashion-MNIST the MNIST-trained prior is $3.10\pm0.30$\,dB below TV-ADMM (range
$-3.43$ to $-2.83$) and on the phantom $4.33\pm0.12$\,dB below. Training with $\sigma_{\max}=0.5$ instead of $0.3$ leaves the Fashion-MNIST gap unchanged ($-3.12\pm0.31$\,dB) and
reduces the phantom deficit by $0.4$\,dB ($-3.94\pm0.02$\,dB). With three seeds the standard
deviations are themselves imprecise, so we read them as an indication of scale rather than as interval estimates.
The sign of every shift effect in Table~\ref{tab:seeds} is the same for all seeds.

\paragraph{Noise-level grid.} At $\sigma=0.2$ the tuned $\sigma_d$ is $0.3$ for all six MNIST-trained denoisers even though values
up to $0.5$ were available, so the choice of $\sigma_d$ in E1 was not limited by the edge of its grid. The wider-range
denoiser (m05) is better at high noise ($+1.44\pm0.11$ versus $+0.97\pm0.11$\,dB over TV at $\sigma=0.2$, ranges not overlapping)
and slightly worse at $\sigma=0.05$ ($+3.86$ versus $+4.01$\,dB in the noise sweep, and $+3.68\pm0.13$ versus $+3.84\pm0.03$\,dB in
Table~\ref{tab:seeds}). The in-distribution gap at $\sigma=0.05$ therefore depends on $\sigma_{\max}$ by a few tenths of a dB; the
headline gap of about $3.4$\,dB over box-constrained TV (E1--E2) uses the m03 family.

\paragraph{Matched-prior control.} The control isolates the cause of the reversal. A prior trained on Fashion-MNIST beats TV on
Fashion-MNIST by $+2.21\pm0.01$\,dB (total-error ratio $0.63$), although by less than the MNIST prior on MNIST
($+3.84$\,dB), possibly because Fashion-MNIST images are less sparse (untested). The control is clean in the following sense: the
tuned hyperparameters are identical in all nine runs ($\sigma_d=0.1$, $k=5$, $\lambda_{\rm TV}=5.6\times10^{-3}$), including the three
tuned on Fashion-MNIST validation data, and the TV-ADMM PSNR is therefore identical for every prior on every test set. The
only difference between the MNIST-trained and the Fashion-trained pipelines is thus the denoiser weights. Whether re-tuning $(\sigma_d,k)$ could rescue the MNIST-trained prior on Fashion-MNIST is examined in E10 (Section~\ref{sec:oppoint-results}).
Transfer is asymmetric:
the Fashion-trained prior still beats TV on MNIST ($+1.38$\,dB) and EMNIST letters ($+0.44$\,dB) and is close to TV
on the phantom ($-0.18\pm0.08$\,dB against TV-ADMM, $-0.05$\,dB against TV-box; error ratio $1.05$), whereas the MNIST-trained prior loses by $4.3$\,dB on the phantom.
The injection statistic $\iota$ (Section~\ref{sec:subspace}) on the phantom follows the prior: $2.8$--$3.4$ for MNIST-trained denoisers versus
$0.30$--$0.35$ for Fashion-trained ones, with identical noise and solver settings; on Fashion-MNIST itself it is $0.75$--$0.83$
versus $0.40$--$0.42$. Whether the broader prior helps because it is richer or because Fashion-MNIST resembles the phantom
more cannot be decided with two training distributions.

\paragraph{Box-constrained TV.} For m03 and f03 we repeated the comparison against TV with the constraint $0\le x\le1$
(E2b control; m05 was not rerun). The gaps move by at most $0.5$\,dB and no gap changes sign: m03 $+3.36$ (MNIST), $+1.72$ (EMNIST), $-3.13$ (Fashion), $-4.55$\,dB (phantom); f03 $+1.44$, $+0.38$, $+2.17$,
$-0.05$\,dB. The constraint helps TV most on in-distribution MNIST (TV-box $20.69$ versus $20.21$\,dB for the MNIST-tuned
pipeline), which is the source of the reduction of the in-distribution gap from $3.84$ to $3.36$\,dB. The Fashion-tuned TV-box values differ slightly from the
MNIST-tuned ones because its weight was re-tuned on Fashion-MNIST validation data. The qualitative conclusions of this subsection
(reversal under shift, recovery with a matched prior, asymmetric transfer) hold for both TV variants.

\begin{table}[htbp]
\centering\small\setlength{\tabcolsep}{3.5pt}
\caption{Distribution shift with three denoisers per prior. Mean test PSNR in dB over seeds; gap = PnP$-$TV, mean $\pm$ s.d.\ over
seeds, against TV-ADMM and against box-constrained TV (TV-box; m03 and f03 only; for f03 the box gap has no s.d.\ entry). Bicubic fine-grid data, $\sigma=0.05$, 20 angles;
each prior's pipeline is tuned on the validation set of its own training distribution.}\label{tab:seeds}
\begin{tabular}{lcccccccc}
\toprule
 & \multicolumn{2}{c}{TV} & \multicolumn{3}{c}{MNIST-trained (m03)} & \multicolumn{3}{c}{Fashion-trained (f03)} \\
\cmidrule(lr){2-3}\cmidrule(lr){4-6}\cmidrule(lr){7-9}
Test set & ADMM & box & PnP & gap & gap (box) & PnP & gap & gap (box)\\
\midrule
MNIST          & 20.21 & 20.69 & 24.05 & $+3.84\pm0.03$ & $+3.36\pm0.03$ & 21.60 & $+1.38\pm0.09$ & $+1.44$\\
EMNIST letters & 20.28 & 20.35 & 22.07 & $+1.79\pm0.09$ & $+1.72\pm0.09$ & 20.73 & $+0.44\pm0.09$ & $+0.38$\\
Fashion-MNIST  & 19.42 & 19.45 & 16.32 & $-3.10\pm0.30$ & $-3.13\pm0.30$ & 21.63 & $+2.21\pm0.01$ & $+2.17$\\
Shepp--Logan   & 19.28 & 19.50 & 14.96 & $-4.33\pm0.12$ & $-4.55\pm0.12$ & 19.10 & $-0.18\pm0.08$ & $-0.05$\\
\bottomrule
\end{tabular}

\smallskip
\raggedright\footnotesize The TV-box column is the MNIST-tuned TV-box used with m03. The f03 gap (box) is computed against the
Fashion-tuned TV-box ($20.16$, $20.34$, $19.46$, $19.15$\,dB, top to bottom). m05 gives
gaps of $+3.68\pm0.13$, $+1.64\pm0.02$, $-3.12\pm0.31$, $-3.94\pm0.02$\,dB (same order of test sets).
\end{table}

\begin{figure}[htbp]
\centering
\includegraphics[width=\linewidth]{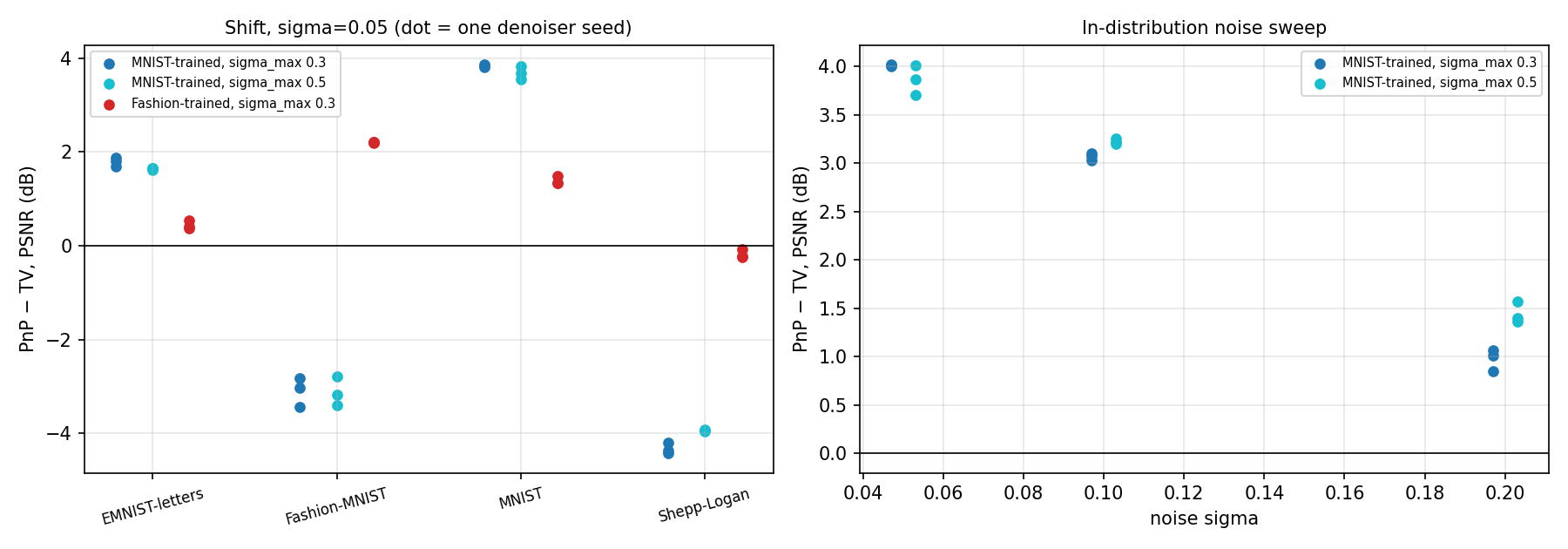}
\caption{Left: PnP$-$TV-ADMM gap for each denoiser seed (dots) on the four test sets, for three priors. Right: gap in the
in-distribution noise sweep for MNIST-trained denoisers with $\sigma_{\max}=0.3$ and $0.5$. Produced by
\texttt{summarize\_step4.py} from \texttt{results/step4/}.}\label{fig:seeds}
\end{figure}

\subsection{E9: Lesion insertion}\label{sec:lesion-results}
\begin{figure}[htbp]
\centering
\includegraphics[width=\linewidth]{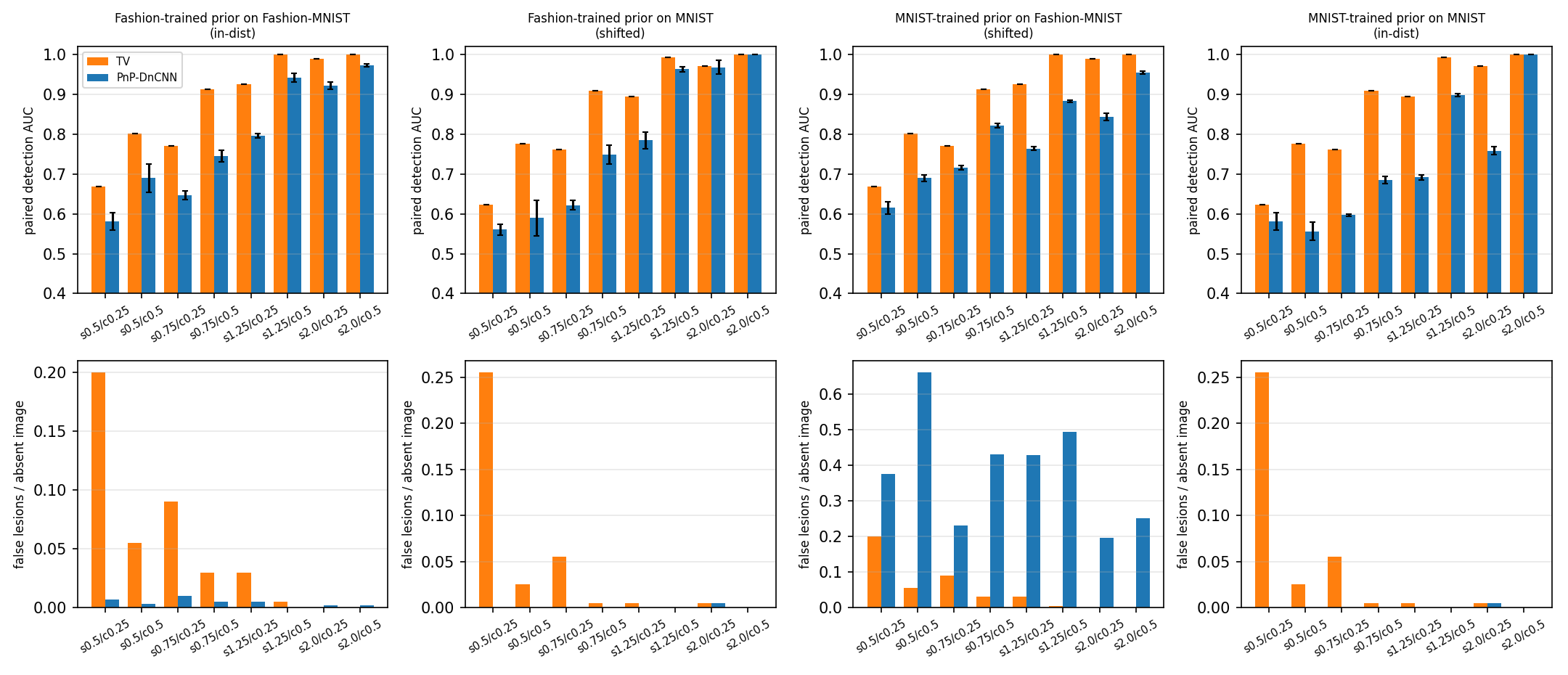}
\caption{Paired detection AUC (top) and false lesions per lesion-absent image (bottom) for TV and PnP-DnCNN in the four
prior/test-set combinations and the eight lesion types (mean over three denoisers; bars: s.d.\ over seeds). TV here is the ADMM
implementation; TV-box values are in Table~\ref{tab:lesion} and \texttt{results/control/}.}\label{fig:lesion-summary}
\end{figure}

Table~\ref{tab:lesion} and Figure~\ref{fig:lesion-summary} summarise the lesion experiment (three denoisers per prior,
protocol of Section~\ref{sec:lesion}; $157$--$197$ valid images per lesion type). TV denotes the ADMM implementation used throughout
the paper; TV-box is the same comparison with the constraint $0\le x\le1$, re-tuned on the same validation split (m03 and f03 families,
same lesion locations and noise seeds). Three observations summarise the eight lesion types.

\emph{TV detects small lesions better than PnP, in distribution and for a matched prior.} For the MNIST-trained prior on MNIST
the paired AUC of PnP is $0.04$--$0.23$ below TV's (for example $0.69$ versus $0.91$ for $\sigma_l=0.75$, $c=0.5$), although
PnP is about $4$\,dB better in PSNR on the same images; for the matched Fashion-trained prior the deficit is $0.03$--$0.17$. The
two methods tie only for the largest high-contrast lesion. Against TV-box the deficit is the same size ($0.00$--$0.22$ on MNIST,
$0.03$--$0.15$ for the matched prior), so the box constraint does not explain it. FBP and Tikhonov also detect small lesions better
than PnP (paired AUC $0.92$ at $\sigma_l=0.75$, $c=0.5$ on MNIST for FBP), at the cost of many false blobs (up to $6$ per image
for FBP).

\begin{figure}[htbp]
\centering
\includegraphics[width=0.85\linewidth]{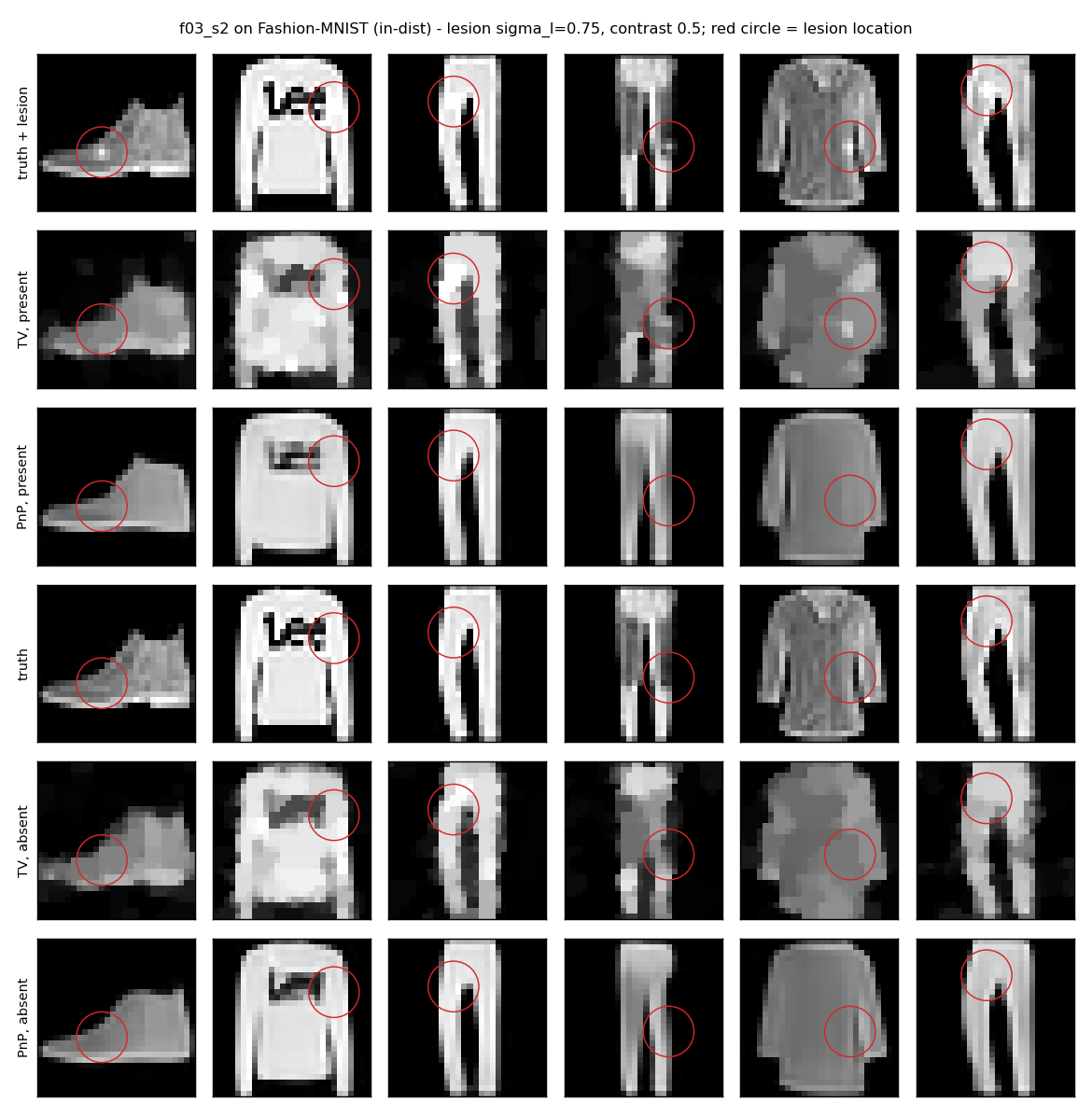}
\caption{Matched Fashion-trained prior on Fashion-MNIST ($\sigma_l=0.75$, $c=0.5$). Rows: ground truth with lesion, TV and
PnP reconstructions of the lesion-present image; ground truth, TV and PnP reconstructions of the lesion-absent image. The
red circle marks the lesion location. In these examples PnP produces the cleaner images but removes the lesion.
}\label{fig:lesion-examples}
\end{figure}

\emph{In distribution PnP attenuates lesions and creates almost no false lesions.} Its retention is $0.12$--$0.78$ for the MNIST prior on MNIST
and $0.16$--$0.49$ for the matched prior (TV: $0.29$--$0.88$ and $0.38$--$0.83$; TV-box: $0.31$--$0.92$ and $0.33$--$0.82$), and it
produces at most $0.01$ false lesions per image against up to $0.26$ for TV and, for the MNIST-tuned TV-box, up to $0.49$
(smallest low-contrast lesion; $0.075$ for the Fashion-tuned TV-box). The reconstructions are cleaner than TV's but the inserted lesion is often absent
(Figure~\ref{fig:lesion-examples}), so a visually clean image does not imply a faithful one.

\emph{Under prior mismatch PnP creates false lesions, but box-constrained TV does too for the smallest lesions.} For the MNIST-trained
prior on Fashion-MNIST the retention is $0.78$--$1.04$ (the lesion survives, although invented contrast may contribute) but there are $0.20$--$0.66$ false lesions per image
against $\le0.20$ for TV, and the paired AUC remains below TV's by $0.05$--$0.16$. The matched Fashion-trained prior on the same
images produces at most $0.01$. The box-constrained TV is a stricter comparison here: its false-lesion rate on the shifted images is $0.005$--$0.48$ per image, and for
the smallest low-contrast lesion type ($\sigma_l=0.5$, $c=0.25$) it exceeds PnP's ($0.48$ versus $0.38$). For the other seven lesion types PnP's rate is higher, by a factor
of $1.4$ to $50$. The claim that a mismatched learned prior creates more false lesions than the classical regulariser therefore holds for TV-box on seven of eight lesion types, not on all of them.
Lesion observability (Section~\ref{sec:lesion}) does not explain these differences: the observability terciles formed within each
lesion type have mean observability $0.88$, $0.92$ and $0.94$, and PnP's paired AUC does not increase with observability (on MNIST
it falls from $0.82$ in the lowest to $0.67$ in the highest tercile); smaller lesions are harder for all methods, but this is
confounded with size.

\begin{table}[htbp]
\centering\small\setlength{\tabcolsep}{3pt}
\caption{Lesion insertion, mean over three denoisers per prior ($157$--$197$ valid test images per lesion type, $\sigma=0.05$, 20 angles, bicubic finer-grid data), for two
of the eight lesion types ($\sigma_l/c$ = lesion size in px / contrast). pAUC: paired AUC; Ret.: retention; FL: false lesions per
lesion-absent image; TV-box: TV with $0\le x\le1$. All methods use their PSNR-tuned settings.}\label{tab:lesion}
\begin{tabular}{llccccccccc}
\toprule
 & & \multicolumn{3}{c}{pAUC} & \multicolumn{3}{c}{Ret.} & \multicolumn{3}{c}{FL}\\
\cmidrule(lr){3-5}\cmidrule(lr){6-8}\cmidrule(lr){9-11}
Prior $\to$ test set & $\sigma_l/c$ & TV & box & PnP & TV & box & PnP & TV & box & PnP\\
\midrule
MNIST $\to$ MNIST (in-dist.)       & 0.75/0.5  & 0.91 & 0.90 & 0.69 & 0.59 & 0.65 & 0.28 & 0.005 & 0.010 & 0.000\\
                                   & 1.25/0.25 & 0.90 & 0.90 & 0.69 & 0.62 & 0.67 & 0.30 & 0.005 & 0.010 & 0.000\\
MNIST $\to$ Fashion (shifted)      & 0.75/0.5  & 0.91 & 0.92 & 0.82 & 0.63 & 0.74 & 0.99 & 0.030 & 0.090 & 0.430\\
                                   & 1.25/0.25 & 0.93 & 0.91 & 0.76 & 0.59 & 0.69 & 0.87 & 0.030 & 0.095 & 0.428\\
Fashion $\to$ Fashion (in-dist.)   & 0.75/0.5  & 0.91 & 0.90 & 0.75 & 0.63 & 0.55 & 0.22 & 0.030 & 0.025 & 0.005\\
                                   & 1.25/0.25 & 0.93 & 0.93 & 0.80 & 0.59 & 0.54 & 0.23 & 0.030 & 0.025 & 0.005\\
Fashion $\to$ MNIST (shifted)      & 0.75/0.5  & 0.91 & 0.90 & 0.75 & 0.59 & 0.55 & 0.35 & 0.005 & 0.000 & 0.000\\
                                   & 1.25/0.25 & 0.90 & 0.90 & 0.78 & 0.62 & 0.60 & 0.36 & 0.005 & 0.000 & 0.000\\
\bottomrule
\end{tabular}
\end{table}

\paragraph{Scope.} The observer is a location-known local-contrast statistic and the lesions are additive Gaussian blobs on
$28\times28$ digits, so these results say how each \emph{reconstruction} preserves or creates a simple signal, not how a human or a
model observer would perform on clinical images. False-lesion counts depend on the response threshold $\theta$, which scales with the
nominal lesion, so they are compared across methods within a lesion type, not across types. No paired intervals were computed for the AUC differences; with
$157$--$197$ valid images the binomial standard error of a paired AUC is about $0.02$--$0.04$, so differences below about $0.05$ should be read as ties.

\subsection{E10: Can re-tuning recover lesion detectability?}\label{sec:oppoint-results}
\begin{figure}[htbp]
\centering
\includegraphics[width=\linewidth]{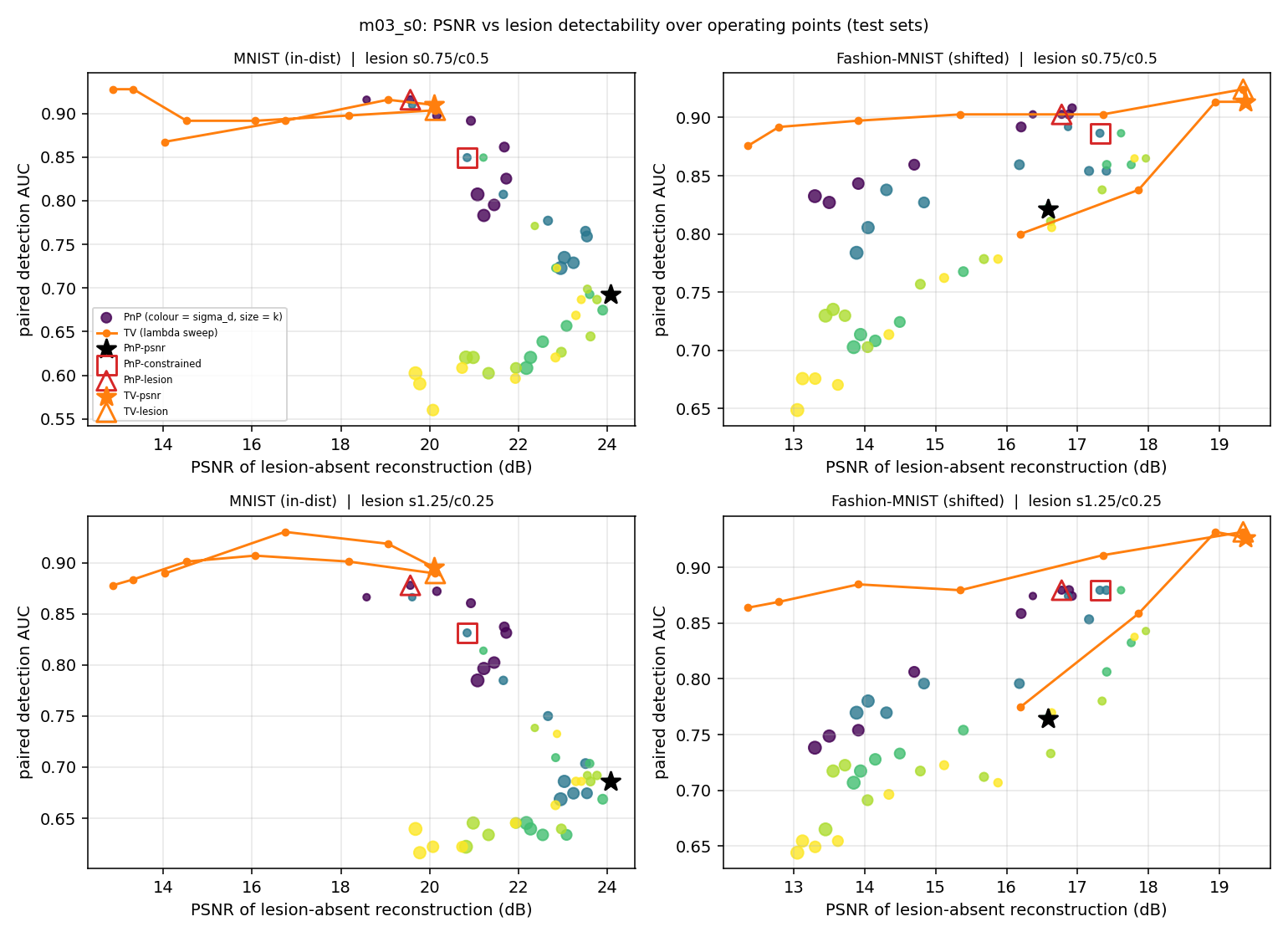}
\caption{PSNR of the lesion-absent reconstruction versus paired detection AUC over all operating points of PnP (colour:
$\sigma_d$; marker size: iterations) and TV (line: weight sweep) for the MNIST-trained denoiser. Markers show the selected
operating points. Left: in distribution; right: Fashion-MNIST (shifted).}\label{fig:oppoint}
\end{figure}

\begin{figure}[htbp]
\centering
\includegraphics[width=\linewidth]{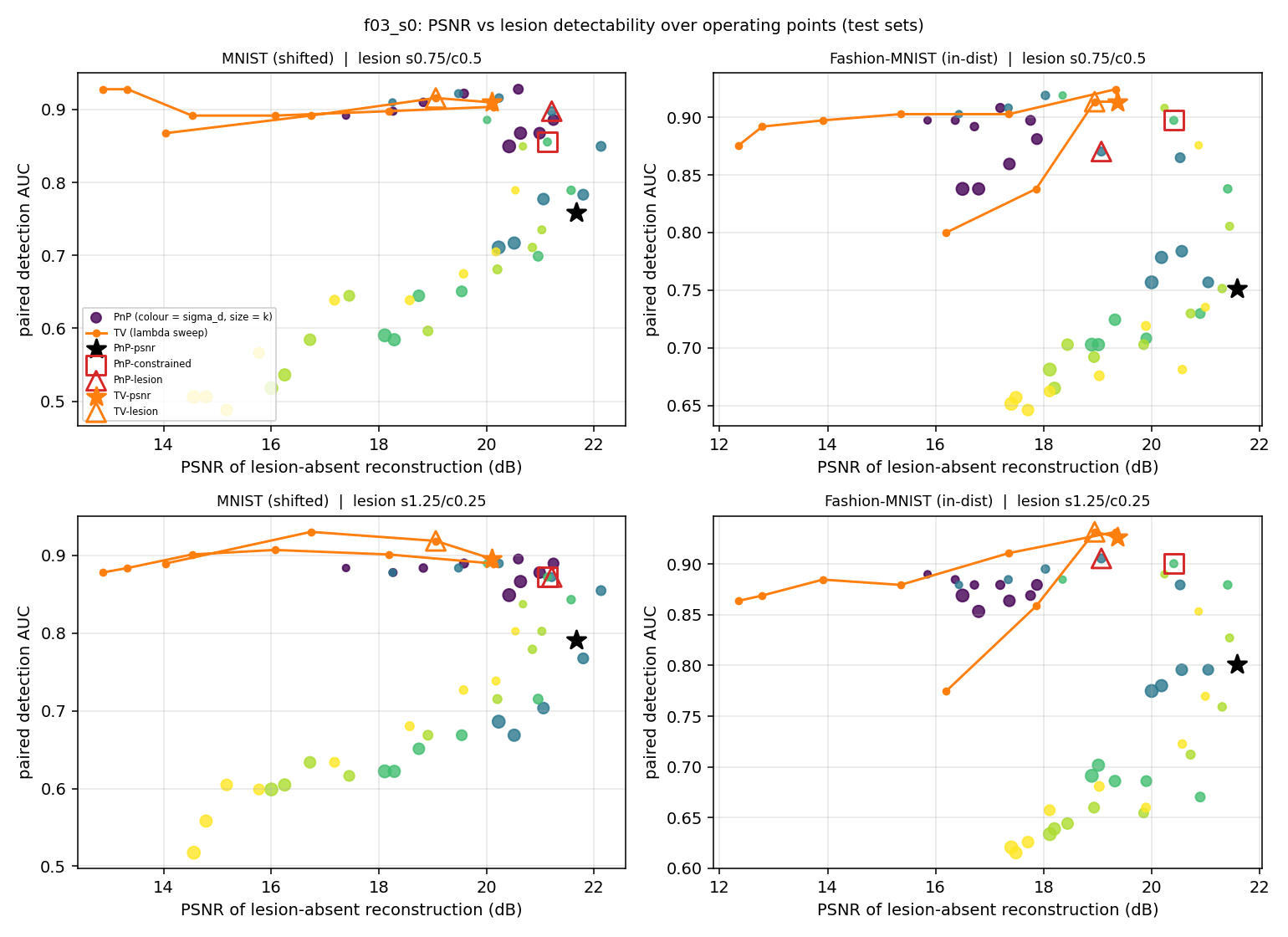}
\caption{As Figure~\ref{fig:oppoint} for the Fashion-trained denoiser. Left: MNIST (shifted); right: Fashion-MNIST (in distribution).}\label{fig:oppoint-f}
\end{figure}

The PSNR-tuned operating point of PnP is not necessarily the best one for lesion detection, so we sweep operating points
(Section~\ref{sec:lesion}) and compare three rules for PnP---PSNR-tuned, best paired AUC on the selection set (``lesion''),
and best paired AUC among points whose selection-set PSNR is at least that of tuned TV (``constrained'')---with TV. All rules are
applied on held-out in-distribution validation images and then reported on the test sets, so no point is chosen on the evaluation images
or with knowledge of the shifted distribution. This experiment uses one denoiser per prior (seed 0 of the m03 and f03 families), the
ADMM implementation of TV, and the two lesion types $\sigma_l=0.75$, $c=0.5$ and $\sigma_l=1.25$, $c=0.25$; neither seed variation nor
the box-constrained TV was examined here. Table~\ref{tab:oppoint} and Figures~\ref{fig:oppoint} and~\ref{fig:oppoint-f} show the results.

\paragraph{MNIST-trained prior, in distribution.} The constrained point keeps a $+0.7$\,dB PSNR advantage over TV (the PSNR-tuned
advantage was $+4.0$\,dB) but its paired AUC is still $0.06$ below TV's (for both lesion types), and its false-lesion rate
rises to 12 times TV's ($0.060$ versus $0.005$ per image). The lesion-tuned point matches TV's detection ($0.916$ versus $0.910$) but has
a lower PSNR than TV ($19.6$ versus $20.1$\,dB) and about 60 times TV's false-lesion rate ($0.32$ versus $0.005$ per image). Tuning TV for
detection instead changes it very little (PSNR $20.13$\,dB, pAUC $0.904$), so TV has no comparable trade-off in this grid. As an upper bound
that uses the test set, the best of the $45$ PnP settings among those with PSNR at least TV's reaches a paired AUC of $0.898$ at $0.28$
false lesions per image, so no rule could have closed the gap for this prior without a large false-lesion cost.

\paragraph{Fashion-trained prior.} In distribution (Fashion-MNIST) the constrained point keeps $+1.0$\,dB over TV with a paired AUC
$0.02$--$0.03$ below TV's and $0.085$ false lesions per image against $0.03$ for TV. A difference of this size is comparable to the
binomial standard error of a paired AUC with $170$--$190$ valid images ($\approx0.02$--$0.03$), so we describe it as close to TV's, not as equal. On the shifted MNIST images the constrained point keeps $+1.0$\,dB, with a paired AUC $0.02$--$0.06$ below TV's and
$0.015$ false lesions per image against $0.005$.

\paragraph{Under mismatch.} Operating points chosen in distribution do not protect the MNIST-trained prior on Fashion-MNIST: all
three rules are $2.1$--$2.8$\,dB below TV in PSNR and produce $0.40$--$0.51$ false lesions per image against $0.03$. This is not an
artefact of the selection rule: none of the $45$ PnP settings in the grid reaches TV's PSNR on the shifted images (the best is $17.96$
versus $19.37$\,dB), even when judged on the test set itself.

\paragraph{Summary.} PSNR, detectability and false lesions trade against each other for PnP. Only the Fashion-trained prior leaves an operating point at which PnP stays about
$1$\,dB above TV in PSNR with a detection within $0.02$--$0.06$ AUC of TV's and few false lesions, and it does so on both Fashion-MNIST (in distribution) and MNIST (shifted); for the MNIST-trained prior no such point exists on either test set.
We hypothesise that the difference between the priors lies in the variety of their training distributions (E8) rather than in whether a test set is formally in distribution; with one denoiser per prior and two training sets this is a statement about two networks and is not tested.

\begin{table}[htbp]
\centering\small\setlength{\tabcolsep}{4pt}
\caption{Operating points for lesion type $\sigma_l=0.75$, $c=0.5$ on the test sets (one denoiser per prior; points chosen on
held-out in-distribution validation images of the prior's own training distribution). PSNR is that of the lesion-absent reconstruction. ``TV, lesion'' is TV with the weight
that maximises paired AUC on the selection set, which differs between priors, so its row can differ for the same test set.}\label{tab:oppoint}
\begin{tabular}{llccc}
\toprule
Prior $\to$ test set & Operating point & PSNR (dB) & pAUC & FL\\
\midrule
MNIST $\to$ MNIST (in-dist.)     & TV (PSNR-tuned)  & 20.11 & 0.910 & 0.005\\
                                 & TV, lesion       & 20.13 & 0.904 & 0.005\\
                                 & PnP, PSNR-tuned  & 24.07 & 0.693 & 0.000\\
                                 & PnP, constrained & 20.84 & 0.849 & 0.060\\
                                 & PnP, lesion      & 19.57 & 0.916 & 0.315\\
MNIST $\to$ Fashion (shifted)    & TV (PSNR-tuned)  & 19.37 & 0.914 & 0.030\\
                                 & TV, lesion       & 19.33 & 0.924 & 0.035\\
                                 & PnP, PSNR-tuned  & 16.58 & 0.822 & 0.420\\
                                 & PnP, constrained & 17.32 & 0.886 & 0.405\\
                                 & PnP, lesion      & 16.78 & 0.903 & 0.505\\
Fashion $\to$ Fashion (in-dist.) & TV (PSNR-tuned)  & 19.37 & 0.914 & 0.030\\
                                 & TV, lesion       & 18.94 & 0.914 & 0.010\\
                                 & PnP, PSNR-tuned  & 21.58 & 0.751 & 0.005\\
                                 & PnP, constrained & 20.41 & 0.897 & 0.085\\
                                 & PnP, lesion      & 19.07 & 0.870 & 0.255\\
Fashion $\to$ MNIST (shifted)    & TV (PSNR-tuned)  & 20.11 & 0.910 & 0.005\\
                                 & TV, lesion       & 19.06 & 0.916 & 0.000\\
                                 & PnP, PSNR-tuned  & 21.67 & 0.759 & 0.000\\
                                 & PnP, constrained & 21.13 & 0.855 & 0.015\\
                                 & PnP, lesion      & 21.21 & 0.898 & 0.040\\
\bottomrule
\end{tabular}
\end{table}

\subsection{E11: The matrix-free operator reproduces the dense pipeline}\label{sec:mf-results}
Table~\ref{tab:mf} compares the two implementations on the same 200 noisy test images ($\sigma=0.05$, 20 angles, bicubic
finer-grid data). The comparison uses fixed, untuned settings (Tikhonov and TV weight $10^{-2}$; PnP with $\sigma_d=0.1$ and $k=25$), so its PSNR values
are not comparable to the tuned values of E1--E2; only the agreement between the two implementations matters here.
The projections agree to $3\times10^{-7}$ in relative norm, and the three reconstruction methods agree to
$\le10^{-5}$ (Tikhonov, TV) and $3\times10^{-4}$ (PnP) in relative norm, with identical PSNR to three decimals. The slightly larger PnP difference
is plausible given that the denoiser is applied $25$ times to slightly different iterates, but we did not isolate its source.

The randomized SVD reproduces the singular values of the dense SVD to within $1.1\times10^{-4}\,s_1$ for the first $400$ and
$6.2\times10^{-5}\,s_1$ for the first $600$ (the number used to build the subspace in E7).
It finds the same $465$-dimensional well-determined
subspace $\{i:s_i/s_1\ge0.05\}$ (so $319$ directions are weakly determined). The resulting strong/weak error split differs from the dense one by at most
$2.3\times10^{-4}$ of the error norm for each of the three methods (Tikhonov, TV-ADMM, PnP), far below the $1\%$ tolerance of the unit test; the strong subspace holds
$91\%$, $82\%$ and $77\%$ of the squared error of the three reconstructions, respectively. These checks are limited by single-precision arithmetic on the GPU.

At this size the dense implementation is much faster for Tikhonov and TV-ADMM (Table~\ref{tab:mf}): the matrix-free TV-ADMM
performs $100$ ADMM iterations with up to $30$ conjugate-gradient steps each, i.e.\ up to $3\times10^{3}$ forward--adjoint pairs per
image, and the adjoint is evaluated by reverse-mode differentiation. PnP is only $1.2$ times slower because the denoiser dominates its cost.
The timings are single runs on one GPU (RTX 3050 Ti) and should be read as orders of magnitude. The matrix-free operator is therefore a requirement for larger images, where the dense matrix cannot be stored,
rather than an optimisation at $28\times28$.

\begin{table}[htbp]
\centering\small\setlength{\tabcolsep}{4pt}
\caption{Dense versus matrix-free pipeline (200 images, fixed untuned settings, single GPU run). Rel.\ diff.: relative difference between
the two reconstructions. Peak GPU memory for the whole comparison: 184\,MB.}\label{tab:mf}
\begin{tabular}{lcccc}
\toprule
Method & Rel.\ diff. & PSNR (dense = MF) & Time dense (s) & Time MF (s)\\
\midrule
Tikhonov & $4.6\times10^{-6}$ & 16.36 & 0.08 & 0.26\\
TV-ADMM  & $9.7\times10^{-6}$ & 19.01 & 0.16 & 7.50\\
PnP ($k=25$) & $2.8\times10^{-4}$ & 23.10 & 0.76 & 0.94\\
\bottomrule
\end{tabular}
\end{table}

\subsection{E12: Poisson dose sweep}\label{sec:dose-results}
\begin{figure}[htbp]
\centering
\includegraphics[width=\linewidth]{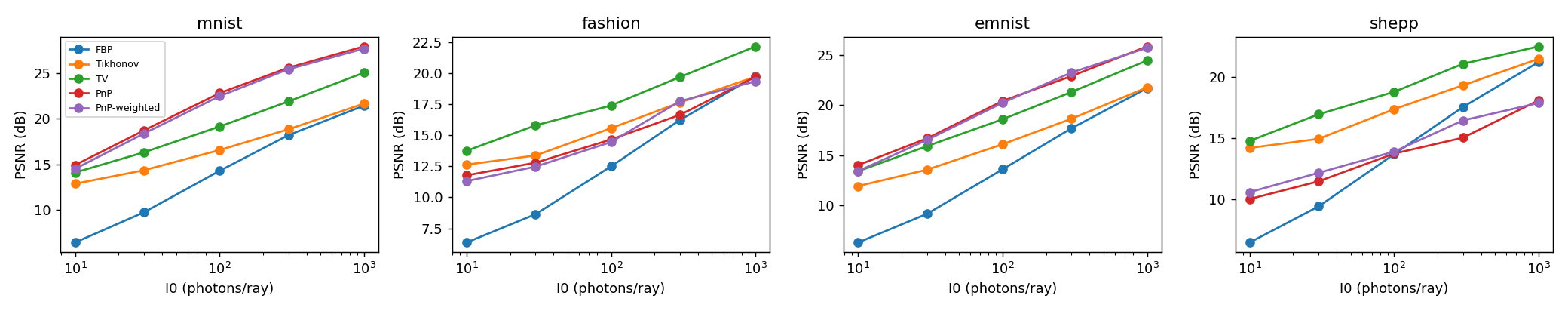}
\caption{PSNR versus photon count for FBP, Tikhonov, TV, PnP and the Poisson-weighted PnP on the four test sets (MNIST-trained denoiser,
hyperparameters tuned on MNIST validation images at each dose).}\label{fig:dose}
\end{figure}

Table~\ref{tab:dose} and Figure~\ref{fig:dose} give the PnP--TV difference over the dose range (200 test images per set; the
Shepp--Logan phantom is one image with one noise draw per dose, so it has no interval). Settings are those of Section~\ref{sec:poisson}: tuned on MNIST
validation images at each dose, unweighted data term, ADMM implementation of TV, and one MNIST-trained denoiser. The qualitative picture of
Stage~1 is reproduced; every interval for the 200-image sets excludes zero by at least $6$ standard errors.

\emph{Prior-matched data.} On MNIST the learned prior beats TV at every dose, by $+3.7$\,dB at $I_0=100$--$300$ (effective $\sigma\approx0.03$--$0.06$), against $+3.84$\,dB at Gaussian $\sigma=0.05$,
and for 99--100\% of the images at $I_0\ge100$. The advantage shrinks at both ends of the dose range, to $+0.8$\,dB at $I_0=10$, of the same order as the Gaussian result at $\sigma=0.2$
($+0.97$\,dB, E8). EMNIST keeps a smaller advantage ($+0.6$ to $+1.8$\,dB). These gaps are against TV without the box constraint; that constraint was not tested
under Poisson noise (in the Gaussian MNIST case it lowered the gap by $0.5$\,dB, E8).

\emph{Shifted data.} The MNIST-trained prior is below TV on Fashion-MNIST ($-2.0$ to $-3.1$\,dB; PnP wins on only 6--14\% of the images)
and on the phantom ($-4.4$ to $-6.1$\,dB), as in Stage~1 ($-3.10$ and $-4.33$\,dB). At $I_0=1000$ the prior is no better than FBP on Fashion-MNIST
($19.72$ versus $19.77$\,dB), and on the phantom it is $3.1$\,dB below FBP ($18.07$ versus $21.20$\,dB).
PnP's error in the weakly determined subspace (the complement of the well-determined one) is $0.68$--$0.99$ times TV's on MNIST
but $1.04$--$1.19$ times on Fashion-MNIST and $1.73$--$1.79$ times on the phantom (for $I_0\le300$; $0.90$ at $I_0=1000$), so the mismatch error again sits partly in the weakly determined directions. These ratios are less favourable to PnP than the
Stage~1 values and are not directly comparable, since the noise model, the tuned settings and the subspace definition differ. The gap on Fashion-MNIST is not monotone in dose
($-2.45$, $-3.05$, $-2.74$, $-3.01$, $-1.98$\,dB from high to low dose); with frozen MNIST-tuned settings this reflects, at least in part, the choice of $(\sigma_d,k)$ (see the weighted variant below), so
we read the Fashion-MNIST range as ``$2$ to $3$\,dB below TV'' and not as a dose trend.

\emph{Structural similarity.} In SSIM the picture differs on Fashion-MNIST (Table~\ref{tab:dosessim}): the PnP--TV difference is between $-0.014$ and $+0.054$ and PnP wins on
$47$--$67$\% of the images, so the $2$--$3$\,dB PSNR reversal has no counterpart in SSIM. On the phantom SSIM is lower for PnP at every dose ($-0.19$ to $-0.31$). On MNIST and EMNIST PnP is better in SSIM at every dose.
The reversal under shift is therefore a PSNR result for Fashion-MNIST and a PSNR and SSIM result for the phantom.

\emph{Weighted data term.} Using the Poisson weights in PnP gave no consistent benefit: the weighted variant was $0.14$--$0.59$\,dB below the unweighted one in 13 of the 15 non-phantom
cells, and above it in two (EMNIST $+0.35$\,dB and Fashion-MNIST $+1.10$\,dB at $I_0=300$), where the tuned $(\sigma_d,k)$ differed ($(0.05,10)$ unweighted versus $(0.1,3)$ weighted). The
$1.1$\,dB on Fashion-MNIST shows that the shifted-data gap depends on how the frozen settings were chosen: the weighted variant would reduce that gap at $I_0=300$ from $-3.05$ to about $-1.95$\,dB.
We report the unweighted scheme, as it is the one tuned on the same footing as in Stage~1.

\begin{table}[htbp]
\centering\small\setlength{\tabcolsep}{4pt}
\caption{PnP$-$TV PSNR difference (dB, mean $\pm$ standard error of the paired difference over 200 images) by photon count $I_0$, with the corresponding
effective sinogram noise standard deviation (measured on the MNIST images; signal dependence makes it about $15$--$20\%$ higher on Fashion-MNIST); and, in brackets, the ratio of PnP's to TV's reconstruction error in the complement of the well-determined subspace. Phantom: one image.}\label{tab:dose}
\resizebox{\linewidth}{!}{%
\begin{tabular}{lccccc}
\toprule
$I_0$ (photons/ray) & 1000 & 300 & 100 & 30 & 10\\
effective $\sigma$ & 0.018 & 0.032 & 0.056 & 0.106 & 0.198\\
\midrule
MNIST (train dist.) & $+2.87\pm.07$ (0.70) & $+3.68\pm.08$ (0.68) & $+3.65\pm.08$ (0.71) & $+2.39\pm.12$ (0.88) & $+0.82\pm.12$ (0.99)\\
EMNIST              & $+1.40\pm.09$ (0.75) & $+1.59\pm.08$ (0.83) & $+1.82\pm.08$ (0.86) & $+0.77\pm.11$ (1.02) & $+0.62\pm.10$ (1.00)\\
Fashion-MNIST       & $-2.45\pm.13$ (1.04) & $-3.05\pm.16$ (1.18) & $-2.74\pm.16$ (1.18) & $-3.01\pm.14$ (1.19) & $-1.98\pm.13$ (1.07)\\
Shepp--Logan        & $-4.41$ (0.90) & $-6.05$ (1.79) & $-5.05$ (1.73) & $-5.49$ (1.77) & $-4.75$ (1.75)\\
\bottomrule
\end{tabular}}
\end{table}

\begin{table}[htbp]
\centering\small\setlength{\tabcolsep}{4pt}
\caption{PnP$-$TV SSIM difference (mean $\pm$ standard error over 200 images; share of images won by PnP in brackets) for the same runs as Table~\ref{tab:dose}.}\label{tab:dosessim}
\resizebox{\linewidth}{!}{%
\begin{tabular}{lccccc}
\toprule
$I_0$ (photons/ray) & 1000 & 300 & 100 & 30 & 10\\
\midrule
MNIST (train dist.) & $+.089\pm.002$ (1.00) & $+.170\pm.005$ (1.00) & $+.244\pm.006$ (1.00) & $+.279\pm.008$ (1.00) & $+.267\pm.009$ (0.99)\\
EMNIST              & $+.039\pm.002$ (0.96) & $+.081\pm.003$ (0.99) & $+.123\pm.004$ (1.00) & $+.136\pm.006$ (0.98) & $+.166\pm.007$ (0.98)\\
Fashion-MNIST       & $-.014\pm.007$ (0.47) & $-.005\pm.009$ (0.50) & $+.026\pm.010$ (0.55) & $+.011\pm.009$ (0.52) & $+.054\pm.009$ (0.67)\\
Shepp--Logan        & $-.216$ & $-.306$ & $-.189$ & $-.245$ & $-.206$\\
\bottomrule
\end{tabular}}
\end{table}

\subsection{E13: Lesion insertion under Poisson noise}\label{sec:lesion-poisson}
\begin{figure}[htbp]
\centering
\includegraphics[width=\linewidth]{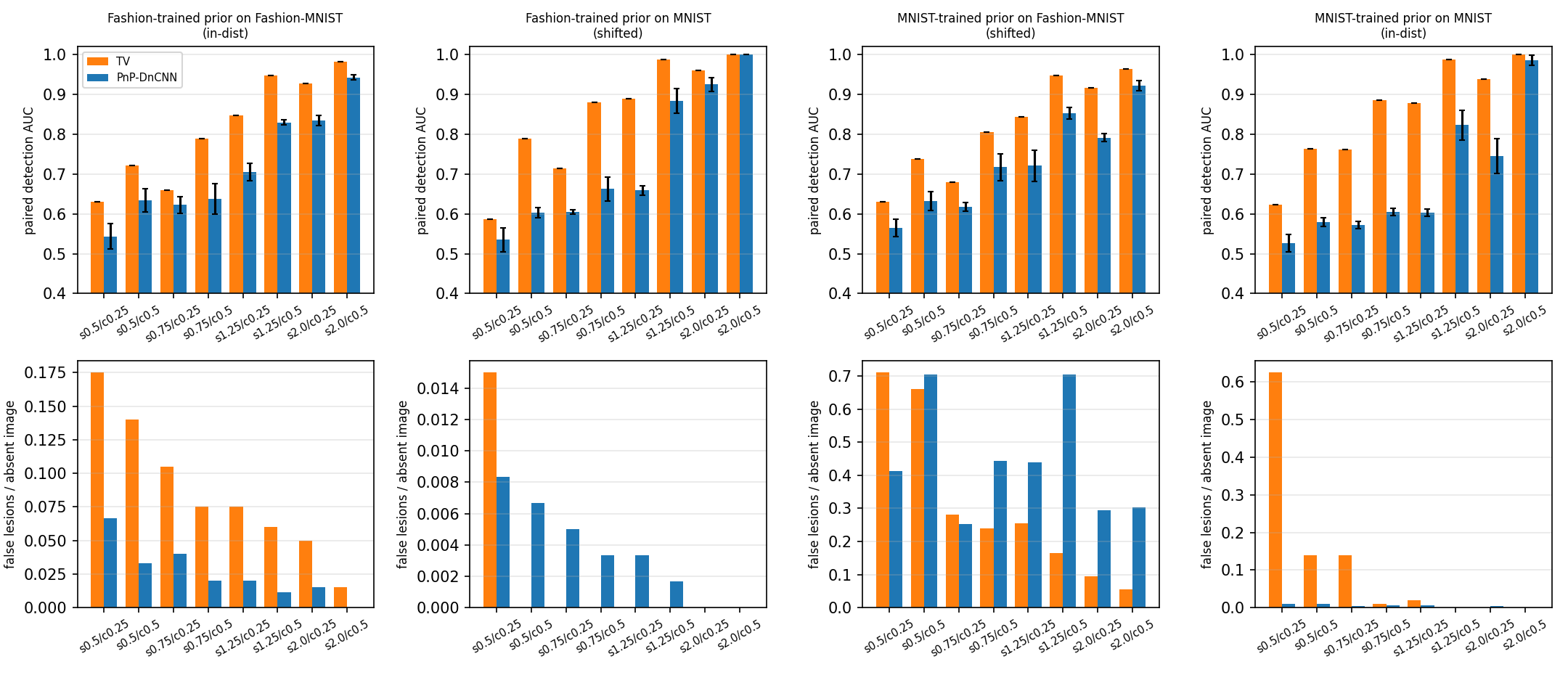}\\[2pt]
\includegraphics[width=\linewidth]{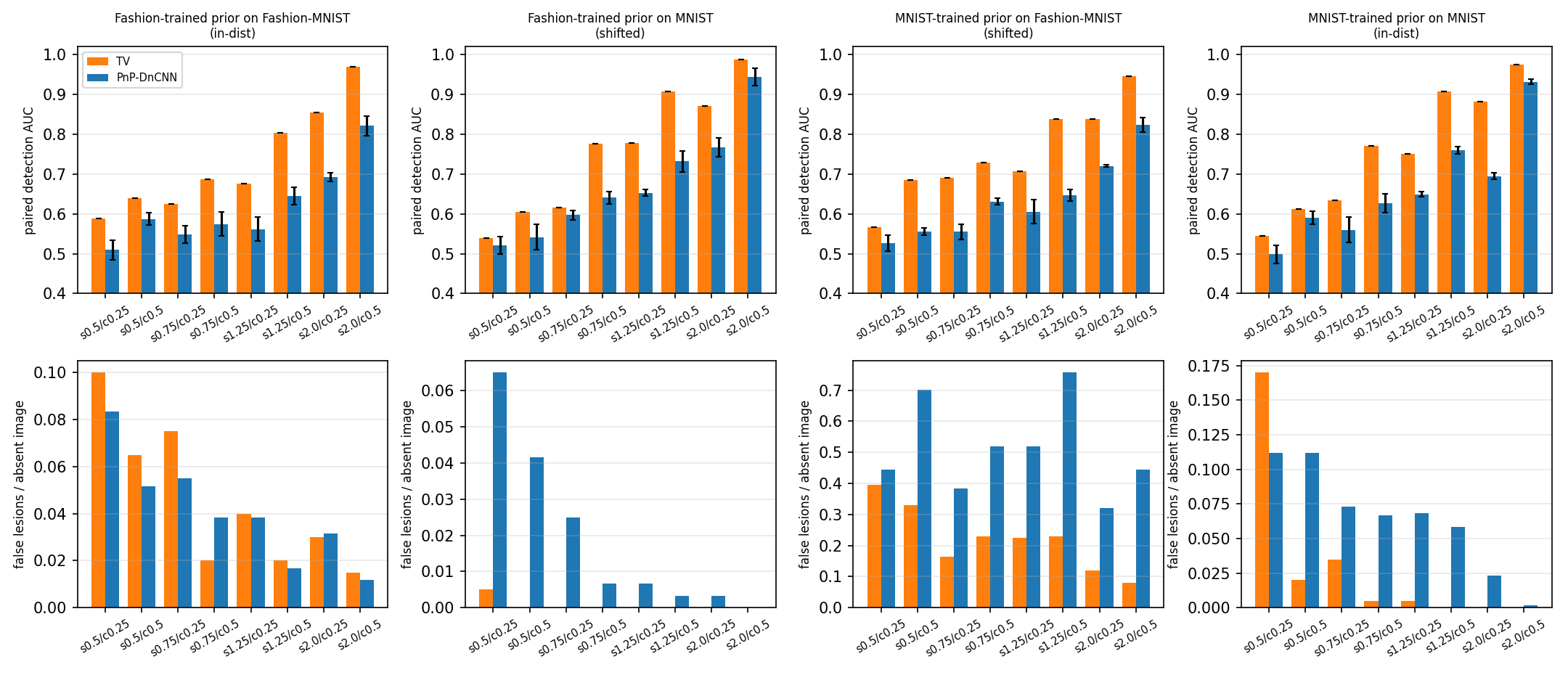}
\caption{Paired detection AUC (top rows) and false lesions per lesion-absent image (bottom rows) under Poisson noise, $I_0=100$ (upper panel) and $I_0=30$ (lower panel). Bars: mean over three
denoisers; error bars: s.d.\ over seeds.}\label{fig:lesion-poisson}
\end{figure}

We repeated the lesion experiment (Section~\ref{sec:lesion}; same lesions, observer, locations and validity filter; three denoisers per prior, 200 images per set) with
Poisson data at $I_0=100$ and $I_0=30$ (effective $\sigma\approx0.06$ and $0.1$) and the matrix-free operator, tuning each method on its prior's own validation distribution
(Table~\ref{tab:lesion-poisson}, Figure~\ref{fig:lesion-poisson}). TV is the ADMM implementation without the box constraint; the box-constrained control of E9 was run for Gaussian noise
only. Only these two doses were run; the other three doses of E12 were not.

\emph{Attenuation persists.} The paired AUC of PnP is below TV's in all four prior/test-set combinations and at both doses (differences of $-0.01$ to $-0.28$ at $I_0=100$, with one tie at $1.0$ for the largest high-contrast lesion, and $-0.02$ to $-0.19$ at $I_0=30$).
For the MNIST prior on MNIST ($\sigma_l=0.75$, $c=0.5$) the paired AUC is $0.60$ against $0.89$ for TV and the retention is $0.29$ against $0.60$ at $I_0=100$,
the same pattern as with Gaussian noise (AUC $0.69$ against $0.91$; retention $0.28$ against $0.59$), although PnP is $3.7$\,dB better in PSNR on the same images. At $I_0=30$ the gap narrows
($0.63$ against $0.77$) because TV's AUC falls while PnP's does not.

\emph{False lesions under mismatch persist.} For the MNIST-trained prior on Fashion-MNIST, PnP produces more false lesions than TV for lesions with $\sigma_l\ge0.75$ in 11 of the 12 dose/lesion-type cells, by factors of $1.7$--$5.5$ ($0.25$--$0.70$ against $0.06$--$0.28$ per image at $I_0=100$; the exception is $0.25$ against $0.28$). The matched Fashion-trained prior, on its own data and on MNIST, stays at $\le0.09$ false lesions per image. Retention above that of TV
for the mismatched prior should not be read as fidelity: invented contrast may contribute to it, although we did not separate the two.

\emph{New at the lower dose: false lesions in distribution.} For the MNIST prior on MNIST at $I_0=30$, PnP produces $0.058$--$0.073$ false lesions per image for the four lesion types with $0.75\le\sigma_l\le1.25$, against $0.000$--$0.035$ for TV,
whereas at $I_0=100$ it produces at most $0.01$ for any type. At high noise the in-distribution prior therefore starts to create lesion-like structure as well as to attenuate the inserted lesion, although at a rate far below that under mismatch.

\emph{TV has a failure of its own.} For the smallest low-contrast lesion at $I_0=100$ the TV reconstruction contains $0.63$ false lesions per image (noise blobs), against $0.01$ for PnP. The statement that PnP creates more false lesions than TV under mismatch therefore applies to lesions of $\sigma_l\ge0.75$\,px. Under a matched prior it is supported only at $I_0=30$, at low rates (at most about $0.07$ per image; Table~\ref{tab:lesion-poisson}), and not at $I_0=100$.
As in E9, where box-constrained TV also produced false lesions for the smallest lesion type, the false-lesion rate of TV depends on its implementation, so
the TV counts here are those of ADMM-TV and may be different for other TV solvers.

Uncertainty is limited by the design: the error bars in Figure~\ref{fig:lesion-poisson} show the variation over three denoisers only, TV is a single run, and with about 170 valid lesions per cell the sampling error of a paired AUC is of the order of $\pm0.03$, so differences below that, such as the ties for the
largest lesions, do not separate the methods. A tercile analysis of PnP's paired AUC by lesion observability showed a decrease with observability on MNIST images and no trend on Fashion-MNIST images (the same pattern as in E9); we do not interpret it, and one untested possibility is the background around the lesion.

\begin{table}[htbp]
\centering\small\setlength{\tabcolsep}{3pt}
\caption{Lesion insertion under Poisson noise, mean over three denoisers per prior (up to 200 test images; 20 angles; bicubic finer-grid data), for two of the eight lesion types.
pAUC: paired AUC; Ret.: retention; FL: false lesions per lesion-absent image. Each method is tuned on its own prior's validation distribution at the given dose.}\label{tab:lesion-poisson}
\resizebox{\linewidth}{!}{%
\begin{tabular}{llcccccccccccc}
\toprule
 & & \multicolumn{6}{c}{$I_0=100$} & \multicolumn{6}{c}{$I_0=30$}\\
\cmidrule(lr){3-8}\cmidrule(lr){9-14}
 & & \multicolumn{2}{c}{pAUC} & \multicolumn{2}{c}{Ret.} & \multicolumn{2}{c}{FL} & \multicolumn{2}{c}{pAUC} & \multicolumn{2}{c}{Ret.} & \multicolumn{2}{c}{FL}\\
\cmidrule(lr){3-4}\cmidrule(lr){5-6}\cmidrule(lr){7-8}\cmidrule(lr){9-10}\cmidrule(lr){11-12}\cmidrule(lr){13-14}
Prior $\to$ test set & $\sigma_l/c$ & TV & PnP & TV & PnP & TV & PnP & TV & PnP & TV & PnP & TV & PnP\\
\midrule
MNIST $\to$ MNIST    & 0.75/0.5  & .89 & .60 & .60 & .29 & .010 & .007 & .77 & .63 & .48 & .34 & .005 & .067\\
                     & 1.25/0.25 & .88 & .60 & .60 & .27 & .020 & .007 & .75 & .65 & .46 & .46 & .005 & .068\\
MNIST $\to$ Fashion  & 0.75/0.5  & .81 & .72 & .64 & .84 & .240 & .443 & .73 & .63 & .42 & .57 & .230 & .518\\
                     & 1.25/0.25 & .84 & .72 & .65 & .84 & .255 & .438 & .71 & .61 & .42 & .54 & .225 & .518\\
Fashion $\to$ Fashion & 0.75/0.5 & .79 & .64 & .46 & .23 & .075 & .020 & .69 & .58 & .25 & .11 & .020 & .038\\
                     & 1.25/0.25 & .85 & .71 & .48 & .25 & .075 & .020 & .68 & .56 & .27 & .16 & .040 & .038\\
Fashion $\to$ MNIST  & 0.75/0.5  & .88 & .66 & .48 & .31 & .000 & .003 & .78 & .64 & .33 & .27 & .000 & .007\\
                     & 1.25/0.25 & .89 & .66 & .49 & .33 & .000 & .003 & .78 & .65 & .34 & .34 & .000 & .007\\
\bottomrule
\end{tabular}}
\end{table}

\subsection{E14: Data, noise model and simulator}\label{sec:e14}
\begin{figure}[htbp]
\centering
\includegraphics[width=\linewidth]{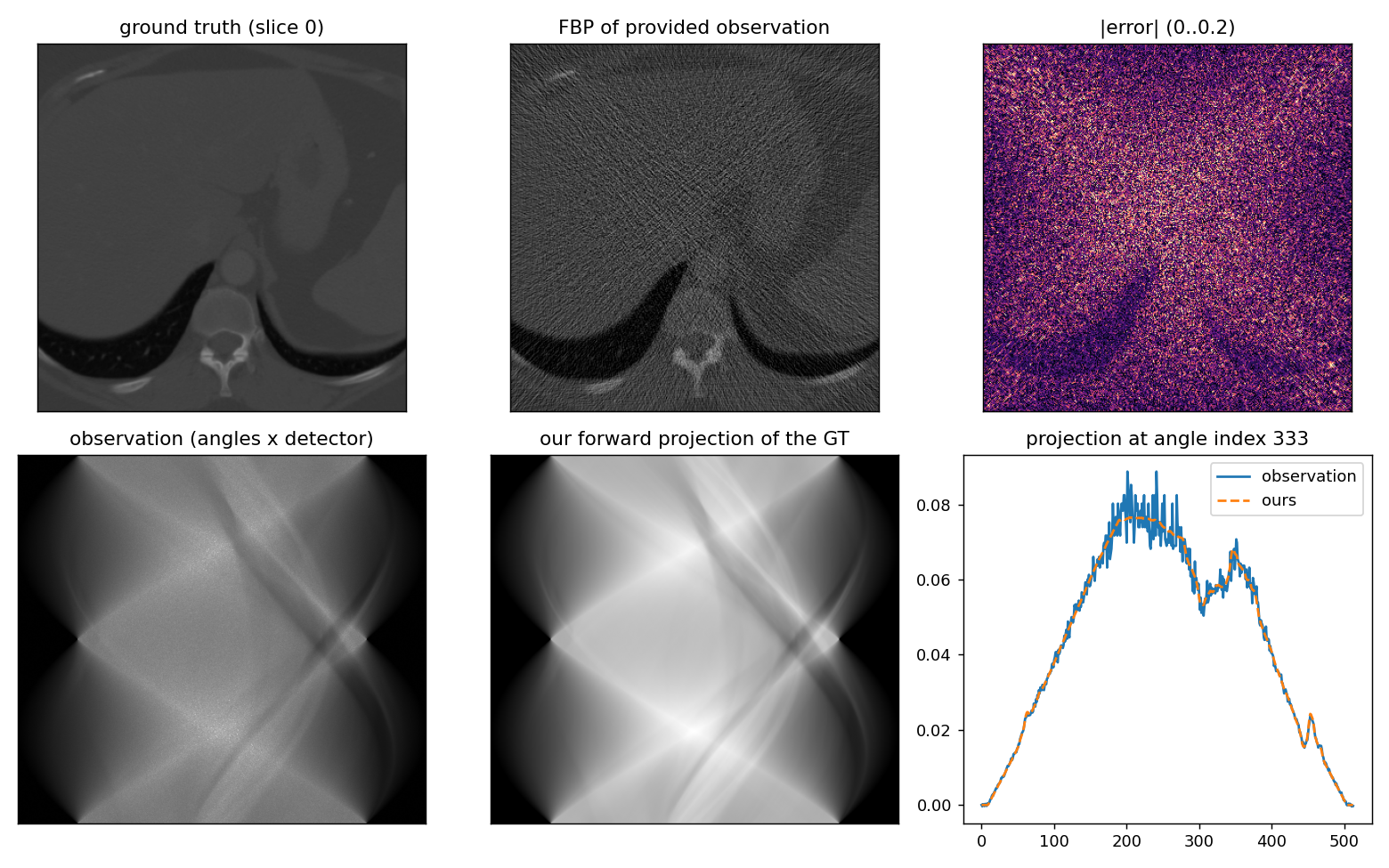}
\caption{Geometry check on one slice (slice 0). Top: ground truth, FBP of the supplied observation, and the absolute error of that FBP. Bottom: the supplied sinogram, our forward projection of the ground truth, and one projection (angle index 333) of the observation (solid) and of our projection (dashed). The observation is the noisy version of our noise-free projection, with matching structure. The fit statistics (relative residual $0.038$, correlation $0.9979$, with a fitted least-squares scale) are pooled over 8 evenly spaced test slices (Section~\ref{sec:lodopab}); the figure shows only the first of them.}\label{fig:geom}
\end{figure}

The geometry fit, the noise-model checks and the dose thinning are described in Sections~\ref{sec:lodopab}--\ref{sec:sim3}; Figure~\ref{fig:geom} illustrates the geometry fit for one test slice. Hyperparameters are tuned on \emph{real} (thinned) observations of the validation slices, whereas every reported comparison is evaluated on data \emph{simulated} on a finer grid from the test slices. Table~\ref{tab:simcheck} therefore compares the TV PSNR obtained from the real (thinned) observations with that obtained from our simulation of the same slices. On the first 8 test slices the two agree within $0.15$\,dB in all nine conditions (one noise realisation each; mean difference $+0.03$\,dB, simulated minus real). On all 40 slices at $K=60$ the TV values are $31.91$ (real) and $32.04$\,dB (simulated) at full dose and $29.82$ and $29.83$\,dB at quarter dose; for FBP-Hann they are $30.65$ and $30.59$\,dB, and $28.73$ and $28.66$\,dB.

The differences are small compared with the method differences studied below ($\le2.3$\,dB), but they are not zero and their sign depends on the method: the simulation is slightly \emph{easier} than the real data for TV at full dose ($+0.07$ to $+0.15$\,dB at the three $K$ on 8 slices; $+0.13$\,dB at $K=60$ on 40 slices) and slightly \emph{harder} for FBP-Hann ($-0.06$\,dB at $K=60$ on 40 slices). At the two lower doses the differences are within $\pm0.07$\,dB. A higher variance of the real counts at high count levels (dispersion index, Section~\ref{sec:sim3}) would make the real data harder for every method, which does not match the TV sign, so we do not interpret these differences. Eight slices and a single noise draw do not allow a test, and we computed no interval. The check does \emph{not} show that the \emph{difference} between PnP and TV is the same on real and simulated data; it covers the TV baseline (and FBP-Hann at $K=60$) only, so we state the PnP--TV comparisons for the simulated data.

\begin{table}[htbp]
\centering\small\setlength{\tabcolsep}{5pt}
\caption{TV PSNR (dB) from real observations (thinned to the dose, subsampled to $K$ angles) and from the finer-grid simulation of the same slices (first 8 test slices).}\label{tab:simcheck}
\begin{tabular}{llccccccccc}
\toprule
$K$ & & \multicolumn{3}{c}{30} & \multicolumn{3}{c}{60} & \multicolumn{3}{c}{120}\\
\cmidrule(lr){3-5}\cmidrule(lr){6-8}\cmidrule(lr){9-11}
dose & & 1 & 0.25 & 0.1 & 1 & 0.25 & 0.1 & 1 & 0.25 & 0.1\\
\midrule
real & & 29.34 & 27.95 & 26.57 & 30.15 & 28.40 & 26.73 & 30.61 & 28.54 & 27.89\\
simulated & & 29.41 & 27.90 & 26.52 & 30.27 & 28.45 & 26.72 & 30.76 & 28.55 & 27.85\\
\bottomrule
\end{tabular}
\end{table}

\begin{table}[htbp]
\centering\small\setlength{\tabcolsep}{4pt}
\caption{Tuned parameters (20 validation slices per condition): Hann cut-off, TV weight, and $(\sigma_d,k)$ of PnP-const and PnP-TVinit.
}\label{tab:params3}
\begin{tabular}{llcccc}
\toprule
$K$ & dose & Hann & TV $\lambda$ & PnP-const $(\sigma_d,k)$ & PnP-TVinit $(\sigma_d,k)$\\
\midrule
30  & 1    & 0.2  & 0.003 & (0.02, 12) & (0.02, 4)\\
30  & 0.25 & 0.16 & 0.01  & (0.05, 8)  & (0.08, 4)\\
30  & 0.1  & 0.12 & 0.03  & (0.08, 12) & (0.05, 8)\\
60  & 1    & 0.3  & 0.003 & (0.02, 8)  & (0.02, 2)\\
60  & 0.25 & 0.2  & 0.01  & (0.08, 4)  & (0.02, 8)\\
60  & 0.1  & 0.12 & 0.03  & (0.05, 12) & (0.02, 12)\\
120 & 1    & 0.4  & 0.003 & (0.05, 2)  & (0.0039, 8)$^\dagger$\\
120 & 0.25 & 0.2  & 0.01  & (0.02, 12) & (0.0062, 8)\\
120 & 0.1  & 0.16 & 0.01  & (0.05, 8)  & (0.128, 1)$^{\dagger,\ddagger}$\\
\bottomrule
\end{tabular}\\[2pt]
{\footnotesize $^\dagger$ optimum on the boundary of the grid after the automatic extensions. $^\ddagger$ $\sigma_d$ above the training range $(0,0.1)$ of the denoiser.}
\end{table}

\subsection{E15: Learned prior versus TV on real slices at $K=60$}\label{sec:e15}
Table~\ref{tab:e15} compares PnP variants with TV on the 40 test slices at $K=60$ (one slice per patient; data simulated on a finer grid, Section~\ref{sec:sim3}); all methods are tuned on the same 20 validation slices.
TV is the box-constrained primal--dual solver of Section~\ref{sec:methods3} (150 iterations), so the comparison is already against the stronger TV of E2b. Besides PnP-const and PnP-TVinit (Section~\ref{sec:methods3}), the table contains
two variants introduced to test whether the PnP set-up, not the prior, is the weak point: \emph{PnP-anneal}, in which $\sigma_d$ decreases geometrically from $\sigma_0\in\{0.05,0.1\}$ to $\sigma_1\in\{0.01,0.03\}$ over $T\in\{20,40\}$ iterations (FBP-Hann start), and \emph{TV+denoise}, a single denoiser pass
on the TV image with tuned $\sigma$. Each variant is run with the 8-layer denoiser (d8) and with a deeper one (d17, 17 layers, trained for 20\,000 steps).

PnP-const is slightly but significantly below TV ($-0.17$ and $-0.10$\,dB at full and quarter dose) and PnP-TVinit is at TV level
at full dose ($+0.03$\,dB, interval $[-0.00,+0.06]$) and $+0.21$\,dB above it at quarter dose. The deeper denoiser does not help (PnP-const $-0.30$ and $-0.25$\,dB). Annealing $\sigma_d$ is worse than constant $\sigma_d$ at full dose ($-0.65$\,dB versus TV) and better at quarter dose ($+0.12$\,dB, but not for the deeper denoiser: $+0.05$, interval including zero).
One denoiser pass on the TV image gives TV (within $\pm0.05$\,dB with d8 and $\pm0.02$\,dB with d17). SSIM does not show the quarter-dose gain: PnP-TVinit [d8] is $0.003$ above TV at quarter dose and $0.002$ below it at full dose, and PnP-const is below TV by $0.013$ (full dose) and $0.016$ (quarter dose).

\emph{How far to trust the best variant.} The gain of PnP-TVinit at quarter dose is the largest positive entry of the table, but three cautions apply. (i)~PnP-TVinit and the other variants were introduced \emph{after} the first PnP-const result was below TV (Section~\ref{sec:history3}), so the best of several variants is
selected in part on the test slices and its $+0.21$\,dB is optimistic; no multiplicity correction is applied to the 16 intervals of the table. (ii)~PnP-TVinit performs more iterations than TV (150 TV iterations plus $k$ PnP iterations); the control is TV-long (E16), which is itself above TV at $K=60$ in the same noise draws ($+0.08$ and $+0.05$\,dB, Table~\ref{tab:grid3}), so the margin of PnP-TVinit over TV-long is smaller than over TV ($-0.03$ and $+0.16$\,dB, Table~\ref{tab:acc3}). (iii)~$0.2$\,dB is a small difference, and the paired intervals reflect sampling over 40 patients with one noise realisation each, not variation over denoiser training runs.
In short, on clinical-size slices at $K=60$ the learned prior is at best competitive with TV in PSNR rather than $\approx4$\,dB better as in the $28\times28$ experiments. This is not a like-for-like contrast: image size, data, noise model and the match between prior and test data all differ, and we do not claim to have isolated the reason.

\begin{table}[htbp]
\centering\small\setlength{\tabcolsep}{4pt}
\caption{$K=60$, 40 test slices: PSNR (dB) and SSIM, and the paired PSNR difference to TV (mean, 95\% bootstrap interval over patients; $^*$ excludes 0). d8: 8-layer denoiser; d17: 17-layer denoiser.}\label{tab:e15}
\resizebox{\linewidth}{!}{%
\begin{tabular}{lcccccc}
\toprule
 & \multicolumn{3}{c}{dose 1} & \multicolumn{3}{c}{dose 0.25}\\
\cmidrule(lr){2-4}\cmidrule(lr){5-7}
Method & PSNR & SSIM & $-$TV & PSNR & SSIM & $-$TV\\
\midrule
FBP-Hann            & 30.65 & 0.738 & ${-}1.27^{*}$ & 28.73 & 0.699 & ${-}1.09^{*}$\\
TV                  & 31.91 & 0.786 & -- & 29.82 & 0.750 & --\\
PnP-const [d8]      & 31.74 & 0.773 & ${-}0.17^{*}$ {\scriptsize$[{-}0.21,{-}0.13]$} & 29.72 & 0.734 & ${-}0.10^{*}$ {\scriptsize$[{-}0.18,{-}0.03]$}\\
PnP-TVinit [d8]     & 31.94 & 0.784 & $+0.03$ {\scriptsize$[{-}0.00,+0.06]$} & 30.03 & 0.753 & $+0.21^{*}$ {\scriptsize$[+0.17,+0.25]$}\\
PnP-anneal [d8]     & 31.26 & 0.769 & ${-}0.65^{*}$ {\scriptsize$[{-}0.77,{-}0.52]$} & 29.94 & 0.750 & $+0.12^{*}$ {\scriptsize$[+0.05,+0.18]$}\\
TV+denoise [d8]     & 31.91 & 0.785 & $+0.00$ {\scriptsize$[{-}0.02,+0.03]$} & 29.77 & 0.747 & ${-}0.05$ {\scriptsize$[{-}0.10,+0.01]$}\\
PnP-const [d17]     & 31.61 & 0.777 & ${-}0.30^{*}$ {\scriptsize$[{-}0.35,{-}0.26]$} & 29.57 & 0.737 & ${-}0.25^{*}$ {\scriptsize$[{-}0.31,{-}0.19]$}\\
PnP-TVinit [d17]    & 31.89 & 0.785 & ${-}0.02$ {\scriptsize$[{-}0.06,+0.03]$} & 29.97 & 0.755 & $+0.15^{*}$ {\scriptsize$[+0.12,+0.18]$}\\
PnP-anneal [d17]    & 31.20 & 0.772 & ${-}0.71^{*}$ {\scriptsize$[{-}0.85,{-}0.58]$} & 29.87 & 0.753 & $+0.05$ {\scriptsize$[{-}0.01,+0.12]$}\\
TV+denoise [d17]    & 31.90 & 0.786 & ${-}0.02$ {\scriptsize$[{-}0.05,+0.02]$} & 29.84 & 0.752 & $+0.02^{*}$ {\scriptsize$[+0.00,+0.04]$}\\
\bottomrule
\end{tabular}}
\end{table}

\subsection{E16: Accuracy over the dose $\times$ angle grid}\label{sec:e16}
Table~\ref{tab:grid3} gives the mean PSNR and SSIM of the five methods in the nine conditions of the lesion-free arm of the lesion experiment (same 40 slices in every condition, one noise realisation per condition), and Table~\ref{tab:acc3} the paired differences. The top row of Figure~\ref{fig:grid3} (shown with E17) displays the PSNR and SSIM differences of Table~\ref{tab:acc3} as heat maps over the dose $\times$ angle grid.
TV-long (300 instead of 150 iterations, with a tuned weight) is the classical comparator we emphasise, because it has more iterations than TV and thus controls for the iteration count in which PnP differs from TV. It is not matched to PnP in cost or in tuned degrees of freedom (PnP-const uses at most 12 iterations, PnP-TVinit at most $150+12$; Table~\ref{tab:params3}), and it was introduced after the first lesion results (Section~\ref{sec:history3}). No correction for multiplicity is applied to the $36$ intervals of Table~\ref{tab:acc3}, and since the conditions share the same slices they are not independent.

\emph{What holds in all nine conditions.} PnP-const and TV-long beat FBP-Hann in PSNR (PnP-const by $0.92$--$1.36$\,dB, TV-long by $0.81$--$2.28$\,dB) and in SSIM.

\emph{PnP-const against TV-long.} PnP-const is \emph{not} better than TV-long on average: its PSNR difference lies between $-0.93$ and $+0.33$\,dB (mean over the nine conditions $-0.13$\,dB), significantly positive in three conditions ($K=30$ and $60$ at dose 0.1, and $K=120$ at full dose), significantly negative in four ($K=30$, $60$ at full dose; $K=60$ at quarter dose; $K=120$ at dose 0.1) and not significant in two.
Its SSIM is below TV-long's in eight of nine conditions, significantly so in seven, and above it in one ($+0.001$, not significant). The sign does not follow dose or angle count monotonically: the low-dose, few-view advantage at $K=30$ and $60$ is reversed at $K=120$, dose 0.1. We therefore do not summarise the grid as a trend.

\emph{PnP-TVinit.} PnP-TVinit has the highest mean PSNR in seven of nine conditions, by at most $0.49$\,dB over TV-long
(significantly higher in five, lower in two: $K=30$, full dose, $-0.15$\,dB, and $K=60$, full dose, $-0.03$\,dB), but the highest (or tied-highest) mean SSIM in only five. Some of its cells are not denoiser effects (Section~\ref{sec:methods3}): at $K=120$, full dose, the tuned $\sigma_d=0.0039$ means almost no denoising, so
the $+0.41$\,dB over TV-long (and $+0.28$\,dB over TV) comes from the additional data-consistency iterations started at the TV solution, not from the learned prior; at $K=120$, dose 0.1, the tuned $k=1$ is a single step with $\sigma_d$ above the denoiser's training range; and at $K=120$, quarter dose, $\sigma_d=0.0062$ is similarly small ($+0.35$\,dB over TV-long).
The largest margin, $+0.49$\,dB at $K=60$, dose 0.1, is in a cell with interior optimum.

\emph{Scale of the differences.} TV is the same method on the same slices in the $K=60$ cells here and in E15, yet its mean PSNR differs by $0.13$\,dB at full dose ($32.04$ versus $31.91$\,dB) and $0.01$\,dB at quarter dose, while PnP-const $-$ TV changes by only $0.02$\,dB at full dose ($-0.19$ versus $-0.17$\,dB). Differences between methods are thus more stable than absolute levels, but margins of a few tenths of a dB are within what a different noise
realisation can move, and we do not rank methods on differences below that.
TV-long is slightly worse than TV at $K=120$, full dose ($32.50$ vs.\ $32.64$\,dB), which suggests that the 150-iteration TV is itself a tuned object and that the weight grid is coarse.

The ranking by PSNR is not the ranking by lesion detectability (E17): TV-long is $0.81$--$2.28$\,dB better than FBP-Hann in PSNR but has a lower paired AUC in all nine conditions, so the accuracy results of this subsection should not be read as evidence about diagnostic usefulness.

\begin{table}[htbp]
\centering\small\setlength{\tabcolsep}{3pt}
\caption{Mean PSNR (dB, upper block) and SSIM (lower block) of the lesion-free reconstructions, 40 patients (best in each condition in bold; ties in bold). Columns: $K$ angles / dose.}\label{tab:grid3}
\resizebox{\linewidth}{!}{%
\begin{tabular}{lccccccccc}
\toprule
 & \multicolumn{3}{c}{$K=30$} & \multicolumn{3}{c}{$K=60$} & \multicolumn{3}{c}{$K=120$}\\
\cmidrule(lr){2-4}\cmidrule(lr){5-7}\cmidrule(lr){8-10}
Method & 1 & 0.25 & 0.1 & 1 & 0.25 & 0.1 & 1 & 0.25 & 0.1\\
\midrule
\multicolumn{10}{l}{PSNR (dB)}\\
FBP-Hann & 29.02 & 27.72 & 26.63 & 30.59 & 28.66 & 27.11 & 31.70 & 29.18 & 27.94 \\
TV & 30.98 & 29.15 & 27.69 & 32.04 & 29.83 & 27.93 & 32.64 & 30.07 & 29.13 \\
TV-long & \textbf{31.31} & 29.21 & 27.72 & \textbf{32.12} & 29.88 & 27.98 & 32.50 & 30.11 & 29.21 \\
PnP-const & 30.38 & 29.05 & 27.86 & 31.85 & 29.68 & 28.31 & 32.64 & 30.23 & 28.86 \\
PnP-TVinit & 31.15 & \textbf{29.22} & \textbf{27.96} & 32.09 & \textbf{30.04} & \textbf{28.47} & \textbf{32.92} & \textbf{30.46} & \textbf{29.26} \\
\midrule
\multicolumn{10}{l}{SSIM}\\
FBP-Hann & 0.692 & 0.666 & 0.655 & 0.738 & 0.699 & 0.680 & 0.763 & 0.725 & 0.693 \\
TV & 0.765 & 0.733 & 0.710 & 0.788 & 0.750 & 0.717 & 0.799 & 0.756 & 0.741 \\
TV-long & \textbf{0.773} & \textbf{0.737} & 0.713 & \textbf{0.790} & \textbf{0.753} & 0.720 & 0.797 & 0.758 & \textbf{0.745} \\
PnP-const & 0.745 & 0.726 & 0.707 & 0.775 & 0.733 & 0.721 & 0.793 & 0.749 & 0.731 \\
PnP-TVinit & 0.767 & 0.730 & \textbf{0.721} & 0.786 & \textbf{0.753} & \textbf{0.731} & \textbf{0.805} & \textbf{0.761} & 0.740 \\
\bottomrule
\end{tabular}}
\end{table}

\begin{table}[htbp]
\centering\small\setlength{\tabcolsep}{3pt}
\caption{Paired differences over patients (mean, 95\% bootstrap interval; $^*$ excludes 0), 40 patients.}\label{tab:acc3}
\resizebox{\linewidth}{!}{%
\begin{tabular}{llcccc}
\toprule
$K$ & dose & PSNR: PnP-const $-$ TV-long & SSIM: PnP-const $-$ TV-long & PSNR: PnP-TVinit $-$ TV-long & PSNR: PnP-const $-$ FBP-Hann\\
\midrule
30 & 1 & ${-}0.93^{*}$ {\scriptsize$[{-}1.09,{-}0.75]$} & ${-}0.027^{*}$ {\scriptsize$[{-}0.035,{-}0.021]$} & ${-}0.15^{*}$ {\scriptsize$[{-}0.24,{-}0.06]$} & $+1.36^{*}$ {\scriptsize$[+1.15,+1.58]$} \\
30 & 0.25 & ${-}0.15$ {\scriptsize$[{-}0.36,+0.08]$} & ${-}0.011^{*}$ {\scriptsize$[{-}0.019,{-}0.002]$} & $+0.01$ {\scriptsize$[{-}0.19,+0.24]$} & $+1.33^{*}$ {\scriptsize$[+1.06,+1.62]$} \\
30 & 0.1 & $+0.13^{*}$ {\scriptsize$[+0.03,+0.23]$} & ${-}0.006$ {\scriptsize$[{-}0.014,+0.001]$} & $+0.24^{*}$ {\scriptsize$[+0.19,+0.29]$} & $+1.22^{*}$ {\scriptsize$[+0.98,+1.48]$} \\
60 & 1 & ${-}0.28^{*}$ {\scriptsize$[{-}0.33,{-}0.22]$} & ${-}0.015^{*}$ {\scriptsize$[{-}0.019,{-}0.011]$} & ${-}0.03^{*}$ {\scriptsize$[{-}0.06,{-}0.00]$} & $+1.25^{*}$ {\scriptsize$[+0.97,+1.56]$} \\
60 & 0.25 & ${-}0.20^{*}$ {\scriptsize$[{-}0.28,{-}0.13]$} & ${-}0.020^{*}$ {\scriptsize$[{-}0.027,{-}0.012]$} & $+0.16^{*}$ {\scriptsize$[+0.11,+0.21]$} & $+1.02^{*}$ {\scriptsize$[+0.65,+1.41]$} \\
60 & 0.1 & $+0.33^{*}$ {\scriptsize$[+0.23,+0.43]$} & $+0.001$ {\scriptsize$[{-}0.003,+0.005]$} & $+0.49^{*}$ {\scriptsize$[+0.37,+0.62]$} & $+1.20^{*}$ {\scriptsize$[+1.04,+1.37]$} \\
120 & 1 & $+0.14^{*}$ {\scriptsize$[+0.05,+0.24]$} & ${-}0.004^{*}$ {\scriptsize$[{-}0.007,{-}0.001]$} & $+0.41^{*}$ {\scriptsize$[+0.32,+0.50]$} & $+0.95^{*}$ {\scriptsize$[+0.54,+1.39]$} \\
120 & 0.25 & $+0.13$ {\scriptsize$[{-}0.02,+0.27]$} & ${-}0.009^{*}$ {\scriptsize$[{-}0.013,{-}0.006]$} & $+0.35^{*}$ {\scriptsize$[+0.25,+0.45]$} & $+1.05^{*}$ {\scriptsize$[+0.89,+1.22]$} \\
120 & 0.1 & ${-}0.34^{*}$ {\scriptsize$[{-}0.47,{-}0.22]$} & ${-}0.013^{*}$ {\scriptsize$[{-}0.020,{-}0.007]$} & $+0.06$ {\scriptsize$[{-}0.07,+0.19]$} & $+0.92^{*}$ {\scriptsize$[+0.73,+1.13]$} \\
\bottomrule
\end{tabular}}
\end{table}

\subsection{E17: Lesion detection}\label{sec:e17}
\begin{figure}[htbp]
\centering
\includegraphics[width=\linewidth]{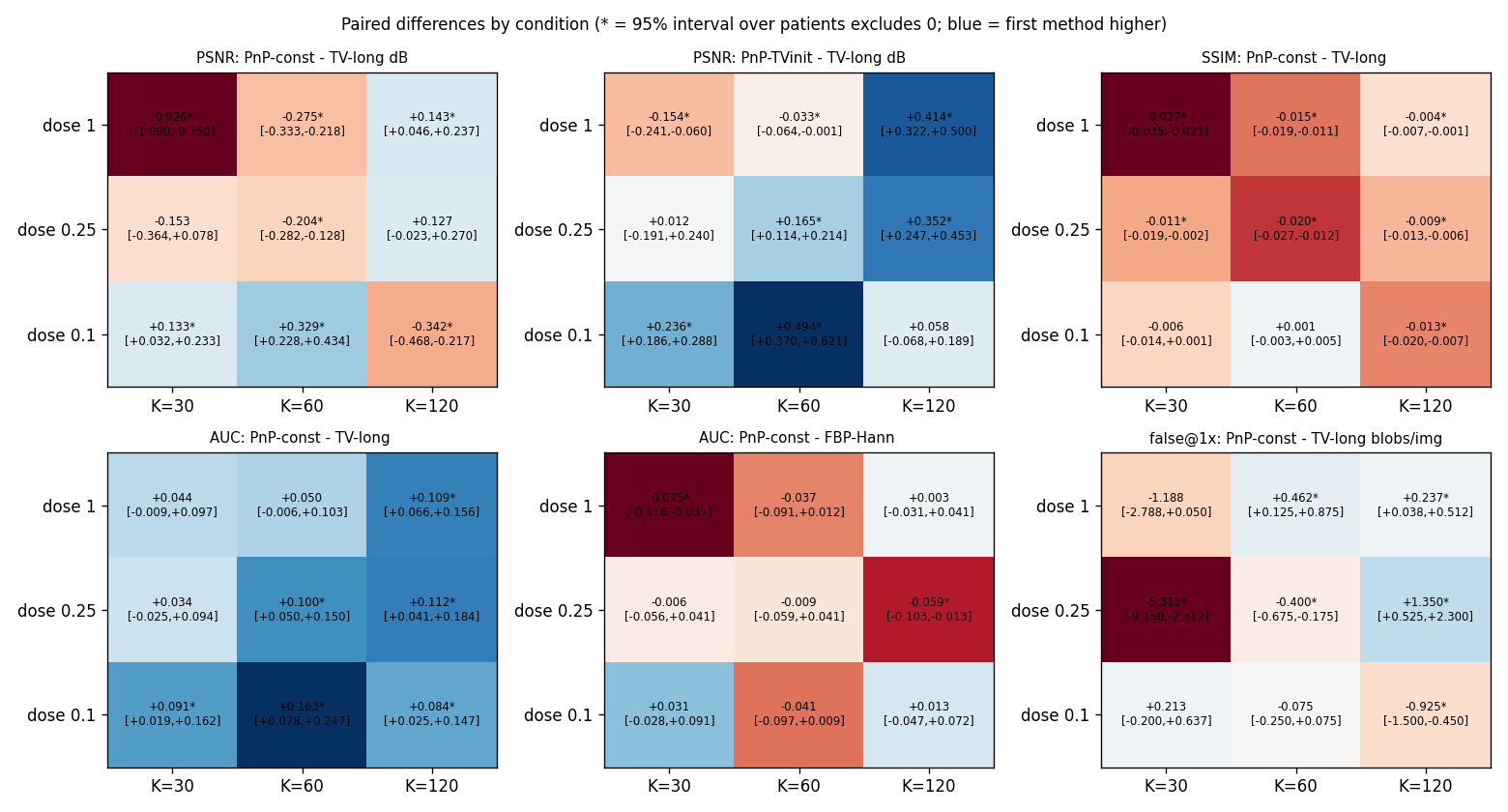}
\caption{Paired differences over the nine dose $\times$ angle conditions (cell: mean, 95\% interval over 40 patients; $^*$ excludes 0; blue: the first method is higher). Top row: PSNR (PnP-const and PnP-TVinit against TV-long) and SSIM (E16); bottom row: paired AUC of PnP-const against TV-long and against FBP-Hann (E17), and the difference in false blobs per slice at $\gamma=1$ (E18).}\label{fig:grid3}
\end{figure}

\emph{Grid (Table~\ref{tab:det3}, Figure~\ref{fig:grid3}).} Pooled over the two lesion sizes (contrast $0.06$; up to 160 lesions per size), PnP-const has a higher paired AUC than TV-long in all nine conditions, by $0.03$ to $0.16$, with intervals excluding zero in six (all three $K=120$ conditions, $K=60$ at doses $0.25$ and $0.1$, $K=30$ at dose $0.1$); it is not significant at full dose for $K=30$ and $60$.
Against FBP-Hann, however, PnP-const is never significantly better (three positive, six negative or zero differences) and is significantly worse in two conditions ($K=30$, full dose, $-0.075$; $K=120$, quarter dose, $-0.059$): the learned prior is about level with FBP-Hann and thus avoids the detection deficit of PSNR-tuned TV, but does not add to FBP's detectability.
TV-long, tuned for PSNR, has a lower AUC than FBP-Hann in all nine conditions (significantly in seven, down to $-0.20$ at $K=60$, dose $0.1$) while being $0.81$--$2.28$\,dB better in PSNR: PSNR and detectability rank the methods in opposite orders. PnP-TVinit exceeds TV-long in AUC in all nine conditions (significantly in seven; $+0.02$ to $+0.17$).
The choice of comparator matters: against the 150-iteration TV the advantage of PnP-const is $0.000$ and $+0.006$ at full dose for $K=60$ and $30$, because at those conditions TV-long has a lower AUC than TV (at $K=60$, full dose, by $0.05$). The conclusion ``PnP beats TV in detection'' is
therefore specific to a PSNR-tuned TV with the iteration count of TV-long; we did not tune TV, nor PnP, for detection on real slices (in the Gaussian experiments of E10 the detection-tuned TV changed very little, but that does not transfer automatically).

\emph{Retention.} At $K=60$ and full dose the regularised methods retain only $7$--$32\%$ of the local-contrast difference of the inserted lesion (FBP-Hann $31$--$67\%$); the detection scores of TV and PnP are therefore obtained mainly from noise suppression, not from preserving lesion contrast.

\emph{Earlier analysis at $K=60$ (E17b).} With four lesion types (two sizes, contrasts $0.03$ and $0.06$; 160 lesions per type pooled), re-tuned on 20 slices, the pooled AUC difference PnP-const $-$ TV-long is $+0.023$ ($[-0.013,+0.061]$) at full dose and $+0.073$ ($[+0.023,+0.128]$) at quarter dose, in line with the grid values ($+0.050$ and $+0.100$; different lesion sets and noise). At both doses PnP-const is level with FBP-Hann ($-0.008$, $+0.016$; not significant) and TV is below it ($-0.037$, not significant; $-0.056$, $[-0.11,-0.00]$).
At $\sigma_l=4$\,px and $c=0.06$ all methods reach AUC $0.95$--$0.96$ at full dose, so the large lesions do not discriminate between methods at that dose (ceiling).

\emph{Contrast with the $28\times28$ experiments.} In E9 the learned prior had a \emph{lower} paired AUC than TV in distribution ($-0.04$ to $-0.23$), at a PSNR advantage of about $4$\,dB. On real slices the sign is reversed: PnP-const is above TV-long in all nine conditions and above TV in eight of nine (significantly in seven),
at a PSNR that differs from TV-long's by between $-0.9$ and $+0.3$\,dB. We did not test why. Candidate explanations that this study cannot separate are the lesion scale relative to the pixel grid ($\sigma_l=2$--$4$\,px here against $0.5$--$2$\,px), the noise texture of real slices against the digit backgrounds, and the different amount of smoothing that the
PSNR-tuned operating points produce; the observer, which sums a disc against an annulus, rewards any method that suppresses background fluctuations at the lesion scale. The result is a property of this prior, observer and lesion type, not a general ranking of learned and classical regularisers.

\emph{Scope.} The lesions have a contrast of $0.06$ on the $\mu/\mu_{\max}$ scale, which is high compared with clinically low-contrast lesions (Section~\ref{sec:observer3}), the observer knows the lesion location, there is one noise realisation per condition and 40 patients, and the AUC of a single cell has a precision of about $\pm0.05$. A detection advantage says nothing about false structures; these are the subject of E18.

\begin{table}[htbp]
\centering\small\setlength{\tabcolsep}{3pt}
\caption{Paired AUC differences ($\sigma_l\in\{2,4\}$\,px pooled, $c=0.06$, 40 patients; mean, 95\% interval; $^*$ excludes 0).}\label{tab:det3}
\resizebox{\linewidth}{!}{%
\begin{tabular}{llccccc}
\toprule
$K$ & dose & PnP-const $-$ TV-long & PnP-const $-$ TV & PnP-const $-$ FBP-Hann & TV-long $-$ FBP-Hann & PnP-TVinit $-$ TV-long\\
\midrule
30 & 1 & $+0.044$ {\scriptsize$[{-}0.009,+0.097]$} & $+0.006$ {\scriptsize$[{-}0.050,+0.059]$} & ${-}0.075^{*}$ {\scriptsize$[{-}0.116,{-}0.037]$} & ${-}0.119^{*}$ {\scriptsize$[{-}0.169,{-}0.072]$} & $+0.066^{*}$ {\scriptsize$[+0.019,+0.113]$} \\
30 & 0.25 & $+0.034$ {\scriptsize$[{-}0.025,+0.094]$} & $+0.078^{*}$ {\scriptsize$[+0.013,+0.144]$} & ${-}0.006$ {\scriptsize$[{-}0.056,+0.041]$} & ${-}0.041$ {\scriptsize$[{-}0.103,+0.025]$} & $+0.069^{*}$ {\scriptsize$[+0.006,+0.131]$} \\
30 & 0.1 & $+0.091^{*}$ {\scriptsize$[+0.019,+0.162]$} & $+0.137^{*}$ {\scriptsize$[+0.059,+0.216]$} & $+0.031$ {\scriptsize$[{-}0.028,+0.091]$} & ${-}0.059$ {\scriptsize$[{-}0.141,+0.019]$} & $+0.053$ {\scriptsize$[{-}0.022,+0.128]$} \\
60 & 1 & $+0.050$ {\scriptsize$[{-}0.006,+0.103]$} & $+0.000$ {\scriptsize$[{-}0.050,+0.050]$} & ${-}0.037$ {\scriptsize$[{-}0.091,+0.012]$} & ${-}0.087^{*}$ {\scriptsize$[{-}0.144,{-}0.031]$} & $+0.069^{*}$ {\scriptsize$[+0.031,+0.109]$} \\
60 & 0.25 & $+0.100^{*}$ {\scriptsize$[+0.050,+0.150]$} & $+0.078^{*}$ {\scriptsize$[+0.028,+0.131]$} & ${-}0.009$ {\scriptsize$[{-}0.059,+0.041]$} & ${-}0.109^{*}$ {\scriptsize$[{-}0.175,{-}0.047]$} & $+0.084^{*}$ {\scriptsize$[+0.031,+0.141]$} \\
60 & 0.1 & $+0.163^{*}$ {\scriptsize$[+0.078,+0.247]$} & $+0.144^{*}$ {\scriptsize$[+0.072,+0.219]$} & ${-}0.041$ {\scriptsize$[{-}0.097,+0.009]$} & ${-}0.203^{*}$ {\scriptsize$[{-}0.291,{-}0.119]$} & $+0.156^{*}$ {\scriptsize$[+0.084,+0.231]$} \\
120 & 1 & $+0.109^{*}$ {\scriptsize$[+0.066,+0.156]$} & $+0.084^{*}$ {\scriptsize$[+0.044,+0.125]$} & $+0.003$ {\scriptsize$[{-}0.031,+0.041]$} & ${-}0.106^{*}$ {\scriptsize$[{-}0.162,{-}0.053]$} & $+0.091^{*}$ {\scriptsize$[+0.050,+0.137]$} \\
120 & 0.25 & $+0.112^{*}$ {\scriptsize$[+0.041,+0.184]$} & $+0.081^{*}$ {\scriptsize$[+0.006,+0.156]$} & ${-}0.059^{*}$ {\scriptsize$[{-}0.103,{-}0.013]$} & ${-}0.172^{*}$ {\scriptsize$[{-}0.247,{-}0.094]$} & $+0.166^{*}$ {\scriptsize$[+0.109,+0.222]$} \\
120 & 0.1 & $+0.084^{*}$ {\scriptsize$[+0.025,+0.147]$} & $+0.084^{*}$ {\scriptsize$[+0.028,+0.147]$} & $+0.013$ {\scriptsize$[{-}0.047,+0.072]$} & ${-}0.072^{*}$ {\scriptsize$[{-}0.128,{-}0.016]$} & $+0.022$ {\scriptsize$[{-}0.041,+0.081]$} \\
\bottomrule
\end{tabular}}
\end{table}

\subsection{E18: False structures on real slices}\label{sec:e18}
Table~\ref{tab:absfb3} (CT rows) gives the mean number of false blobs per lesion-free slice at $\gamma=1$ (a response of at least one nominal lesion) for the four conditions that are also used in the shift study. A false blob is defined in Section~\ref{sec:observer3}; because the ground truth of LoDoPaB-CT is itself a reconstruction, ``false'' means unsupported by this reference and not absent from the patient.
In these four conditions PnP-const makes $0.24$--$1.38$ such blobs per slice and TV-long $0.00$--$6.06$ (FBP-Hann $5.8$--$26.8$), so the regularised methods make far fewer than FBP-Hann, but the counts of TV-long are not small in every condition.
Over the nine conditions the paired difference PnP-const $-$ TV-long at $\gamma=1$ (Table~\ref{tab:fb3}) is significantly positive in three ($K=60$ full dose: $+0.46$; $K=120$ full and quarter dose: $+0.24$, $+1.35$), significantly negative in three ($K=30$ quarter dose: $-5.31$; $K=60$ quarter dose: $-0.40$; $K=120$ dose 0.1: $-0.93$) and not significant in three, with no consistent pattern in dose or angle count: at $K=120$ the sign changes from positive to negative as the dose decreases, at $K=30$ it is negative in two of three conditions.
At $\gamma=1.5$ the excess of PnP-const over TV-long is at most $0.45$ blobs per slice ($K=120$, quarter dose). In the $K=60$ experiment (E17b, $\gamma=0.5$) PnP-const has fewer false blobs than TV at both doses ($-1.7$ at full dose, $-6.6$ at quarter dose). Counts are not comparable with E9: there the lesion-free images are digits on an empty background and the learned prior produced essentially none in distribution, whereas here the background is real anatomy, which contains many structures of lesion size.

\begin{figure}[p]
\centering
\includegraphics[height=0.9\textheight,keepaspectratio]{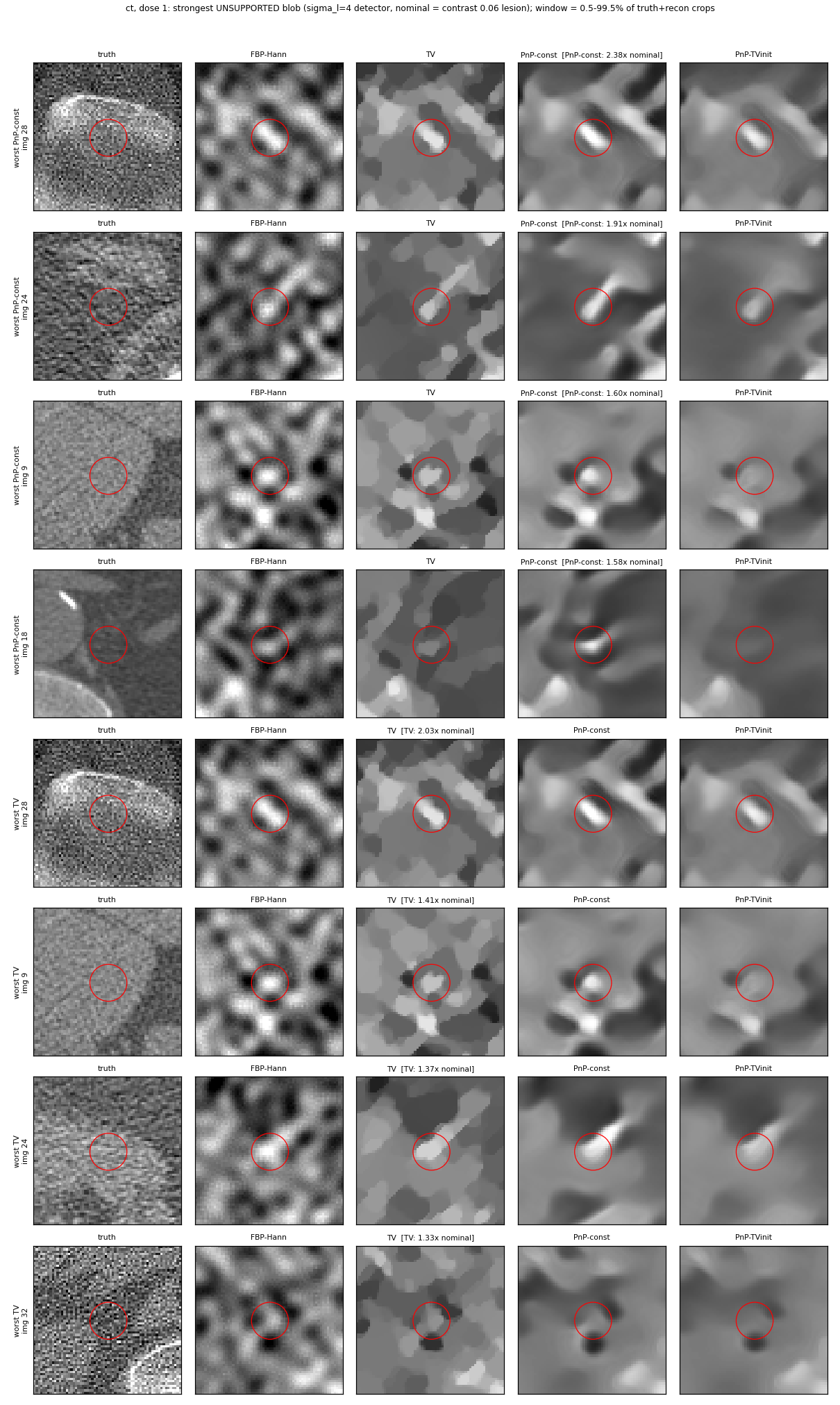}
\caption{Strongest unsupported blob per slice at full dose ($K=60$; $\sigma_l=4$\,px detector) for the four slices with the largest false-blob response in PnP-const (top four rows) and in TV (bottom four rows). Columns: ground truth crop, FBP-Hann, TV, PnP-const, PnP-TVinit; red circle: location of the blob; labels give the response as a multiple of the nominal lesion response.}\label{fig:gal1}
\end{figure}

\begin{figure}[p]
\centering
\includegraphics[height=0.9\textheight,keepaspectratio]{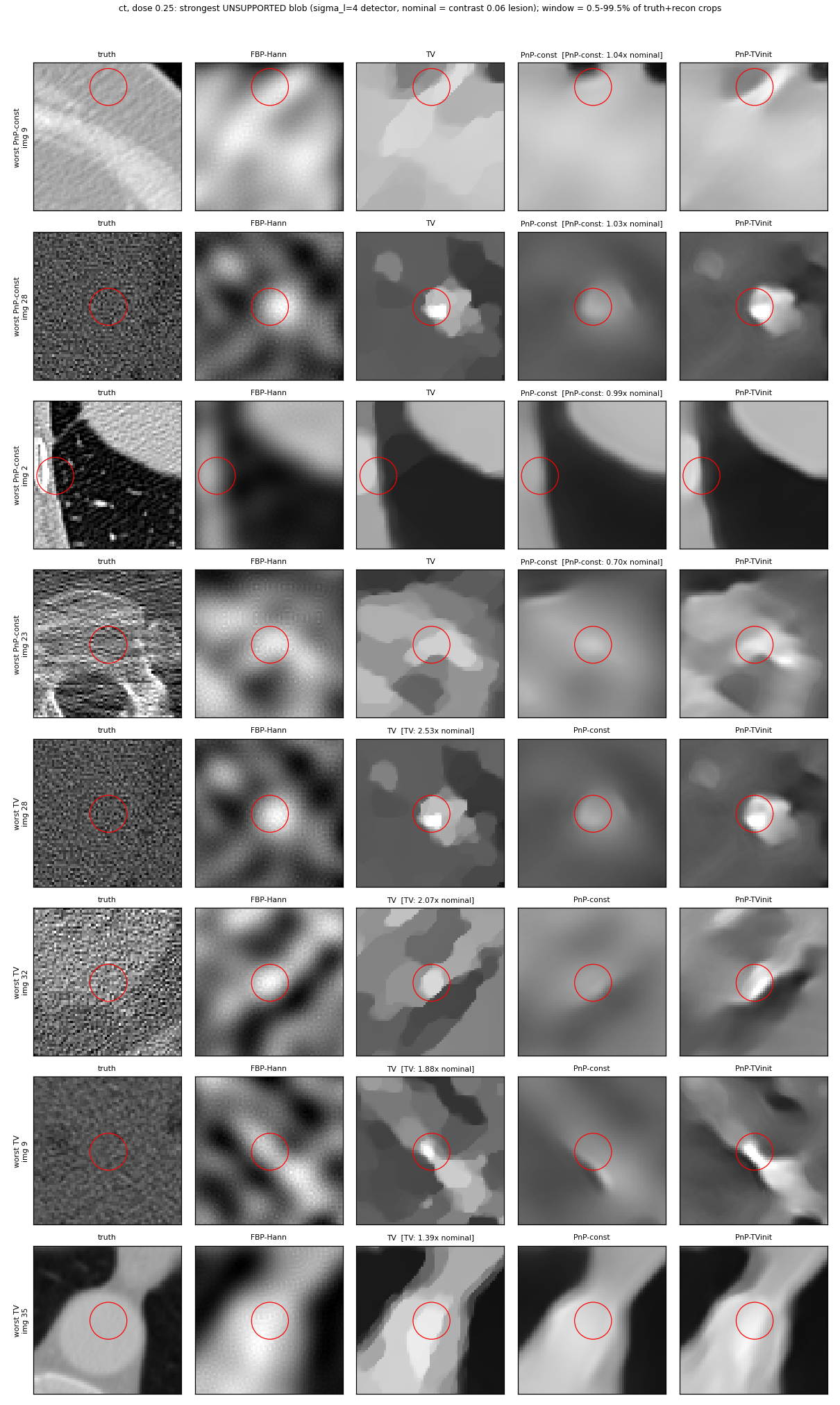}
\caption{As Figure~\ref{fig:gal1} at quarter dose.}\label{fig:gal025}
\end{figure}

The galleries of the strongest unsupported blob per slice (Figures~\ref{fig:gal1} and~\ref{fig:gal025}) show that the largest false blobs occur at the same location in FBP-Hann and in TV as well as in PnP (for example the slices 28 and 9 at full dose and most rows at quarter dose), so they appear to be driven by the data (few views, noise, streaks) and not by the prior. Only a few look specific to the prior: a crescent in slice 18 at full dose
(PnP-const only) and streaks in the PnP-TVinit image of slice 32 at quarter dose. We therefore find \emph{no evidence that the prior creates many lesion-like structures in distribution}, but we cannot exclude rare ones: 40 slices and one noise draw per condition do not bound a rare event, and the galleries show the worst cases selected by the methods' own responses, not a random sample.

\begin{table}[htbp]
\centering\small\setlength{\tabcolsep}{3pt}
\caption{Paired differences in false blobs per lesion-free slice on real CT slices (40 patients; mean, 95\% interval; $^*$ excludes 0), pooled over $\sigma_l\in\{2,4\}$; $\gamma$: threshold in units of the nominal lesion response.}\label{tab:fb3}
\resizebox{\linewidth}{!}{%
\begin{tabular}{llccc}
\toprule
$K$ & dose & $\gamma=1$: PnP-const $-$ TV-long & $\gamma=1$: PnP-const $-$ TV & $\gamma=1.5$: PnP-const $-$ TV-long\\
\midrule
30 & 1 & ${-}1.19$ {\scriptsize$[{-}2.79,+0.05]$} & ${-}0.54$ {\scriptsize$[{-}1.57,+0.36]$} & ${-}0.23$ {\scriptsize$[{-}0.64,+0.11]$} \\
30 & 0.25 & ${-}5.31^{*}$ {\scriptsize$[{-}9.15,{-}2.31]$} & ${-}4.81^{*}$ {\scriptsize$[{-}8.41,{-}2.00]$} & ${-}2.42^{*}$ {\scriptsize$[{-}4.56,{-}0.82]$} \\
30 & 0.1 & $+0.21$ {\scriptsize$[{-}0.20,+0.64]$} & $+0.31$ {\scriptsize$[{-}0.12,+0.76]$} & $+0.00$ {\scriptsize$[{-}0.14,+0.14]$} \\
60 & 1 & $+0.46^{*}$ {\scriptsize$[+0.12,+0.88]$} & $+0.19^{*}$ {\scriptsize$[+0.01,+0.40]$} & $+0.12^{*}$ {\scriptsize$[+0.03,+0.25]$} \\
60 & 0.25 & ${-}0.40^{*}$ {\scriptsize$[{-}0.68,{-}0.17]$} & ${-}0.60^{*}$ {\scriptsize$[{-}1.11,{-}0.19]$} & ${-}0.11^{*}$ {\scriptsize$[{-}0.25,{-}0.01]$} \\
60 & 0.1 & ${-}0.07$ {\scriptsize$[{-}0.25,+0.07]$} & ${-}0.05$ {\scriptsize$[{-}0.20,+0.09]$} & ${-}0.01$ {\scriptsize$[{-}0.06,+0.04]$} \\
120 & 1 & $+0.24^{*}$ {\scriptsize$[+0.04,+0.51]$} & $+0.11^{*}$ {\scriptsize$[+0.01,+0.25]$} & $+0.06$ {\scriptsize$[+0.00,+0.15]$} \\
120 & 0.25 & $+1.35^{*}$ {\scriptsize$[+0.53,+2.30]$} & $+1.35^{*}$ {\scriptsize$[+0.54,+2.29]$} & $+0.45^{*}$ {\scriptsize$[+0.16,+0.81]$} \\
120 & 0.1 & ${-}0.93^{*}$ {\scriptsize$[{-}1.50,{-}0.45]$} & ${-}0.88^{*}$ {\scriptsize$[{-}1.48,{-}0.39]$} & ${-}0.11^{*}$ {\scriptsize$[{-}0.26,{-}0.01]$} \\
\bottomrule
\end{tabular}}
\end{table}

\subsection{E19: Distribution shift to non-anatomical phantoms}\label{sec:e19}
The CT-trained prior and all tuned parameters are applied, without retuning, to 20 random piecewise-constant phantoms (ellipses, rectangles and a stripe or checker block on a faint disc, $\mu/\mu_{\max}$ scale; nothing like a CT slice, but the kind of image TV favours; same phantoms in every condition). This is a deliberately TV-friendly shift, so it tests how much of the learned prior's behaviour survives, not how it fares under realistic shifts such as another scanner or anatomy. The same lesion protocol is used. The shift was first run at $K=60$ (doses 1 and 0.25) and
then extended to $K=30$ and $120$; the $K=60$ analysis used four lesion types, the others two, so the designs differ slightly.

\emph{Accuracy and detection (Table~\ref{tab:shift3a}).} On the phantoms PnP-const is below TV-long in PSNR in all six conditions ($-0.11$ to $-0.75$\,dB; significantly in five of the six, the exception being $K=30$, full dose) and in SSIM in all four conditions where it was stored ($-0.095$ to $-0.190$, all significant, against $-0.004$ to $-0.027$ in the corresponding CT conditions).
At quarter dose its SSIM is also below FBP-Hann's on the phantoms ($-0.045$ at $K=30$ and $-0.041$ at $K=120$, both significant).
The detection advantage over TV-long is significant in one of six shift conditions ($K=120$, full dose: $+0.075$, $[+0.013,+0.137]$), against three of the six corresponding CT conditions ($K=60$ quarter dose, $K=120$ at both doses);
the shift effect (phantom minus CT) is significantly negative only at $K=120$, quarter dose ($-0.175$, $[-0.297,-0.056]$). The CT advantage is therefore not reliably transferred to the shifted images, but with 20 phantoms the interval widths of $0.08$--$0.19$ do not allow us to say that it is lost either.
In PSNR the deficit against TV-long on the phantoms is at most $0.75$\,dB, much smaller than the $3$--$4$\,dB reversals of the $28\times28$ experiments (E4, E8); relative to CT it is larger in some conditions and smaller in others (for example $-0.11$\,dB on phantoms against $-0.93$\,dB on CT at $K=30$, full dose; $-0.75$ against $-0.20$\,dB at $K=60$, quarter dose). In SSIM the deficit is large ($-0.095$ to $-0.190$).

\begin{figure}[p]
\centering
\includegraphics[height=0.9\textheight,keepaspectratio]{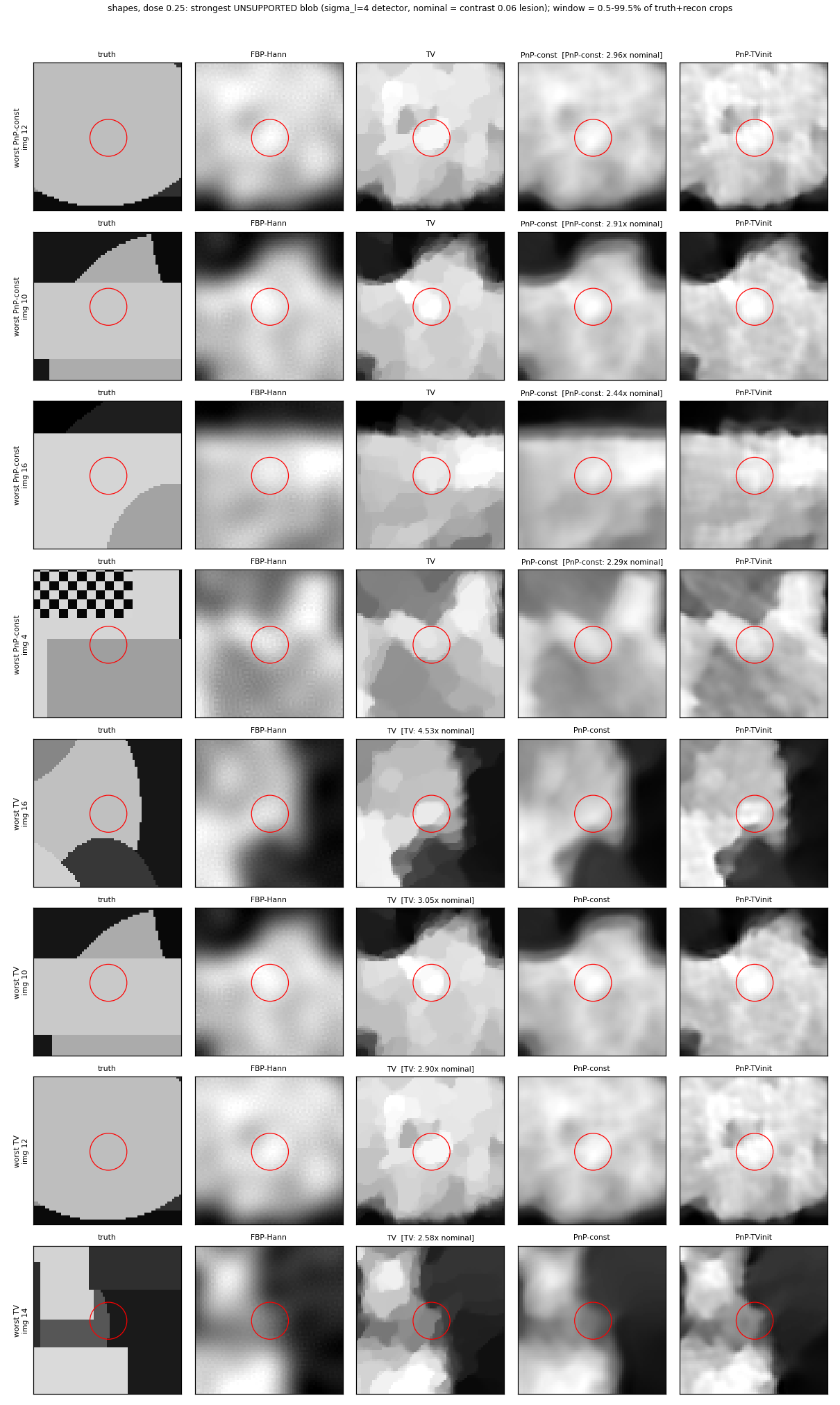}
\caption{Strongest unsupported blob per phantom at quarter dose ($K=60$; $\sigma_l=4$\,px detector) for the four phantoms with the largest false-blob response in PnP-const (top four rows) and in TV (bottom four rows). Columns: ground truth crop, FBP-Hann, TV, PnP-const, PnP-TVinit; red circle: location of the blob; labels give the response as a multiple of the nominal lesion response.}\label{fig:galshape}
\end{figure}

\emph{False blobs (Tables~\ref{tab:shift3a}--\ref{tab:absfb3}).} The sign of PnP-const $-$ TV at $\gamma=1$ follows the number of angles: negative at $K=30$ (not significant for $\sigma_l=4$ at quarter dose) and at full-dose $K=60$, mixed at $K=60$ quarter dose ($\sigma_l=2$: $-7.5$; $\sigma_l=4$: $+9.1$), and positive at $K=120$ ($+9.0$ to $+13.0$ per phantom at both doses; at $\gamma=1.5$ against TV-long $+10.2$ at full dose and $+3.8$ at quarter dose, both significant).
The absolute counts show why (Table~\ref{tab:absfb3}): the count of TV-long varies from $3.2$ to $83.9$ per phantom over the four conditions, that of PnP-const only from $12.8$ to $50.3$; at $K=30$ TV-long makes two to three times as many false blobs as FBP-Hann ($30.2$ vs.\ $15.5$ at quarter dose, $83.9$ vs.\ $31.6$ at full dose), at $K=120$ about a third or less ($3.2$ vs.\ $8.7$, i.e.\ $37\%$; $32.1$ vs.\ $118.7$, i.e.\ $27\%$).
So the ``excess'' of PnP-const at $K=120$ is relative to a TV that is unusually clean there. Relative to \emph{FBP-Hann} the picture is more informative: at full dose PnP-const has fewer false blobs on the phantoms ($-7.1$ at $K=30$, $-68.4$ at $K=120$, both significant), but at quarter dose it has more at both $K$ ($+1.0$, not significant, and $+4.0$ [$+2.4,+5.9$] at $\gamma=1$;
$+2.2$ [$+1.1,+3.4$] and $+2.7$ [$+1.4,+4.4$] at $\gamma=1.5$), in contrast to real slices, where it has significantly fewer than FBP-Hann in both quarter-dose conditions ($-9.0$ and $-4.4$). In this sense the shift turns the prior from suppressing to adding false blobs relative to the unregularised reconstruction at quarter dose, by a few blobs per phantom; this is significant at $K=120$ at both strengths and at $K=30$ only at $\gamma=1.5$. The pooled counts hide a dependence on lesion size: at $\gamma=1$, for $\sigma_l=2$ PnP-const has more false blobs than FBP-Hann on the phantoms in three of the four conditions ($+10.1$, $+9.7$ and $+10.6$ at $K=30$ full and quarter dose and $K=120$ quarter dose, all significant), for $\sigma_l=4$ fewer in all four ($-24.3$, $-7.7$, $-30.9$, $-2.5$).
At $K=120$, full dose, PnP-TVinit with almost no denoising ($\sigma_d=0.0039$) also produces $14.0$ more blobs than TV-long ($[+7.4,+22.0]$ at $\gamma=1$, for all $\sigma_l$), so at least part of the excess at $K=120$ is not due to the denoiser; we did not isolate the cause.
In the gallery of the strongest unsupported blob per phantom (Figure~\ref{fig:galshape}, $K=60$, quarter dose) the worst blobs again occur at the same location in FBP-Hann, TV and PnP, mostly in uniform regions of the phantom, with TV reaching up to $4.6$ times the nominal lesion response and the four worst PnP-const blobs between about $2.3$ and $3$ times; in these examples the prior does not create blobs that the data do not already suggest.

\begin{table}[htbp]
\centering\small\setlength{\tabcolsep}{3pt}
\caption{Distribution shift to phantoms (20 phantoms; mean, 95\% interval; $^*$ excludes 0). Left: false blobs per phantom at $\gamma=1$, PnP-const $-$ TV, by lesion size. Right: PnP-const $-$ TV-long on the phantoms: pooled paired AUC, PSNR (dB) and SSIM.
$K=60$ rows: AUC from four lesion types (E17b), PSNR interval from a bootstrap over the 20 phantoms, SSIM not stored.}\label{tab:shift3a}
\resizebox{\linewidth}{!}{%
\begin{tabular}{llcccccc}
\toprule
 & & \multicolumn{2}{c}{false blobs, $\gamma=1$ (vs TV)} & \multicolumn{3}{c}{vs TV-long}\\
\cmidrule(lr){3-4}\cmidrule(lr){5-7}
$K$ & dose & $\sigma_l=2$ & $\sigma_l=4$ & AUC & PSNR & SSIM\\
\midrule
30 & 1 & ${-}70.7^{*}$ {\scriptsize$[{-}107.0,{-}38.7]$} & ${-}31.8^{*}$ {\scriptsize$[{-}48.7,{-}16.8]$} & ${-}0.037$ {\scriptsize$[{-}0.113,+0.037]$} & ${-}0.11$ {\scriptsize$[{-}0.37,+0.18]$} & ${-}0.095^{*}$ {\scriptsize$[{-}0.125,{-}0.063]$} \\
30 & 0.25 & ${-}21.1^{*}$ {\scriptsize$[{-}34.3,{-}10.3]$} & ${-}1.8$ {\scriptsize$[{-}5.8,+1.8]$} & $+0.056$ {\scriptsize$[{-}0.031,+0.144]$} & ${-}0.53^{*}$ {\scriptsize$[{-}0.64,{-}0.40]$} & ${-}0.190^{*}$ {\scriptsize$[{-}0.204,{-}0.172]$} \\
60 & 1 & ${-}20.1^{*}$ {\scriptsize$[{-}34.4,{-}7.6]$} & ${-}6.5^{*}$ {\scriptsize$[{-}12.6,{-}1.4]$} & ${-}0.019$ {\scriptsize$[{-}0.06,+0.02]$} & ${-}0.30^{*}$ {\scriptsize$[{-}0.44,{-}0.14]$} & n/a \\
60 & 0.25 & ${-}7.5^{*}$ {\scriptsize$[{-}12.9,{-}2.8]$} & $+9.1^{*}$ {\scriptsize$[+6.0,+12.8]$} & $+0.050$ {\scriptsize$[{-}0.01,+0.11]$} & ${-}0.75^{*}$ {\scriptsize$[{-}0.82,{-}0.67]$} & n/a \\
120 & 1 & $+13.0^{*}$ {\scriptsize$[+8.8,+17.5]$} & $+9.2^{*}$ {\scriptsize$[+5.4,+13.2]$} & $+0.075^{*}$ {\scriptsize$[+0.013,+0.137]$} & ${-}0.51^{*}$ {\scriptsize$[{-}0.56,{-}0.45]$} & ${-}0.179^{*}$ {\scriptsize$[{-}0.190,{-}0.168]$} \\
120 & 0.25 & $+10.4^{*}$ {\scriptsize$[+5.9,+16.0]$} & $+9.0^{*}$ {\scriptsize$[+5.9,+12.3]$} & ${-}0.062$ {\scriptsize$[{-}0.162,+0.031]$} & ${-}0.62^{*}$ {\scriptsize$[{-}0.68,{-}0.57]$} & ${-}0.155^{*}$ {\scriptsize$[{-}0.162,{-}0.149]$} \\
\bottomrule
\end{tabular}}
\end{table}

\begin{table}[htbp]
\centering\small\setlength{\tabcolsep}{4pt}
\caption{Mean false blobs per lesion-free image at $\gamma=1$ (mean of $\sigma_l=2$ and $4$) on real CT slices and on the phantoms, and PnP-const $-$ FBP-Hann (mean, 95\% interval over patients or phantoms).}\label{tab:absfb3}
\begin{tabular}{llcccccc}
\toprule
set & $K$, dose & FBP-Hann & TV & TV-long & PnP-const & PnP-TVinit & PnP-const $-$ FBP\\
\midrule
CT     & 30, 0.25  & 9.78 & 5.56 & 6.06 & 0.75 & 0.64 & $-9.03^{*}$ {\scriptsize$[-13.21,-5.20]$}\\
CT     & 30, 1     & 5.78 & 1.04 & 1.69 & 0.50 & 0.32 & $-5.28^{*}$ {\scriptsize$[-7.54,-3.29]$}\\
CT     & 120, 0.25 & 5.79 & 0.02 & 0.02 & 1.38 & 0.39 & $-4.41^{*}$ {\scriptsize$[-7.34,-2.00]$}\\
CT     & 120, 1    & 26.79 & 0.12 & 0.00 & 0.24 & 0.18 & $-26.55^{*}$ {\scriptsize$[-43.93,-12.63]$}\\
phantom & 30, 0.25  & 15.48 & 27.95 & 30.20 & 16.50 & 11.65 & $+1.03$ {\scriptsize$[-0.68,+2.75]$}\\
phantom & 30, 1     & 31.60 & 75.72 & 83.90 & 24.48 & 57.30 & $-7.13^{*}$ {\scriptsize$[-10.68,-4.02]$}\\
phantom & 120, 0.25 & 8.72 & 3.08 & 3.18 & 12.75 & 11.48 & $+4.03^{*}$ {\scriptsize$[+2.35,+5.90]$}\\
phantom & 120, 1    & 118.68 & 39.20 & 32.08 & 50.30 & 46.12 & $-68.38^{*}$ {\scriptsize$[-92.68,-45.23]$}\\
\bottomrule
\end{tabular}
\end{table}

\section{Discussion}
\label{sec:discussion}

We organise the discussion around the three guiding questions, then state what the evidence says about mechanisms and about evaluation practice. All statements refer to one architecture (a noise-conditioned DnCNN inside PnP-PGD), two training distributions on $28\times28$ digits and one CT training run; the Limitations section lists the limits.

\paragraph{(1) Does the learned prior keep its advantage over TV when the image distribution changes?}
Not for far shifts. On $28\times28$ digits the in-distribution advantage is about $3.4$\,dB over box-constrained TV ($+3.8$\,dB over TV-ADMM; Sections~\ref{sec:e1e2} and~\ref{sec:seeds-results}). It halves for a near shift (EMNIST letters, $+1.9$\,dB over TV-ADMM) and reverses for an MNIST-trained prior on Fashion-MNIST ($-3.1\pm0.3$\,dB over TV-ADMM, three seeds) and on the Shepp--Logan phantom ($-4.3\pm0.1$\,dB), with the same signs against box-constrained TV (Sections~\ref{sec:e4} and~\ref{sec:seeds-results}). The reversal persists under Poisson noise ($-2.0$ to $-3.1$\,dB and $-4.4$ to $-6.1$\,dB; Section~\ref{sec:dose-results}), although in SSIM it appears on the phantom but not on Fashion-MNIST.
A prior trained on Fashion-MNIST beats TV on Fashion-MNIST ($+2.2$\,dB) and, with the same tuned hyperparameters, on MNIST ($+1.4$\,dB) and EMNIST letters ($+0.4$\,dB), and is close to TV on the phantom ($-0.2$\,dB). What matters is therefore how the training distribution relates to the test distribution; that the Fashion-trained prior transfers better is observed, but its cause (for example a greater variety of grey levels and shapes) is untested.
On real CT slices a CT-trained prior tested on CT has no consistent accuracy advantage over TV-long to begin with. On non-anatomical phantoms its detection advantage over TV-long is significant in one of six conditions, against three of the six corresponding CT conditions, and its SSIM deficit against TV-long grows from $0.004$--$0.027$ to $0.095$--$0.190$, while its PSNR deficit stays below $0.75$\,dB (Section~\ref{sec:e19}).

\paragraph{(2) Can we find lesion-like structure that the ground truth does not support?}
Yes, in some settings. For the MNIST-trained prior on Fashion-MNIST it produces $0.20$--$0.66$ false lesions per image, more than TV on seven of eight lesion types even against the box-constrained TV (Section~\ref{sec:lesion-results}); under Poisson noise the same holds for lesions with $\sigma_l\ge0.75$\,px in 11 of 12 dose and lesion-type cells, against the unconstrained TV only (Section~\ref{sec:lesion-poisson}). On non-anatomical phantoms at quarter dose it adds a few false blobs per phantom relative to FBP-Hann (significant at $K=120$ at both response thresholds, and at $K=30$ only for $\gamma=1.5$) and, at $K=120$, more than TV (Section~\ref{sec:e19}).
In distribution the picture is different. On digits with Gaussian noise it produces at most $0.01$ false lesions per image ($0.06$--$0.07$ for some lesion types at $30$ photons per ray) but attenuates inserted lesions. On real slices it produces $0.24$--$1.38$ false blobs per slice, fewer than FBP-Hann in the four conditions where this was tabulated, and the strongest blobs lie mostly where FBP-Hann and TV also produce them (Section~\ref{sec:e18}).
The false-blob count of TV is itself unstable: it varies $26$-fold across the phantom conditions and for the smallest lesions exceeds that of the learned prior (Sections~\ref{sec:lesion-results},~\ref{sec:lesion-poisson},~\ref{sec:e19}), so ``more than TV'' is a statement about a specific PSNR-tuned TV solver.
We therefore do not claim that the prior creates false structure in general, nor that it is safe: its behaviour moves with noise level, lesion scale and condition, and 40 patients with one noise draw per condition cannot bound rare events.

\paragraph{(3) Do the conclusions hold across noise levels and conditions?}
Only in part. The in-distribution advantage is largest at moderate noise ($+3.7$\,dB at $100$--$300$ photons per ray) and shrinks at both ends of the dose range (Section~\ref{sec:dose-results}). At $\sigma\le0.01$ the digit experiments do not determine the ranking, which changes with the data model by up to $5.5$\,dB (Section~\ref{sec:crime-results}), so we make no claim there. Under angular mismatch the learned prior stays ahead throughout the tested range, but its margin falls from $+3.86$ to $+2.98$\,dB for $4^\circ$ jitter and to $+1.49$\,dB for a $4^\circ$ offset; the latter is mostly a rigid rotation of the reconstruction that a pixel-wise metric penalises (Section~\ref{sec:e3}), so we do not read it as evidence that the learned prior is the less robust method.
On CT the signs of the PSNR, SSIM and false-blob differences against TV-long change with dose and number of angles without a monotone trend (Sections~\ref{sec:e16},~\ref{sec:e18}). The sign of the detection advantage over TV-long is the same in all nine conditions (significant in six), but it is specific to the 300-iteration, PSNR-tuned TV: against the 150-iteration TV it is negligible at full dose for $K=30$ and $60$ ($+0.006$ and $0.000$; Section~\ref{sec:e17}).

\paragraph{Accuracy depends on the operating regime and on the baseline.}
A single accuracy number for a learned reconstruction method is uninformative without its regime, data model and baseline. The headline gap on digits moves by about $0.5$\,dB with the choice between unconstrained and box-constrained TV and, at low noise, changes sign with the data model. On CT it depends on the iteration count given to TV and on tuning details: the apparent deficit of PnP at $K=60$ fell from $0.85$ to $0.17$\,dB at full dose when the tuning set grew from 6 to 20 slices and the starting image was changed (Section~\ref{sec:history3}).
On clinical-size slices the advantage found on digits disappears: PnP-const differs from TV-long by $-0.93$ to $+0.33$\,dB, with a lower SSIM in eight of nine conditions (significantly in seven). Both iterative methods improve on FBP-Hann ($0.9$--$1.4$\,dB for PnP-const, $0.8$--$2.3$\,dB for TV-long), so the PSNR gain over FBP-Hann is not specific to the learned prior (Section~\ref{sec:e16}). The best PnP variant, started from the TV solution, is highest in PSNR in seven of nine conditions by at most $0.49$\,dB, but the variants were introduced after the first PnP-const result, and two of its tuned settings lie on a grid boundary, one of them with almost no denoising ($\sigma_d=0.0039$ at $K=120$, full dose; Table~\ref{tab:params3}; Sections~\ref{sec:e15},~\ref{sec:e16}). TV-long and the PnP variants are not matched in cost or in tuned degrees of freedom.
We did not test why the digit and CT results differ. Candidate explanations are the more varied anatomy, which makes TV a weaker model than for digits, and a small denoiser trained once for $15\,000$ steps on $8\,960$ of $35\,820$ slices (a 17-layer variant did not help at $K=60$), so that the CT result cannot separate the prior from this network.

\paragraph{The iteration count acts as a regularisation parameter.}
In PnP-PGD with an unconstrained denoiser the accuracy peaks after a handful of iterations and then degrades, while the data residual stays at the noise level ($0.994$ to $1.019$ times its expected value between the selected iterate and $k=400$; Section~\ref{sec:e5}). Of the error growth, $74\%$ lies in directions with $0.1\le s_i/s_1<0.3$. Without the denoiser, the gradient step with $\eta=1/L$ would contract the distance to the data-fitting solution in direction $i$ by $(1-(s_i/s_1)^2)^k$, which is slow in these directions. The selected $k$ therefore determines how much of them is left to the denoiser, and it is a regularisation hyperparameter, not an implementation detail.
That the degradation is semi-convergence, in which further iterations fit noise in poorly determined directions, is our interpretation and not a tested mechanism: the selected $k$ is also $5$ at $\sigma=0$ and we did not record the iteration curve on noise-free data. The behaviour is consistent with the observation that convergence guarantees for PnP need constraints on the denoiser~\citep{ryu2019pnp}; we did not measure the Lipschitz constant of ours. In distribution the converged iterate is still better than TV on MNIST, whereas under shift its error grows to $3.7$--$4.8$ times that of TV (Section~\ref{sec:nullspace-results}).

\paragraph{The reversal under shift is a property of the prior, but not of the null space alone.}
The matched-prior control isolates the denoiser: in all runs the tuned hyperparameters are identical and only the weights differ (Section~\ref{sec:seeds-results}). The effect exceeds the seed-to-seed variation ($\le0.3$\,dB) by an order of magnitude and persists under four data models. It is not removed by re-tuning $(\sigma_d,k)$ within our grids: on Fashion-MNIST none of $45$ PnP settings reaches TV's PSNR, for one denoiser (Section~\ref{sec:oppoint-results}).
The error decomposition adds that the reversal is not simply null-space invention: $93\%$ (Fashion-MNIST) and $89\%$ (phantom) of PnP's excess error over TV lies outside the weakly determined subspace, and a signature of a prior that fills that subspace with up to three times the true energy, misaligned with the truth, appears only on the phantom (Section~\ref{sec:nullspace-results}). ``Hallucination'' in the null-space sense is therefore an incomplete description of the failure.

\paragraph{PSNR is not task performance, in either direction.}
In distribution on digits the PSNR-tuned PnP reconstruction is better than TV by about $4$\,dB but detects small inserted lesions less reliably (paired AUC $0.04$--$0.23$ below TV, also below the box-constrained TV), because it attenuates them (Section~\ref{sec:lesion-results}). Re-tuning trades attenuation against false lesions: for the MNIST-trained prior no operating point is competitive with TV on PSNR, detection and false lesions at once, on MNIST or on Fashion-MNIST. The Fashion-trained prior does leave such a point on both Fashion-MNIST and MNIST (about $+1.0$\,dB, a paired AUC $0.02$--$0.06$ below TV and few false lesions). We hypothesise that this reflects the variety of its training distribution and not whether a test set is formally in distribution; with one denoiser per prior this is untested (Section~\ref{sec:oppoint-results}).
On CT the sign is reversed: PnP detects better than TV-long in all nine conditions yet no better than FBP-Hann, and the PSNR-tuned TV-long, $0.8$--$2.3$\,dB better than FBP-Hann in PSNR, detects worse than it in all nine (significantly in seven) (Section~\ref{sec:e17}). A comparison reporting only PSNR or SSIM would have ranked the methods differently from this detection statistic, whichever method it favours.
We did not test why the sign of the detection difference differs between digits and CT; the lesion scale in pixels, the noise texture and the smoothing of the PSNR-tuned operating points are candidates. The CT operating points were tuned for PSNR, the lesions are small and of high contrast, and the observer knows their location, so the conclusions for detection-tuned operating points, low-contrast lesions, other observers and search tasks are untested.

\paragraph{Evaluation protocol changes the answer.}
Besides the choice of TV and its iteration count, three protocol choices affected results by amounts comparable to or larger than some of the differences we interpret. The data model (inverse crime or finer grid) changes the low-noise gap by up to $5.5$\,dB. The seed range of the Fashion-MNIST gap ($-2.83$ to $-3.43$\,dB) is of the size of the $0.5$\,dB box-constraint effect, although small against the shift effect itself. And choosing the lesion operating points on evaluation images would flatter re-tuning: the best of the $45$ settings with PSNR at least TV's, chosen on the test set, reaches a paired AUC of $0.898$, against $0.849$ for the point chosen on held-out validation images (Table~\ref{tab:oppoint}).
The shift sensitivity we observe is of the kind reported for other learned reconstruction methods~\citep{darestani2021shift}; those studies concern other methods and we make no comparison. Replacing the dense operator by a matrix-free projector with an exact adjoint changes no reconstruction at $28\times28$ by more than $3\times10^{-4}$ in relative norm, and replacing Gaussian by Poisson noise leaves the qualitative picture of the digit experiments unchanged (Sections~\ref{sec:mf-results},~\ref{sec:dose-results}); the Poisson-weighted data term showed no consistent benefit, so we claim only that the unweighted scheme is adequate in this setting.
We suggest, as practices and not as tested conclusions, that studies of learned priors report: a baseline with the same constraints and an iteration count tuned (or compute matched) like the learned method's, hyperparameters tuned on held-out data for every method, several training seeds, at least one mismatched and one matched prior, a task-based metric next to PSNR and SSIM, and an explicit data model.

\section{Conclusion}
\label{sec:conclusion}

We compared classical and learned regularization for sparse-view CT, using a noise-conditioned DnCNN in Plug-and-Play proximal gradient against FBP, Tikhonov and TV, under noise, undersampling, forward-model mismatch, distribution shift and a lesion-detection task, first on $28\times28$ digits with Gaussian and Poisson noise and then on real LoDoPaB-CT slices.

On digits the learned prior is about $3.4$\,dB better than box-constrained TV at moderate noise ($+3.8$\,dB over TV-ADMM). The advantage shrinks at both ends of the noise range: at $\sigma=0$ TV is ahead under every data model tested, and at $\sigma=0.01$ the ranking depends on the data model and the TV solver. The advantage halves for a near shift (EMNIST letters) and reverses for a far shift (Fashion-MNIST, Shepp--Logan phantom). The matched-prior control, in which only the denoiser weights change, shows that the reversal is a property of the prior and not of the PnP scheme, and the error decomposition shows that most of the excess error lies in the well-determined directions, not only in the null space.
PSNR is not task performance: in distribution the PSNR-tuned prior attenuates small lesions and detects them less reliably than TV, and under mismatch it creates more false lesions than box-constrained TV for seven of eight lesion types. Re-tuning trades these against each other, and an operating point competitive with TV on PSNR, detection and false lesions exists only for the Fashion-trained prior, which we hypothesise reflects the variety of its training data (untested with two priors).
The number of iterations acts as a regularisation parameter: the data residual stays at the noise level while the error grows in weakly determined directions, which we interpret as semi-convergence but did not test.

On real CT slices the PnP prior does not consistently improve on a well-tuned TV run for 300 iterations in PSNR (differences between $-0.93$ and $+0.33$\,dB over nine dose and angle conditions) and has a lower SSIM in eight of nine; the two are not matched in cost or in tuned parameters. It detects small high-contrast lesions better than that TV in all nine conditions (significant in six, negligible against the 150-iteration TV at full dose for $K=30$ and $60$) but not better than unregularised FBP, whereas the PSNR-tuned TV detects them worse than FBP in all nine. Its false-blob counts on real slices are below those of FBP-Hann in the four conditions tabulated and not systematically different from those of TV-long. Under a shift to non-anatomical phantoms its PSNR and SSIM deficit against TV-long is consistent, its detection advantage is significant in one of six conditions, and its false-blob count relative to TV changes sign with the number of angles.

Our conclusions concern one small denoiser architecture, two training distributions on digits and one CT training run. They are conditional on the regime, the data model, the TV variant, the tuning protocol and a location-known detection statistic at PSNR-tuned operating points, and the CT analysis was developed after seeing results, without a pre-specified primary comparison.
They suggest evaluating learned priors against baselines with the same constraints and a tuned or compute-matched iteration count, with hyperparameters tuned on held-out data, several training seeds, mismatched and matched priors, and task-based metrics next to PSNR and SSIM.
Future work should designate the primary analyses in advance and use all 60 test patients with several noise realisations; add paired intervals for the digit detection differences; check the PnP--TV difference on real data and not only the TV baseline; tune TV and PnP for detection on CT; use a finer TV and Hann grid, a compute-matched TV, several denoiser seeds, stronger denoisers and the full training set; run box-constrained TV in every experiment; and use clinically calibrated lesion contrasts with model observers, re-tuning under forward-model mismatch, an iteration curve on noise-free data, and provably convergent PnP variants.

\section*{Limitations}

\paragraph{Scope.} The digit experiments are intentionally small ($28\times28$) so that a thorough grid is possible; trends need not transfer to clinical resolution. The CT study uses one small denoiser (8 layers, one training run of $15\,000$ steps on $8\,960$ of the $35\,820$ training slices, no seed variation; a 17-layer variant was tested at $K=60$ only), so it cannot separate the prior from this network, and stronger or longer-trained PnP denoisers were not tried. Only two training distributions were compared on digits, so the explanation of the asymmetric transfer between them is a hypothesis. The PnP scheme has no convergence guarantee, we did not measure the Lipschitz constant of the denoiser, and only moderate forward-model mismatches (angles up to $4^\circ$, one jitter draw per level) were evaluated, with hyperparameters frozen at their nominal values.

\paragraph{Baselines and tuning.} PnP has two tuned hyperparameters and TV one (88 against 17 grid points on digits, 40 against 4 on CT), and the TV iteration count is fixed rather than tuned. TV-long, the CT comparator with more iterations, is not matched to PnP in cost or in tuned degrees of freedom, and it was introduced after the first lesion results. The box-constrained TV control (E2b) was run for E1--E2, E4, E6 (inverse-crime and bicubic data only), E8 (the MNIST- and Fashion-trained families, not the second MNIST family) and E9; the TV baseline of E3, E5, E7, E10 and of the Poisson study (E12--E13) is the unconstrained TV-ADMM, so the corresponding gaps may be overstated; in the Gaussian MNIST case the box constraint lowered the gap by about $0.5$\,dB (E8). On CT, TV is tuned on a grid with steps of a factor three and TV-long is sometimes worse than TV. PSNR was the tuning criterion for every method, which favours accuracy over detectability, and no detection-tuned operating point was explored on CT. Two optima of PnP-TVinit lie on the boundary of the grid, and one tuned $\sigma_d$ ($0.128$) lies above the training range $(0,0.1)$ of the CT denoiser. Several tuned PnP-TVinit settings amount to almost no denoising ($\sigma_d=0.0039$ at $K=120$, full dose), so part of its gain over TV comes from additional data-consistency iterations and not from the prior. The PnP variants of E15 were introduced after the first PnP-const result.

\paragraph{Data and simulation.} E3, E5 and the original E4 use data generated with the reconstruction operator (an inverse crime). E3 and E5 were not repeated on finer-grid data. The finer-grid simulation reduces the inverse crime but does not remove it (a mean model error of about $6\%$ of the noise norm at $\sigma=0.05$), and the sensitivity band of Table~\ref{tab:crime} shows that the near noise-free ranking depends on this choice, which we adopted as a convention after seeing all variants. The Poisson study uses a single attenuation scale ($\mu=2$), one angle count and one MNIST-trained denoiser in the dose sweep, and its Shepp--Logan result is one image with one noise draw per dose. The matrix-free operator was validated against the dense operator only at $28\times28$ (agreement limited by single-precision arithmetic, timings from single runs). On CT, hyperparameters are tuned on real thinned observations while every comparison is evaluated on simulated data. The simulator was checked against real data on 8 slices and one noise draw per condition for the TV baseline (and on 40 slices at $K=60$ for TV and FBP-Hann), the sign of the differences differs between methods, and we did not verify that the PnP--TV difference agrees between real and simulated data.
LoDoPaB's ground truth is itself a reconstruction, so ``false'' blobs are only unsupported by this reference.

\paragraph{Statistics.} Seed variation comes from three denoisers per digit configuration, which gives only a crude standard deviation; the CT study has no seed variation. It uses 40 of the 60 test patients (one slice each) and one noise realisation per condition, so a single cell has an AUC uncertainty of about $\pm0.05$, and the conditions share the same patients and are not independent. The digit lesion experiments report no paired intervals (the binomial standard error of a paired AUC is about $0.02$--$0.04$). No analysis was designated primary in advance and no correction for multiplicity was applied. The analysis was developed in steps after seeing results (Section~\ref{sec:history3}): the false-blob definition, the threshold sweep, the PnP variants, TV-long as comparator and the shift extension to $K=30,120$ were not pre-specified, and the Poisson lesion experiment was run at only two of the five doses ($I_0=100$ and $30$). Interval conventions differ between tables (standard errors, bootstrap intervals, standard deviations over seeds) and are stated in each caption.

\paragraph{Task-based evaluation.} The lesions are bright additive Gaussian blobs (on CT of 120--240\,HU under an assumed mapping from $\mu/\mu_{\max}$ to HU), the observer is a location-known local-contrast statistic that rewards suppression of background fluctuations at the lesion scale, false blobs are defined by this statistic and a threshold, and no reader study was done, so false-lesion rates are meaningful for comparing methods but not as absolute rates. The observability analysis is confounded with lesion size and location. The operating-point study (E10) uses one denoiser per prior, two lesion types and a coarse grid, and the galleries of false blobs show the worst cases selected by the methods' own responses, not a random sample.

\paragraph{Shift.} The shift study covers a single kind of shift: digits to other digits and a phantom, and CT slices to 20 random piecewise-constant phantoms, which favour TV; no change of scanner geometry, dose model or anatomy was tested. MNIST is close to binary, whereas Fashion-MNIST and the phantoms contain intermediate grey levels, so structural and intensity statistics are confounded in the digit shifts. The $K=60$ shift analysis used four lesion types and the other conditions two, and SSIM on the phantoms at $K=60$ was not stored. The error-decomposition statistics cannot separate prior-driven structure from amplified noise except in comparisons with identical noise and settings.

\section*{Reproducibility}
\addcontentsline{toc}{section}{Reproducibility}

The development repository is
\url{https://github.com/AboOraig/learned-regularization}. The repository contains a
\texttt{README}, a dependency file (\texttt{requirements.txt})
and 68 unit tests (\texttt{python -m pytest tests}).

\section*{Acknowledgments}

The CT ground truth derives from the LIDC-IDRI database; we acknowledge the National Cancer Institute and
the Foundation for the National Institutes of Health, and their critical role in the creation of this
publicly available database, and the authors of LoDoPaB-CT for the dataset and its simulation code.

\section*{Use of AI tools}

An AI assistant (Claude, Anthropic) was used to help develop the code and the experimental design and, in a later stage, to review and edit the text of this manuscript. The research
questions, the choice of experiments, the interpretation of results and the conclusions are the author's.

\bibliographystyle{plainnat}
\bibliography{references}

\end{document}